\documentclass[12pt,epsfig]{article}

\usepackage{amssymb,epsfig}

\usepackage{psfrag}

\psfrag{kt}{{\footnotesize $k_\perp$}}
\psfrag{=}{{\footnotesize $=$}}
\psfrag{+}{{\footnotesize $+$}}
\psfrag{l}{{\footnotesize $l$}}
\psfrag{p1z'}{{\footnotesize $p_1z'$}}
\psfrag{q1}{{\footnotesize $q_1$}}
\psfrag{p1}{{\footnotesize $p_1$}}
\psfrag{-p1bz'}{{\footnotesize $-p_1{\bar z}'$}}
\psfrag{-q2}{{\footnotesize $-q_2$}}
\psfrag{x1p2}{{\footnotesize $x_1p_2$}}
\psfrag{x2p2}{{\footnotesize $x_2p_2$}}
\psfrag{p2}{{\footnotesize $p_2$}}
\psfrag{p2'}{{\footnotesize $p_2'$}}
\psfrag{kpx1p2}{{\footnotesize $k_\perp, x_1$}}
\psfrag{kpx2p2}{{\footnotesize $k_\perp, x_2$}}
\psfrag{pi}{{\footnotesize $\pi$}}
\psfrag{aa}{{\footnotesize $(a)$}}
\psfrag{bb}{{\footnotesize $(b)$}}
\psfrag{cc}{{\footnotesize $(c)$}}
\psfrag{dd}{{\footnotesize $(d)$}}
\psfrag{ee}{{\footnotesize $(e)$}}
\psfrag{ff}{{\footnotesize $(f)$}}
\psfrag{gg}{{\footnotesize $(g)$}}
\psfrag{Fipi}{{\footnotesize $\phi_\pi(z')$}}
\psfrag{Fchi}{{\footnotesize $F_\xi(x_1)$}}

\psfrag{I2}{{$|{\cal I}|^2$}}
\psfrag{zz}{{$z$}}
\psfrag{zeta}{{$\zeta$}}
\psfrag{TH}{\footnotesize $T_H(z',x_1)$}

\newcommand{\beq}[1]{
\begin{equation}\label{#1}}
\newcommand{\eeq}{\end{equation}}
\newcommand{\bea}[1]{
\begin{eqnarray}\label{#1}}
\newcommand{\eea}{\end{eqnarray}}
\newcommand\re[1]{(\ref{#1})}
\newcommand{\al}{\alpha}

\renewcommand{\theequation}{\thesection.\arabic{equation}}

\begin{document}

\begin{titlepage}

\begin{flushright}
\begin{tabular}{l}
 hep-ph/0204191
\end{tabular}
\end{flushright}
\vspace{1.5cm}

\begin{center}
{\LARGE \bf
Towards the theory of coherent hard dijet production on hadrons and nuclei}
\vspace{1cm}

{\sc V.M.~Braun}${}^1$,
{\sc D.Yu.~Ivanov}${}^{1,2}$
{\sc A.~Sch\"afer}${}^1$ and
{\sc L.~Szymanowski}${}^{3,4}$
\\[0.5cm]
\vspace*{0.1cm} ${}^1${\it
   Institut f\"ur Theoretische Physik, Universit\"at
   Regensburg, \\ D-93040 Regensburg, Germany
                       } \\[0.2cm]
\vspace*{0.1cm} ${}^2$ {\it
Institute of Mathematics, 630090 Novosibirsk, Russia
                       } \\[0.2cm]
\vspace*{0.1cm} ${}^3${\it
    CPhT, \'Ecole Polytechnique, F-91128 Palaiseau, 
    France\footnote{Unit{\'e} mixte C7644 du CNRS.}
                       } \\[0.2cm]
\vspace*{0.1cm} ${}^4$ {\it
 Soltan Institute for Nuclear Studies,
Hoza 69,\\ 00-681 Warsaw, Poland
                       } \\[1.0cm]

\vspace{0.6cm}  
\bigskip  
\centerline{\large \em \today}  
\bigskip  
\vskip1.2cm
{\bf Abstract:\\[10pt]} \parbox[t]{\textwidth}{
  We carry out a detailed calculation of the cross section of pion 
  diffraction dissociation into two jets with large transverse momenta, 
  originating from a hard gluon exchange between the pion constituents.
  Both the quark and the gluon contribution are considered and in the 
  latter case we present calculations both in covariant and in axial gauges.
  We find that the standard collinear factorization does not 
  hold in this reaction. The structure of non-factorizable contributions 
  is discussed and the results are compared with the experimental data.
  Our conclusion is that the existing theoretical uncertainties do not 
  allow, for the time being,  for a quantitative extraction of the pion 
  distribution amplitude. 
}
  \vskip1cm  
{\em Submitted to Nuclear Physics B }\\[1cm] 
\end{center}

\end{titlepage}

{\tableofcontents}

\newpage

\section{Introduction}  
\setcounter{equation}{0}  

After thirty years of Quantum Chromodynamics, many aspects of 
hadron structure remain poorly understood. The bulk of the existing 
experimental information 
comes from parton distributions  that can be interpreted as one-particle 
probabilities to find quarks, antiquarks and gluons carrying certain    
momentum fractions of the parent hadron. The parton distributions 
are inclusive 
quantities, in the sense that contributions of different parton states are 
summed over, and therefore provide us only with  global information about the 
hadron wave function. It is well known that hard exclusive 
reactions allow, in principle, to separate contributions of different Fock 
states and study the momentum fraction distributions of the 
components with the minimum number of Fock constituents at small transverse 
separations, dubbed distribution amplitudes. In practice, however, 
progress in this direction had been limited, due to both experimental 
difficulties to isolate exclusive amplitudes and theoretical 
problems to formulate their quantitative description.  

Classical applications of QCD to exclusive 
reactions addressed electromagnetic form factors 
\cite{earlyCZ,earlyBL,earlyER}. Nowadays it is almost generally accepted,
however, that the asymptotic behavior of form factors is only achieved   
for very large momentum transfers. In addition, the form factors
involve a convolution (overlap integral) of distribution amplitudes  
which makes it difficult to extract them directly from the data. 
In recent years, the list of applications of QCD factorization to 
hard exclusive reactions have been increased significantly, with 
hard exclusive meson production \cite{BFGMS94,CFS96} and deeply-virtual Compton
scattering (DVCS) \cite{Ji97,Rad96,CF98} providing notable examples.   
This raises hopes that suitable hard processes can be found in which 
hadron distribution amplitudes can be studied in a more direct way.

In particular, coherent diffractive production of dijets by incident 
pions (or photons)  on nuclei has attracted a lot of attention.  
The subject of diffraction is very old. Ever since the classic works
in early 50' on the diffraction breakup of deuterons \cite{dif_classic}
it was known that the momentum distribution of the proton and the neutron
in the final state is close to their momentum distribution as deuteron 
constituents. More recently, the same idea gave rise to the method
of so-called Coulomb Explosion Imaging \cite{VNK89} which is widely 
used to study the high-momentum tails of wave functions of small molecules.
To our knowledge, the pion (and photon) diffraction dissociation into a 
pair of jets with large transverse momentum on a nucleon target 
was first discussed in \cite{KDR80}. In Ref.~\cite{BBGG81} the possibility
to use this process to probe the nuclear filtering of pion components 
with a small transverse size was suggested. The A-dependence and the 
$q_\perp^2$-dependence of the 
coherent dijet cross section was first calculated 
in \cite{FMS93}. In the same work it was argued that the jet
distribution with respect to  the longitudinal momentum fraction has to 
follow the quark momentum distribution in the pion and hence provides 
a direct measurement
of the pion distribution amplitude. Recent experimental data 
by the E791 collaboration \cite{E791a,E791b} indeed confirm the strong 
A-dependence which is  a signature for color transparency, and are consistent 
with the predicted $\sim 1/q_\perp^8$
dependence on the jet transverse momentum. Moreover, the jet longitudinal 
momentum fraction distribution turns out to be consistent with the 
$\sim z^2(1-z)^2$ shape corresponding to the asymptotic pion distribution
amplitude which is also supported by an independent measurement of the pion 
transition form factor $\pi\gamma\gamma^*$ \cite{CLEO}.       
  
After these first successes, one naturally asks whether the 
QCD description of coherent dijet production can be made fully quantitative. 
Two recent papers \cite{NSS99} and \cite{FMS00} address this question, 
with contradictory conclusions. In the present work we attempt to clarify 
the situation and develop a perturbative QCD framework for the description
of coherent dijet production that would be in line with other known 
applications of the QCD factorization techniques. Some of the results 
reported in this paper were published earlier in a 
letter format~\cite{BISS01}. An approach close to ours was suggested 
independently in \cite{Che01}, \cite{CG01}. We disagree with 
\cite{Che01,CG01} on several important issues and the reasons for this 
disagreement will be elucidated in what follows. 

\begin{figure}[t]
\centerline{\epsfxsize7.0cm\epsffile{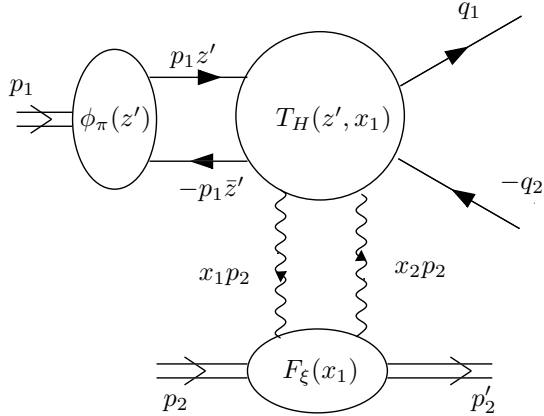}}
\caption[]{\small 
Kinematics of the coherent hard dijet production $\pi\to 2\, {\rm jets}$.  
The hard scattering amplitude $T_H$ contains at least one hard gluon 
exchange.
 }
\label{fig:1}
\end{figure}

The kinematics of the process is shown in Fig.~\ref{fig:1}.
For definiteness, we consider $\pi^-$ scattering from the proton target.
%
%
The momenta of the incoming pion, incoming nucleon and the outgoing nucleon  
are $p_1, p_2$ and $p_2^\prime$, respectively. 
The pion and the nucleon masses are both neglected, 
$p_1^2=0$, $p_2^2=(p_2^\prime)^2=0$. 
We denote the momenta of the outgoing quark and antiquark (jets) 
as $q_1$ and $q_2$, respectively. They are on the mass shell, 
$q_1^2=q_2^2=0$.
We will use the Sudakov decomposition of 4-vectors with respect to 
the momenta of the incoming particles $p_1$ and $p_2$.
For instance, the jet momenta  are decomposed according to
\beq{sudakov}
q_1=zp_1+\frac{q_{1\perp}^2}{zs}p_2+q_{1\perp} \ ,
q_2=\bar zp_1+\frac{q_{2\perp}^2}{\bar zs}p_2+q_{2\perp}
\eeq
such that $z$ is the longitudinal momentum fraction of the quark 
jet in the lab frame.
We will often use the shorthand notation: $\bar u \equiv (1-u)$ for any
longitudinal momentum fraction $u$.
The Dirac spinors for the quark and the antiquark are denoted by   
$\bar u(q_1)$ and $v(q_2)$. 

We are interested in the forward limit, when the transferred momentum 
$t=(p_2-p_2^\prime)^2$ is equal to zero%
\footnote{If the target mass $m$ is taken into account, the momentum 
transfer $t=(p_2-p_2^\prime)^2$ contains a non-vanishing longitudinal 
contribution and is constrained from below by  
$|t|\geq t_0$, where $t_0=({\displaystyle m^2M^4})/{\displaystyle 
(s-m^2)^2}$, 
$M^2$ being the invariant mass of the dijet.}, 
and the transverse momenta of jets compensate each other
$q_{1\perp} \equiv q_{\perp}$, $q_{2\perp}\equiv -q_\perp$.
In this kinematics the invariant mass of the produced $q\bar q$ pair is
equal to 
$
M^2={\displaystyle q_{\perp}^2}/{\displaystyle z\bar z}
$.
The invariant c.m. energy $s=(p_1+p_2)^2=2p_1 p_2$ 
is taken to be much larger than the transverse jet momentum 
$q_{\perp}$. In what follows we often neglect
contributions to the amplitude that are suppressed by powers of 
$q_\perp^2/s$.   

{}From the theoretical point of view, the principal question is whether 
the relevant transverse size of the pion $r_\perp$ remains small,
of the order of the inverse transverse momenta of the jets 
$r_\perp  \sim 1/q_\perp$. 
In this paper we investigate the possibility that the amplitude 
of hard dijet coherent production can be calculated using the 
standard collinear factorization in the form suggested by  Fig.~\ref{fig:1}:
\beq{factor}
{\cal M}_{\pi \to 2\, {\rm jets}}
=\sum_{p=q,\bar q, g} \int\limits^1_0 dz^\prime \int\limits^1_0 dx_1 \,\phi_\pi
(z^\prime,\mu_F^2)\,T^p_H(z^\prime , x_1, \mu_F^2)\,F^p_\zeta(x_1, \mu_F^2)\,.
\eeq     
Here  $\phi_\pi (z^\prime,\mu_F^2)$ is the pion distribution amplitude,
and $F^p_\zeta(x_1, \mu_F^2)$ is the generalized (skewed) parton 
 distribution $p=q,\bar q,g$ 
\cite{early_skewed,Rad96a,Ji97a} in the target nucleon or nucleus; 
$x_1$ and $x_2=x_1-\zeta$ are the 
momentum fractions of the emitted and the absorbed partons, respectively.
The asymmetry parameter $\zeta$ is fixed by the process kinematics:
\beq{zeta}
       \zeta = M^2/s ={\displaystyle q_{\perp}^2}/{\displaystyle z\bar z s}\,. 
\eeq
$T_H(z^\prime , x_1, \mu_F^2)$ is the hard scattering amplitude involving at 
least one hard gluon exchange and $\mu_F$ is the (collinear) factorization 
scale. By definition, the pion distribution amplitude only involves 
small momenta, $k_\perp < \mu_F$, and the hard scattering amplitude 
is calculated  neglecting the parton transverse
momenta. In this paper
we present a detailed calculation of the leading-order contribution to  
$T_H(z^\prime , x_1, \mu_F^2)$ corresponding to a single hard gluon exchange.

We consider both the quark and the gluon contribution to the amplitude,
and in both cases find that the corresponding hard kernels $T^q_H$, $T^g_H$ 
diverge as $1/z'^2$ and $1/\bar z'^2$ in the $z'\to0$ and $z'\to1$ limit, 
respectively. This implies that the integration of the pion momentum 
fraction diverges at the end-points and the collinear factorization is,
therefore, broken. Physically this means that the approximation of neglecting
the incident quark transverse momenta becomes insufficient close to the 
end points, similar to what happens e.g. in the heavy quark limit in 
B-decays \cite{SHB80}, albeit for a different reason. One may argue 
that Sudakov-type radiative corrections should suppress the end-point
contributions and try to develop a modified factorization framework, 
by the resummation of soft gluon emission to all orders \cite{Sudakov}.
Alternatively, the end-point singularities may be softened on a nuclear
target because of the color filtering of configurations with a large 
transverse size.
Both possibilities 
are interesting and require detailed studies that go far beyond the tasks
of this paper. We will rather assume that the end-point behavior can be 
tamed in some reasonable way, and examine the consequences.   

In particular, we will analyze the structure of the hard gluon exchange 
contribution in some detail. We will find that the structure of this 
exchange is such that it generates an enhancement by a logarithm 
of the energy in the region $|z'- z| \ll 1$. If only this logarithmic 
contribution is retained, the collinear factorization is restored and the 
longitudinal momentum distribution of the jets to this accuracy indeed 
follows the shape of the pion distribution amplitude \cite{FMS93}. The 
hard gluon exchange can in this case be considered as a part of the 
unintegrated gluon distribution, as advocated in \cite{NSS99}.
Beyond the leading logarithms in energy this proportionality does not hold.
Remarkably enough, we find that the longitudinal momentum fraction 
distribution of the jets for the non-factorizable contribution is calculable,
and turns out to be the same as for the factorizable contribution with 
the asymptotic pion distribution amplitude. We also find, in agreement 
with \cite{CG01} that the quark contribution is significant in the energy 
range of the E-791 experiment and present a new data analysis including all 
contributions.   
On the technical side, we present a detailed study of the light-cone limit 
of the relevant amplitudes, the structure of different absorption
parts and the gauge-dependence. We believe that some of this discussion
is of general interest and relevant for all exclusive processes in the 
diffractive kinematics. 

The presentation is organized as follows. In Sec.~2 we consider the quark
contribution to the coherent dijet production, which is simpler than the 
gluon contribution as it does not involve subtleties related to
the choice of the gauge. We find an end-point singularity in the integration
over the pion quark (antiquark) momentum fraction and trace its origin to
pinching of the integration contour over the partonic momentum fraction
in the so-called Glauber region. We explain this result by analyzing 
the structure of different dispersion parts of the amplitude.
The gluon contribution is considered in Sec.~3. We point out that
there is a subtlety in the definition of the generalized gluon distribution,
and explain how this subtlety is resolved in a more familiar case 
of hard electroproduction of vector mesons. Next, we present a calculation
of the imaginary part of the amplitude of the coherent dijet production
from a quark, in which the above-mentioned problem is avoided. 
Finally, we perform  the calculation of the full amplitude (including
the real part) using axial gauge and not assuming the light-cone 
dominance from the beginning. Similar as in the quark case, we observe 
pinching of singularities between soft gluon interactions in the initial 
and the final states. A detailed numerical analysis and the comparison 
to the E791 data is presented in Sec.~4, while Sec.~5 is reserved
for the conclusions. 

One technical remark is in order. Calculations in the main text are presented
using Radyushkin's definitions \cite{Rad96,Rad96a} 
for the generalized parton distributions. The relation to the more commonly
used  symmetric notation by Ji \cite{Ji97a} is explained in App.~A and 
in App.~B we collect all the final expressions in the symmetric notation.

\section{The Quark Contribution}
\setcounter{equation}{0}

\subsection{The Calculation}

\begin{figure}[t]
\centerline{\epsfxsize7.0cm\epsffile{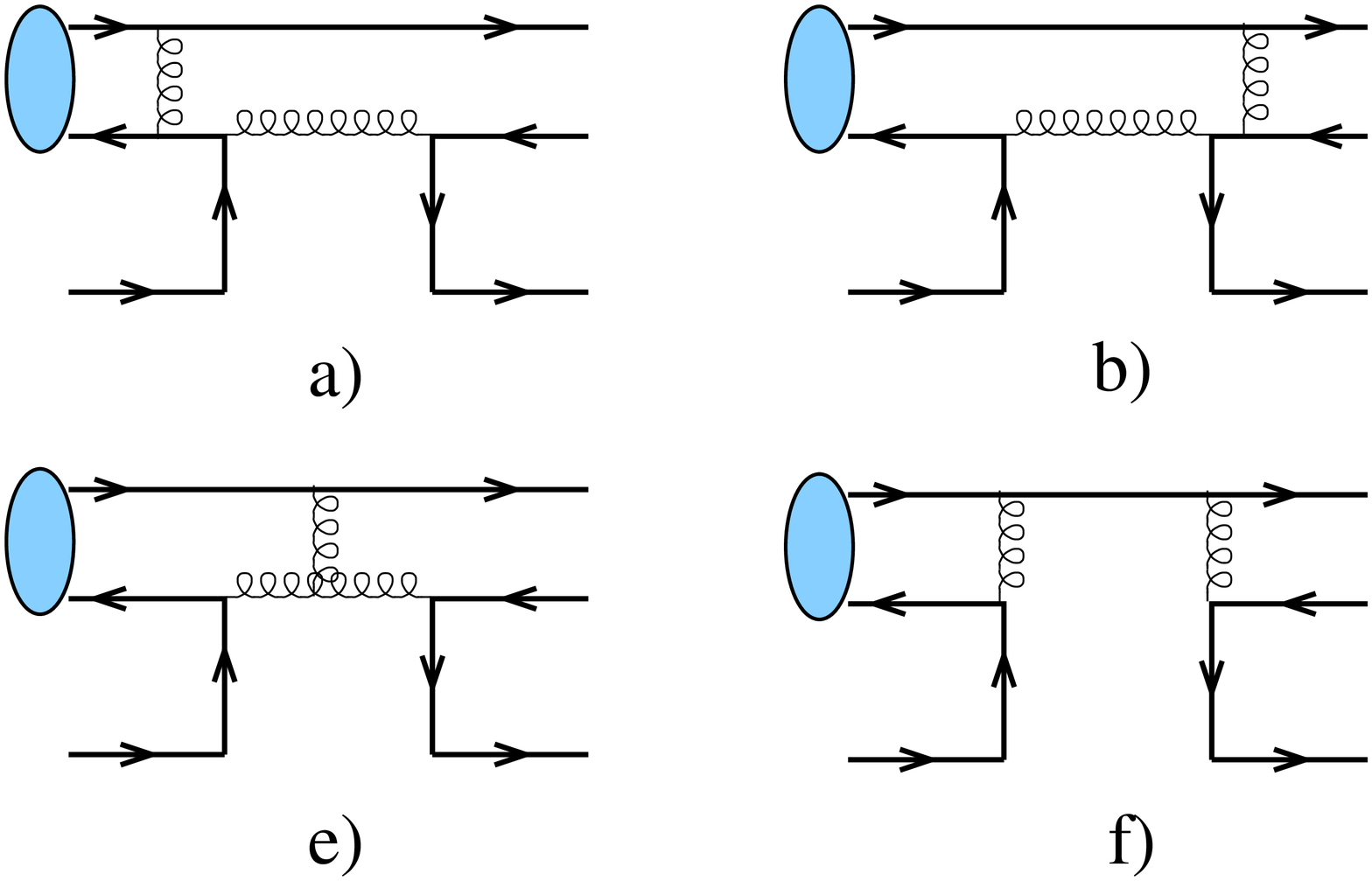}~~~~~~~
            \epsfxsize7.0cm\epsffile{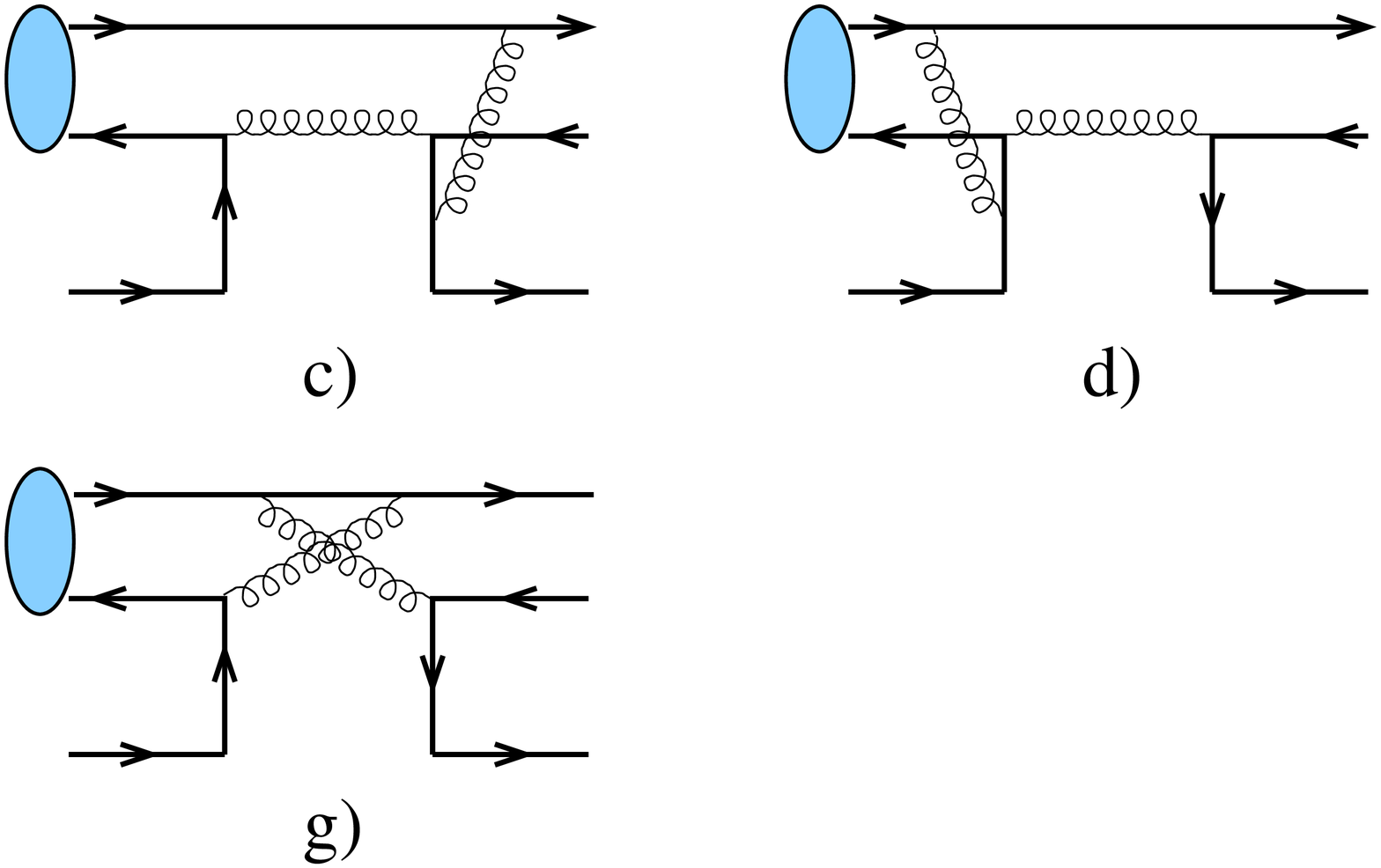}
}
\caption[]{\small 
The $u$-quark (annihilation) contribution to the coherent hard dijet 
 production \\ $\pi^-\to 2\, {\rm jets}$, see text.  
 }
\label{fig:quark_u}
\end{figure}

\begin{figure}[t]
\centerline{\epsfxsize7.0cm\epsffile{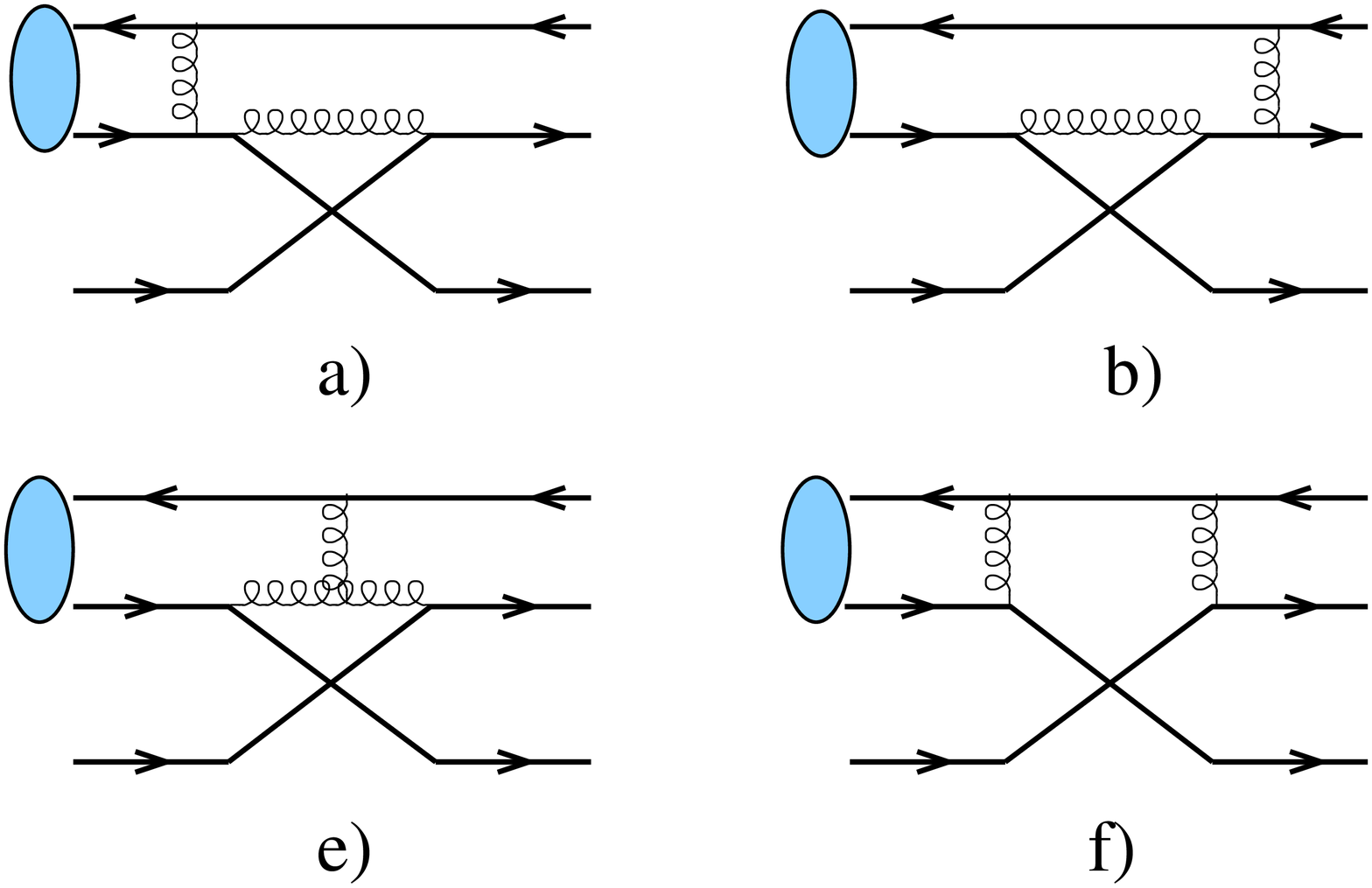}~~~~~~~
            \epsfxsize7.0cm\epsffile{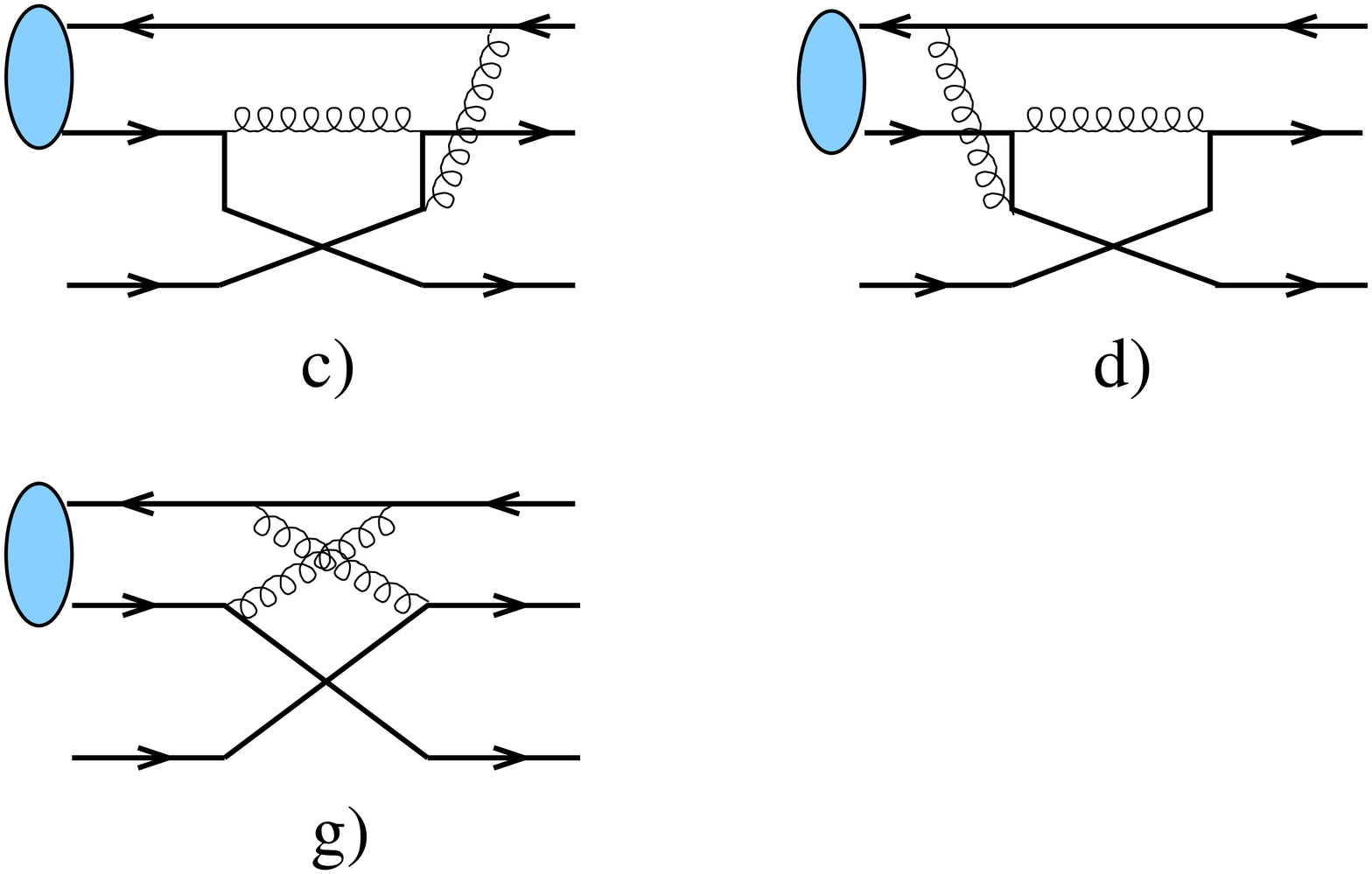}
}
\caption[]{\small 
The $d$-quark (exchange) contribution to the coherent hard dijet 
production \\ $\pi^-\to 2\, {\rm jets}$, see text. Note that the 
quark and the antiquark lines entering the pion blob are interchanged 
compared to Fig.~\ref{fig:quark_u}. 
 }
\label{fig:quark_d}
\end{figure}

\begin{figure}[t]
\centerline{\epsfxsize9.0cm\epsffile{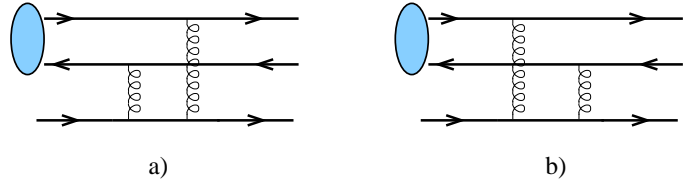}}
\caption[]{\small 
The flavor-singlet quark contribution to the coherent hard dijet 
 production $\pi\to 2\, {\rm jets}$.  
 }
\label{fig:quark_s}
\end{figure}

We begin with the calculation of the quark contribution to the hard dijet
production, which is simpler than the gluon contribution 
as it does not involve subtleties related to
the choice of the gauge. For definiteness we consider $\pi^-$-proton 
scattering. The quark contribution starts at order 
${\cal{O}}(\alpha_s^2)$ and can be decomposed in three topologically different
contributions  shown in Fig.~\ref{fig:quark_u}, Fig.~\ref{fig:quark_d} and  
Fig.~\ref{fig:quark_s}. In what follows we refer to them as the 
$u$-quark annihilation, the 
$d$-quark exchange, and the  gluon exchange (flavor-singlet) contribution,
respectively. All three contributions are 
separately gauge invariant and can be calculated with quarks on the mass 
shell. In this section we work in the Feynman gauge. 
The antiquark contributions do not require a new calculation 
and can be obtained from the corresponding quark contributions by obvious 
substitutions, see below.    
The reaction kinematics and the notation for the momenta are specified in
the Introduction. We define the generalized quark (antiquark) distribution 
as the matrix element of the light-cone operator \cite{Rad96,Rad96a}  
\beq{quark-distr}
\langle p^\prime_2|\bar q(0) \!\not\! y\, q(y)|p_2 \rangle_{y^2 = 0}
 =\bar u(p^\prime_2) \!\not\! y u(p_2)\cdot 
\int\limits^1_0 \!dx_1\, \left[ 
e^{-ix_1(p_2y)} {\cal F}_{\zeta}^q (x_1)- e^{ix_2(p_2y)}
 {\cal F}_{\zeta}^{\bar q} (x_1)\right],
\eeq
where $q = u,d,\ldots$, etc.
Hereafter  we use the notation 
\beq{def:x2}
        x_2 = x_1-\zeta\,,
\eeq
where the asymmetry parameter $\zeta$ is defined as 
$p_2 - p_2^\prime  = \zeta p_2$ and is fixed by the kinematics of the reaction,
see \re{zeta}. The relation to other 
common parametrizations of the generalized parton distributions is 
discussed in Appendix A. 
In turn, the pion distribution amplitude is defined as 
\cite{earlyCZ,earlyBL,earlyER} 
\beq{pion-DA}
 \langle 0|\bar d(0) \!\not\! y \gamma_5 u(y)|\pi^-(p_1) \rangle_{y^2 = 0}
 = i(p_1 y)\,f_\pi \int\limits_0^1 \!dz' \, e^{-iz'(p_1y)} \phi_\pi(z')\,,
\quad \int\limits_0^1 \!dz' \, \phi_\pi(z') =1\,,
\eeq 
where $f_\pi =133$~MeV is the pion decay constant. In both cases, 
\re{quark-distr} and \re{pion-DA}, the insertion of the path-ordered gauge 
factor between the quark field operators is implied. Both distributions 
depend on the factorization scale $\mu_F$ which to the leading 
logarithmic  accuracy has to be 
taken of the order of the transverse momentum of the jets.

%
%

As an example, consider the first diagram in 
Fig.~\ref{fig:quark_u}. The corresponding contribution reads:
\bea{example}
i {\cal M}_{2a} &=& -(ig)^4 (-i)^2 (i)  \int\limits^1_0 dz^\prime 
\frac{\displaystyle i f_\pi \phi_{\pi}(z^\prime)}{\displaystyle 4N_c} 
\int\limits^1_0 dx_1
\frac{\displaystyle {\cal F}^{u}_\zeta (x_1) \sqrt{1-\zeta}}{\displaystyle
2N_c}
\left( t^a t^a t^b t^b \right)_{i j}
\nonumber \\
&&\times
\frac
{\displaystyle 
\bar u(q_1)\gamma_\mu \gamma_5 
\!\not\! p_1 \gamma_\mu (\zeta \!\not\! p_2- \!\not\! q_2)
\gamma_\nu \!\not\! p_2 \gamma_\nu v(q_2) 
}
{\displaystyle 
[(\zeta p_2-q_2)^2+i\epsilon ][(q_1-z^\prime p_1)^2][(q_2+x_2 p_2)^2]
},
\eea
where $\bar u(q_1)$ and $v(q_2)$ are Dirac spinors for the outgoing 
quark and antiquark jets, respectively, and $i,j$ are the jet color indices.
The color state of the target quark is averaged over.
Note that we only show the Feynman $+i\epsilon$ prescription for one of 
the propagators, since the other two are strongly off-shell. 

A simple calculation yields  
\beq{example-res}
{\cal M}_{2a}=\frac{\displaystyle 4\pi^2 \alpha_s^2 i f_\pi}
 {\displaystyle N_c^2 q^2_\perp }  \sqrt{1-\zeta}\,
\bar u(q_1)\gamma_5 \frac{\displaystyle \!\not\! p_2}{\displaystyle s}
v(q_2)\delta_{i j} \int\limits^1_0\! dz^\prime \,\phi_{\pi}(z^\prime)\!\!
\int\limits^1_0 dx_1 
{\cal F}^{u}_\zeta (x_1) \cdot I_{2a}(z,z^\prime ,x_1, x_2)
\eeq
with the coefficient function
\beq{2a}
I_{2a} (z,z^\prime ,x_1, x_2) =- 
\frac{\displaystyle 2C^2_F z}{\displaystyle
z^\prime \bar z (x_2+i\epsilon)} \, ,
\eeq
where $C_F=(N_c^2-1)/(2N_c)$, $N_c$ is the number of colors. 
Calculation of the other diagrams in Fig.~\ref{fig:quark_u} is equally simple. 
The corresponding coefficient functions are 
\bea{u-ann}
I_{2b}(z,z^\prime ,x_1, x_2) &=&
 \frac{\displaystyle 2 C_F^2  z }{\displaystyle 
z^\prime \bar z^\prime  [ x_1 +i\epsilon
]} \, ,
\nonumber\\
I_{2c}(z,z^\prime ,x_1, x_2) &=&
- \frac{\displaystyle  C_F  z (\zeta z-x_1)}{\displaystyle 
N_c z^\prime \bar z^\prime [ x_1 +i\epsilon
][x_1(z-z^\prime)-z\bar z^\prime \zeta +i\epsilon]} \, ,
\nonumber\\
I_{2d} (z,z^\prime ,x_1, x_2) &=&
- \frac{\displaystyle  C_F  z (\zeta z^\prime -x_1)}{\displaystyle 
N_c z^\prime \bar z [ x_2 +i\epsilon
][x_1(z^\prime-z)-z^\prime \bar z \zeta  +i\epsilon]} \, ,
\nonumber\\
I_{2e} (z,z^\prime ,x_1, x_2) &=& 
 \frac{\displaystyle C_F N_c  z (x_1+\zeta (1-z-z^\prime))}{\displaystyle 
z^\prime \bar z^\prime \bar z [x_1 +i\epsilon] [ x_2 +i\epsilon
]} \, ,
\nonumber\\
I_{2f} (z,z^\prime ,x_1, x_2) &=&
- \frac{\displaystyle 2 C_F^2 \zeta  z }{\displaystyle 
\bar z^\prime [ x_1 +i\epsilon] [ x_2 +i\epsilon
]} \, ,
\nonumber\\
I_{2g} (z,z^\prime ,x_1, x_2) &=&
- \frac{\displaystyle  C_F \zeta^2  z (1-z-z^\prime )}{\displaystyle
N_c \bar z^\prime [ x_1 +i\epsilon][x_2 +i\epsilon ]
[x_1(1-z-z^\prime)-\bar z\bar z^\prime \zeta +i\epsilon]} \,,                  
\eea
so that the total contribution of the $u$-quark annihilation is given 
by the expression in \re{example-res} with the coefficient function 
\beq{def:I}
 I(z,z^\prime ,x_1, x_2)=\sum_{i=2a,2b,\dots , 2g} I_i(z,z^\prime ,x_1, x_2)
\, 
\eeq
instead of $I_{2a}$.

%
%

The $d$-quark exchange contributions in Fig.~\ref{fig:quark_d} can be obtained
from the corresponding diagrams in Fig.~\ref{fig:quark_u} by the substitution:
$x_1\to -x_2$, $x_2\to -x_1$, $z\to \bar z$, $z^\prime \to \bar z^\prime$ 
and an overall minus sign. Similarly, the antiquark $\bar d$-annihilation
and $\bar u$-exchange coefficient functions can be obtained from the above 
expressions by the substitution $x_1\to -x_2$, $x_2\to -x_1$ and 
changing the overall sign.
Altogether, we obtain four coefficient functions:
\bea{rules}
I_{u}(z,z^\prime ,x_1, x_2) &=& I(z,z^\prime ,x_1, x_2)\,,
\nonumber \\
I_{d}(z,z^\prime ,x_1, x_2) &=& - I(\bar z,\bar z^\prime ,-x_2, -x_1)\,,
\nonumber \\
I_{\bar u}(z,z^\prime ,x_1, x_2)&=& - I(z,z^\prime ,-x_2, -x_1)\,,
\nonumber \\
I_{\bar d}(z,z^\prime ,x_1, x_2) &=& I(\bar z,\bar z^\prime ,x_1, x_2)\,.
\eea  

%
%

Last but not least, we have to take into account the diagrams in 
Fig.~\ref{fig:quark_s}
corresponding to the flavor-singlet two-gluon exchange. Note that the 
virtuality of both gluons is large, in fact equal to the transverse 
momentum of the jets, and, therefore,
these diagrams are part of the quark coefficient function rather than the 
gluon one. A simple calculation yields
\bea{4a}
I_{4a}(z,z^\prime ,x_1, x_2) &=& 
- \frac{\displaystyle C_F  (\zeta (1+z-z^\prime)-2x_1)}{\displaystyle  
 z^\prime \bar z^\prime \zeta [ x_1(z-z^\prime ) -z\bar z^\prime \zeta  +i\epsilon
]} \, ,
\nonumber\\
I_{4b}(z,z^\prime ,x_1, x_2) &=&
- \frac{\displaystyle C_F  (\zeta
(1-z+z^\prime)-2x_1)}{\displaystyle
 z^\prime \bar z^\prime \zeta [ x_1(z^\prime -z ) -z^\prime \bar z \zeta
 +i\epsilon
]} \,.
\eea
Denoting the sum of them as
\beq{glue}
   I_{\rm glue} = I_{4a} + I_{4b}
\eeq
we obtain the final answer for the leading-order quark contribution 
to the dijet production in the form
\bea{Q-result}
{\cal M}_{\rm quark} &=&\frac{\displaystyle 4\pi^2 \alpha_s^2 
 i f_\pi}{\displaystyle N_c^2 q^2_\perp}\sqrt{1-\zeta}\,
\bar u(q_1)\gamma_5 \frac{\displaystyle \!\not\! p_2}{\displaystyle s}
v(q_2)\, \delta_{i j}\int\limits^1_0 dz^\prime
\, \phi_{\pi}(z^\prime)\!        
                            \int\limits^1_0 dx_1 
\nonumber\\
&&{}\hspace*{2cm}\times\sum_{p} {\cal F}^{p}_\zeta (x_1)\,
\Big[I_{\rm glue}(z,z^\prime, x_1,x_2)+I_{p}(z,z^\prime ,x_1,x_2)\Big]\,,
\eea
where the summation goes over all possible quark-parton species: 
$p=u,\bar u, d, \bar d$, etc. 

\subsection{The End-Point Behavior}

Next,  we have to examine the behavior of the coefficient 
functions $I_{glue}$ and
$I_p$ at the end points $z^\prime \to 0$ and $z^\prime \to 1$
 and check the convergence of the integration in (\ref{Q-result}) over the 
quark momentum fraction $z'$ in the pion.
Convergence is necessary for the self consistency of the 
collinear factorization approach as it ensures that quark transverse 
momenta in the pion can be neglected. 

Consider the gluon-exchange contribution $I_{glue}$ first. 
In this case it is easy to see that $I_{glue}\sim 1/z^\prime$
and $I_{glue}\sim 1/\bar z^\prime$ at 
$z^\prime \to 0$ and $z^\prime \to 1$, respectively.
Since the pion distribution amplitude vanishes linearly at the end-points,
at least at a sufficiently high scale, the corresponding contribution 
to the result in (\ref{Q-result}) is finite.

From  Eqs.~(\ref{2a}) and (\ref{u-ann}) it is seen
that the contributions of the $u$-quark and $\bar d$-antiquark are 
finite as well%
\footnote{We remind that the flavor identification refers to 
 the particular case of $\pi^-$ meson scattering.}.
For the $d$-quark and the $\bar u$-antiquark contributions
the situation is different, however. In this case we find 
\beq{d-limit}
I^{d}(z,z^\prime ,x_1 ,x_2)|_{z^\prime \to 0}= 
\frac{\displaystyle 2i\pi C_F }{\displaystyle N_c}
\frac{\displaystyle \bar z}{\displaystyle z  z^{\prime
2}}\delta (x_2) +{\cal O}(\frac{1}{z^\prime})\, ,
\eeq
and
\beq{au-limit}
I^{\bar u}(z,z^\prime ,x_1 ,x_2)|_{z^\prime \to 1}= 
\frac{\displaystyle 2i\pi C_F }{\displaystyle N_c}
\frac{\displaystyle z}{\displaystyle \bar z \bar z^{\prime
2}}\delta (x_2) +{\cal O}(\frac{1}{\bar z^\prime}) \, .
\eeq
Note that: a) the $\sim 1/z'^2$ ($\sim 1/(\bar z')^2$) behavior leads to the 
logarithmically divergent integral over the pion quark momentum fraction;
 b) this contribution is purely imaginary, and c) this contribution is 
 proportional to $\delta(x_2)=\delta(x_1-\zeta)$, i.e. it is due to 
 parton configurations with vanishing longitudinal momentum fraction
 of one of the quarks in the target proton. In the terminology of 
 \cite{BBL81} this corresponds to the Glauber region.
In the remaining part of this Section we explain this result for the 
example of the $d$-quark contribution in some detail.

%
\begin{figure}[t]
\centerline{\epsfxsize9.0cm\epsffile{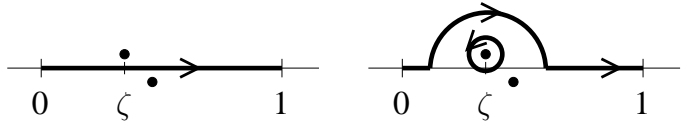}}
\caption[]{\small 
 Deformation of the integration contour in the complex $x_1$ plane 
 in the Glauber region, see text. }
\label{fig:pinch}
\end{figure}
%

Explicit expressions for the diagrams in Fig.~\ref{fig:quark_d} can be 
obtained from Eqs.~(\ref{2a}), (\ref{u-ann}) using the substitution rules 
in (\ref{rules}). One can easily convince oneself that the only 
diagrams capable of  producing  a $\sim 1/z^{\prime 2}$ singularity  
are those in Fig.~\ref{fig:quark_d}c and Fig.~\ref{fig:quark_d}g.   
In particular, the diagram in Fig.~\ref{fig:quark_d}c gives:
\beq{3c}
I_{3c}(z,z^\prime ,x_1, x_2) = -
 \frac{\displaystyle  C_F  \bar z (\bar z+x_2)}{\displaystyle
N_c z^\prime \bar z^\prime [ x_2 - i\epsilon
][x_2(z-z^\prime)-z^\prime \bar z \zeta +i\epsilon]}.
\eeq
It is seen that for finite values of $x_2$ this contribution is proportional 
to $1/ z^\prime$ to the first power. However, as $z^\prime \to 0$ the 
two poles in the Feynman denominators in  Eq.~(\ref{3c}) produce a pinch 
at $x_2=0$ ( $x_1=\zeta$) in the integral over $x_1$, see
Fig.~\ref{fig:pinch}. This pinch is responsible for the singular 
behavior of $I_{3c}$, $I_{3c}\sim 1/z^{\prime 2}$.   To see this, note 
that if the two poles would lie on the same side of the integration 
contour, the singularity at $x_2=0$ could be avoided by the contour 
deformation, following the arguments of Ref.~\cite{CSS85} 
(see also \cite{CF98}). In the presence of the pinch one can also deform 
the contour, but in this case an additional contribution arises
\beq{circle}
I^{\rm pole}_{3c}(z,z^\prime ,x_1 ,x_2)= 
\frac{\displaystyle 2i\pi C_F }{\displaystyle N_c}
\frac{\displaystyle \bar z}{\displaystyle   z^{\prime
2}}\delta (x_2) \ . 
\eeq   
This pole contribution is entirely responsible for the leading asymptotics 
of $I_{3c}$ in the $z^\prime \to 0$ region. 
The calculation of the diagram in Fig.~\ref{fig:quark_d}g is very similar.
In this case also one can deform the contour as it is shown at Fig. 5 
and the leading asymptotics of $I_{3g}$ is 
again given by the pole contribution
\beq{circle3g}
I^{\rm pole}_{3g}(z,z^\prime ,x_1 ,x_2)= 
\frac{\displaystyle 2i\pi C_F }{\displaystyle N_c}
\frac{\displaystyle \bar z^2}{\displaystyle z  z^{\prime
2}}\delta (x_2) \ .
\eeq
The sum of these pole contributions gives the result in (\ref{d-limit}). 
For the $\bar u$-contribution, again, the pinch at $x_1=\zeta$ 
appears at $z^\prime \to 1$ and leads to the result (\ref{au-limit}). 

The pinching of integration contours in the Glauber region indicates a
serious problem with collinear factorization. It is known \cite{CSS83}
that such pinches generally occur between initial and final state interactions
involving soft particle (gluon) exchanges.
In the case of the Drell-Yan production the pinches disappear in the sum 
of all Feynman diagrams for the cross section, which is due, in 
physical terms, to cancelation of the final state interactions 
\cite{CSS85} (see also \cite{CCM86}). In our case the pinch occurs in the 
leading-order contribution and for quark exchange, which is unusual. 
Note that the problem with pinching singularities in the Glauber region
is in principle unrelated to the end-point behavior of the pion 
distribution amplitude and is more general. We will return to this discussion 
in the next section.     
 
\begin{figure}[t]
\centerline{\epsfxsize11.0cm\epsffile{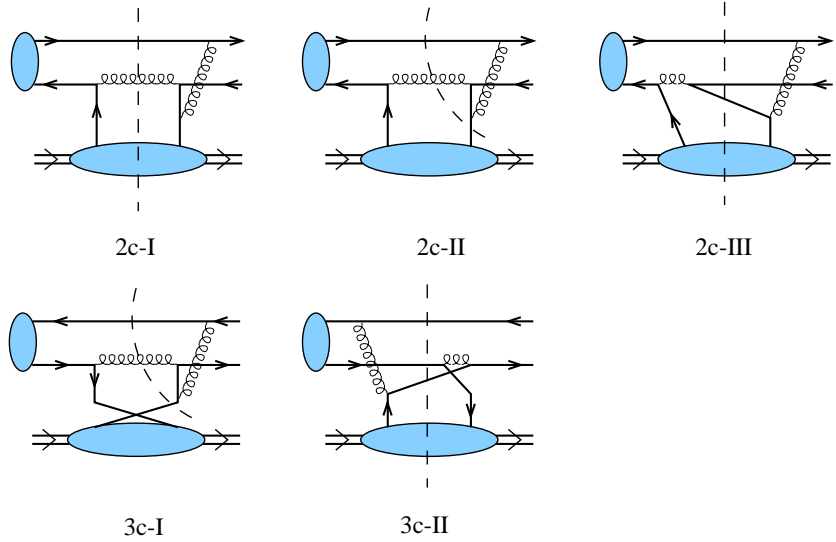}}
\caption[]{\small 
 Different dispersion parts corresponding to the diagrams 
 in Fig.~\ref{fig:quark_u}c and Fig.~\ref{fig:quark_d}c.
 }
\label{fig:quark_cuts}
\end{figure}

Our result in \re{Q-result} differs from the expression for the 
quark contribution obtained in \cite{CG01} by the above discussed pole terms.
The authors of \cite{CG01} have restored the imaginary part of the 
coefficient function by requiring that the physical amplitude only depends
on $s+i\epsilon$  and hence on $\zeta -i \epsilon$ (with the real integration
variable $x_1$). It is easy to see that with this prescription all 
singularities in the denominators are below the integration
contour and do not obstruct the analytic continuation. In this sense, 
this prescription corresponds to the ``true'' light-cone contribution. 
The argument is not correct, however, since apart from the usual $s$-channel
discontinuity the amplitude in question has a  
discontinuity in another invariant variable, $M^2$, 
the invariant mass of the jets.
The full imaginary part is only  restored in their sum and the existence
of two different types of cuts  is reflected
in the structure of Feynman $i\epsilon$ prescription in the propagators. 
To explain this point, consider the different dispersion parts corresponding 
to the diagrams  in Fig.~\ref{fig:quark_u}c and Fig.~\ref{fig:quark_d}c, 
see  Fig.~\ref{fig:quark_cuts}.

%
%

It is obvious that the cut diagrams denoted as 2c-I and 2c-III 
describe the discontinuity in the invariant energy $s$, and the cut diagram 
2c-II corresponds to the discontinuity in the $M^2$-variable. 
Note that in the light-cone limit the $s$-channel
cut 2c-I and the $M^2$-channel cut 2c-II both occur at $x_1=0$, see
(\ref{u-ann}). 
Moreover, their contributions exactly cancel each other. In the usual 
language this cancelation manifests itself in the following way. The  
part of the coefficient function,  $I_{2c}(z,z^\prime, x_1, x_2)$,
which corresponds to the diagram in Fig.~\ref{fig:quark_u}c 
has a pole at $x_1\to 0$.
This pole does not contribute, however,  to the imaginary part of the 
amplitude since the generalized quark distribution function 
vanishes at this point, ${\cal F}_\zeta (x)|_{x\to 0}\to 0$
 \cite{Rad96a,CF98}.  
As the result, we are left with the single s-channel cut 2c-III corresponding 
to the $[x_1(z-z^\prime ) -z\bar z^\prime\zeta +i \epsilon ]$
denominator in $I_{2c}(z,z^\prime, x_1, x_2)$. Note that the Feynman 
$i\epsilon$ prescription to go around the  pole is in agreement with the 
above statement that 
it corresponds to a singularity in the $s$-channel: The sign of $i\epsilon $
can be understood as the substitution $s\to s+i\epsilon$, or $\zeta\to \zeta
-i\epsilon$. 

Now let us turn over to the dispersion parts of the diagram 
corresponding to Fig.~\ref{fig:quark_d}c. In this case there are two 
possible cuts 3c-I and 3c-II that describe the singularities in the 
$M^2$- and $s$-channels, respectively. 
The corresponding coefficient function $I_{3c}(z,z^\prime,
x_1, x_2)$ contain two poles, see eq. (\ref{3c}).
The $[x_2-i\epsilon]=[x_1-\zeta -i\epsilon]$ pole is related to the 
$M^2$-channel discontinuity and its $i\epsilon$ prescription corresponds to 
the substitution  $M^2\to M^2+i\epsilon$, or $\zeta\to \zeta
+ i\epsilon$. On the other hand, the $i\epsilon$ prescription of 
the second denominator  $[x_2(z-z^\prime -z^\prime
) \bar z +i \epsilon]$ can be understood as the substitution  
$s\to s+i\epsilon$, or $\zeta\to \zeta
-i\epsilon$, in agreement with the interpretation 
that this singularity occurs in the  $s$-channel. 
To summarize, we see that for the case of the diagram in
 Fig.~\ref{fig:quark_d}c both dispersion parts contribute in a nontrivial way
and pinching of the integration contour described above in fact occurs 
between the discontinuities in different channels.  It is also seen that
the pinching occurs between the soft quark exchange in the initial 
state (3c-I) and in the final state (3c-II)
with respect to the hard interaction, in agreement with 
the general arguments in \cite{CSS83}.

\section{Gluon Contribution}
\setcounter{equation}{0}

The gluon contribution to hard dijet production is 
more involved because of subtleties related to gauge invariance.
One is tempted to calculate the coefficient function in front of the gluon 
distribution by considering pion scattering from on-shell gluons with 
zero transverse momentum, similar to the above calculation for the quarks.
The result has in this case to be multiplied by the gluon distribution 
in the target in the  physical light-cone gauge $p_1^\mu A_\mu =0$, 
which has the form \cite{CFS96,Rad96a}
\beq{naiveFg}
\langle p^\prime_2|A_{\mu }^a(0) A_{\nu}^a(y)
|p_2 \rangle_{y^2 = 0} =  
- \frac{\bar u(p_2^\prime)\!\not\!p_1\,
u(p_2)}{2(p_1p_2)}
\,g_{\mu\nu}^\perp
\int\limits^1_0\!dx_1\,\frac12
\left[e^{-ix_1(p_2y)} + e^{ix_2(p_2y)}
\right]\frac{{\cal F}_{\zeta}^g (x_1)}{x_1 x_2},
\eeq
where 
\beq{gperp}
      g_{\mu\nu}^\perp = g_{\mu\nu}-
\frac{p_{1\mu}p_{2\nu}+p_{1\nu}p_{2\mu}}{(p_1p_2)}.
\eeq
The difficulty arises because of the factor $1/(x_1x_2)$ which is singular 
within the integration domain. The particular procedure to deal with  this 
singularity is related to the gauge condition for the gluon field 
at time infinity and has to be established by considering carefully 
the light-cone limit of the relevant Feynman diagrams. 
For the classical process of hard vector 
meson production by longitudinally polarized photons 
\cite{BFGMS94,CFS96} the 
correct prescription was formulated by Radyushkin \cite{Rad96a}
\beq{Rad+}
     \frac{1}{x_1 x_2} \rightarrow \frac{1}{(x_1-i\epsilon )(x_2+i\epsilon)}. 
\eeq  
It is not obvious, however,  whether the same substitution is true for
the hard dijet production. In fact, we will argue that no simple 
prescription exists in this case at all. 
In order to find the answer, in this section we will not assume zero gluon 
transverse momentum from the beginning, 
but consider instead the full scattering amplitude of the hard dijet production
from the quark target mediated by the two-gluon exchange, 
cf.~Fig.~\ref{fig:1}. We will identify the (IR divergent) contribution to 
this amplitude corresponding to the region of small gluon transverse 
momenta and try to find a factorized expression for it, as a product 
of the coefficient function times the perturbative 
gluon distribution in a quark. 

For pedagogical reasons, we first consider the simpler case of
hard vector meson production by longitudinally polarized photons. We 
do the calculation in Feynman and in the light-cone  gauge and explain how the 
prescription in (\ref{Rad+}) arises in both cases. Next, we 
present the calculation of the imaginary part of the amplitude for 
hard dijet production in Feynman gauge. This calculation was 
previously reported by us in \cite{BISS01}. Finally, we derive the 
full expression for the amplitude (both real and imaginary part)   
using the axial gauge and compare our results to the work \cite{Che01}.
      
\subsection{Getting Started: Exclusive Vector Meson Production}

\begin{figure}[t]
\centerline{\epsfxsize12.0cm\epsffile{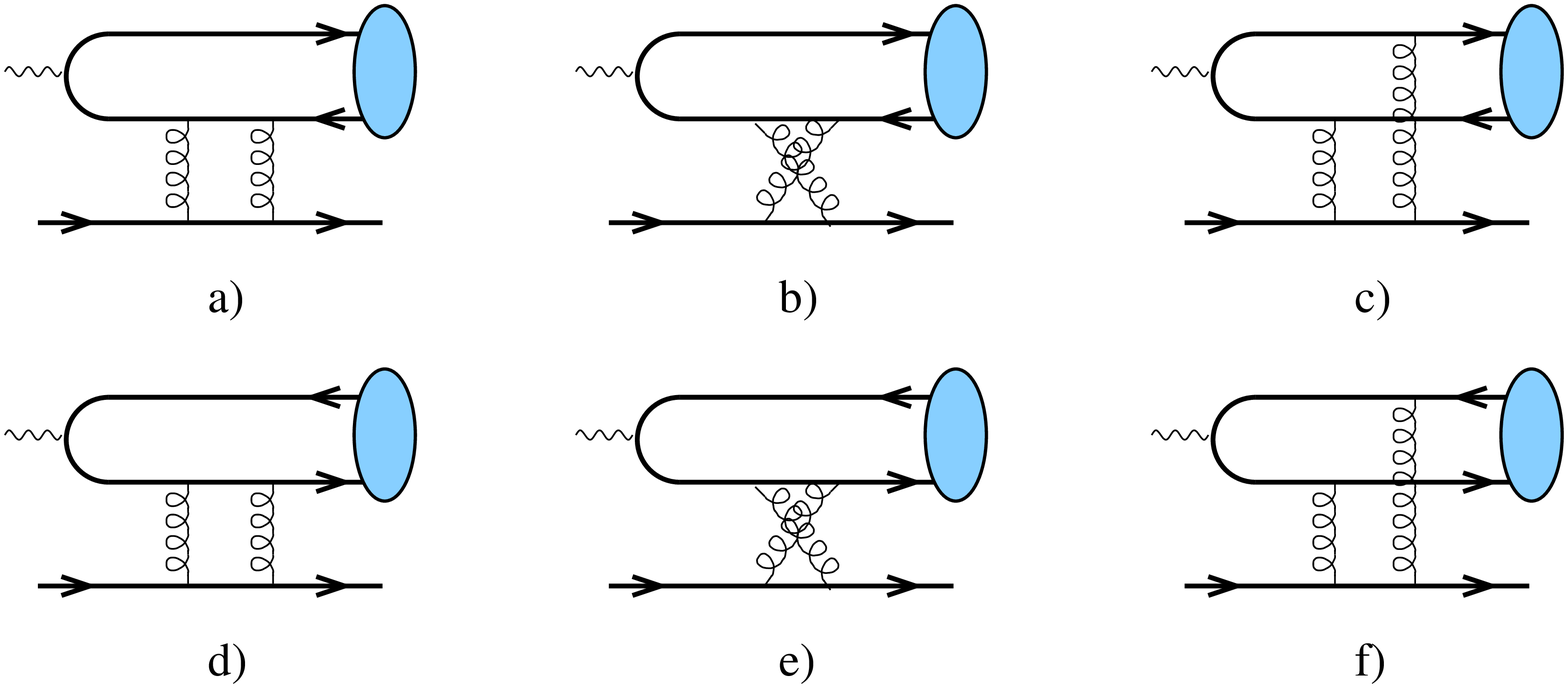}}
\caption[]{\small 
The leading-order gluon contribution to the hard exclusive vector 
meson production $\gamma^* \to \rho^0$.  
 }
\label{fig:vector}
\end{figure}

To the lowest order in $\alpha_s$ the relevant Feynman 
diagrams are shown in Fig.~\ref{fig:vector}.
%
%
The kinematics is similar to that of the dijet production. 
We take $p_2$ and $p_2'$
to be the target quark momenta in the initial and the final state,
respectively. The momentum of the vector meson in the final state is 
denoted by $p_1$. The virtual photon momentum is equal to $q = p_1-
\zeta p_2$ and we use the further notations 
 $Q^2 = -q^2$ and  $\zeta = Q^2/s$ with $s=2 p_1p_2$. 
We take the photon to be longitudinally polarized, with the 
polarization vector
\beq{eL}
e^L_\mu =\frac{1}{Q}p_{1,\mu}+\frac{Q}{s}p_{2,\mu}
 = \frac{1}{Q}q_\mu + 2\frac{Q}{s}p_{2,\mu}\,, \quad
 e^L_\mu q^\mu = 0\,, \quad e_L^2=1\,.
\eeq
Thanks  to the $U(1)$ gauge invariance one can omit the first term 
in the above expression and use a simpler vector
\beq{eLnew}
e^L_\mu\to e^L_\mu =2\frac{Q}{s}p_{2,\mu}\,.
\eeq
The vector meson distribution amplitude is defined by the expression 
similar to (\ref{pion-DA})
\beq{Vdampl}
\langle0|\bar q(0) \!\not\!y\, q(y)|V(p_1)\rangle_{y^2 = 0}
 = f_V\, (p_1y)
\int\limits^1_0 \!dz\, e^{-iz(p_1y)}\, \phi_V (z)\,,
\eeq
where $f_V$ is the corresponding decay constant. 
For simplicity we do not elaborate on the isospin structure and consider 
below the contribution of the single light quark flavor $q$. 
The variable $z$ corresponds
to the momentum fraction carried by the quark. In the calculation of the 
amplitude sandwiched  between the two target quark spinors $\bar u(p'_2)$
and $u(p_2)$ we will assume averaging over the color of the target quark
and make the replacement 
\beq{targetspur}
\bar u (p_2^\prime){\cal M}\, u(p_2)\to
\frac{\sqrt{1-\zeta}}{2}\mbox{ Sp}\left[{\cal M} \!\not\!p_2 \right]
\eeq
picking up contribution of the unpolarized gluon distribution, 
which is dominant at large energies.

For definiteness, consider the contribution of the diagram in 
Fig.~\ref{fig:vector}a. We obtain
\bea{Va}
i {\cal M}_a &=& \sqrt{4\pi\alpha}\, e_q\, g^4 f_V \sqrt{1-\zeta}
\frac{1}{8N_c} C_F 
\int \frac{d^4 k}{(2 \pi)^4} \int\limits^1_0 dz \, \phi_V(z) 
\nonumber \\
&& {}
\times \mbox{Tr}
\left[
\!\not\! p_1 \left( \frac{
2Q}{s}\right)\!\not\! p_2 \frac{(z\!\not\! p_1- \!\not\!
q)}{(zp_1-q)^2+i\epsilon}
 \gamma_{\mu_1} \frac{(\zeta \!\not\! p_2 - \!\not\! k
-\bar z \!\not\! p_1)}{(\zeta p_2 -k -\bar z p_1)^2+i\epsilon}
\gamma_{\mu_2} \right] \, {\cal R}_{\mu_1\mu_2},
\eea
where $e_q$ is the quark electric charge, $\alpha=1/137$ is the fine structure 
constant, and 
\beq{Rfac} 
{\cal R} _{\mu_1\mu_2} =
\frac12 \mbox{Tr}
\left[\!\not\! p_2  \gamma_{\nu_2} 
\frac{(\!\not\! p_2 - \!\not\! k)}{(p_2-k)^2+i\epsilon} 
 \gamma_{\nu_1}\right]
\frac{N_{\mu_1\nu_1}(k) N_{\mu_2\nu_2}(k-\zeta p_2)}
{[k^2+i\epsilon][(k-\zeta p_2)^2+i\epsilon]}
\eeq
is a common factor for all Feynman diagrams in Fig.~\ref{fig:vector}
that describes the emission of the pair of gluons from the target quark line. 
In the expression for ${\cal R}$ we have assumed a general form for 
the gluon propagator
\beq{Dgluon}
   D_{\mu\nu}(k) = \frac{-i}{k^2+i\epsilon} N_{\mu\nu}\,.
\eeq
In order to do the integration over the loop momentum we use the Sudakov
parametrization 
\beq{Sudakov}
k= \alpha p_1 +x_1 p_2 + k_\perp \, , \quad d^4k=\frac{s}{2}d\alpha dx_1
d^2k_\perp\,, \quad k^2 = \alpha x_1 s - k_\perp^2
\eeq
and  take the integral over $\alpha$ by residues.  
After this, the integration over $x_1$ gets confined to the interval
$0\leq x_1\leq 1$ and the  variable $x_1$ acquires the meaning of
the longitudinal momentum fraction. In the Feynman gauge 
$N_{\mu\nu} = g_{\mu\nu}$ we obtain after some algebra 
\bea{Ma1}
{\cal M}_a= \sqrt{4\pi\alpha}\, e_q\, f_V \sqrt{1-\zeta}\,
\frac{Q}{s}\, \frac{C_F g^4}{16\pi^3 N_c}
\, \int\limits^1_0 \!dz\, \phi_V(z) 
 \int d^2k_\perp \int\limits^1_0\! dx_1 \,I_a   
\eea 
with
\bea{a1(0)}
I_a&=&
\frac{1}{k^4_\perp \bar z \zeta (1-\zeta) }
\left[
\frac{\bar z s(2s(1-x_1)^2\bar z+k^2_\perp x_2)
}{
[s\bar z x_2 -k^2_\perp (1-\zeta )/(1-x_1) +i\epsilon]}
\right.\nonumber\\
&&{}\left. \hspace*{2cm} +
\Theta (-x_2)\frac{2s(1-x_1)\bar zx_2+k^2_\perp
(2-x_1-\zeta)}{ x_2\zeta}
\right] \,. 
\eea
The contribution of large transverse momenta $k_\perp \ge Q$ in   
Eq.~(\ref{Ma1}) gives rise to the $\alpha_s^2$ correction to the 
coefficient function in front of the quark distribution in the target 
and is not relevant for our discussion. We concentrate, therefore, 
on the region of small $k_\perp \ll Q$. Notice that the $k_\perp$ integral 
is infrared (IR) divergent. The quadratic  divergence 
$d k_\perp^2/k_\perp^4$ must cancel in the sum of all Feynman diagrams
and the remaining logarithmically divergent integral 
$\sim d k_\perp^2/k_\perp^2$ has to be interpreted as the perturbative 
gluon distribution in a quark. In order to observe factorization one has, 
therefore, to expand the integrand in powers of $k_\perp^2$ and keep two
first terms. The expansion of the denominator in the first term in 
Eq.~(\ref{a1(0)}) in the vicinity of $x_2 =0$ ($x_1=\zeta$) may seem 
dangerous. However, the integral over $x_1$ can be deformed to the complex
plane away from this singularity, and the expansion can be done 
without further problems. We obtain      
\bea{a1}
 I_a &\stackrel{k_\perp^2\to0}{=}& 
 \frac{1}{k^4_\perp \bar z \zeta (1-\zeta)}
\left[
\frac{2s(1-x_1)^2\bar zx_2+k^2_\perp
\left((1-x_1)^2+(1-\zeta)^2\right)
}{(x_2 +i\epsilon)^2}
\right. \nonumber \\
&&
\left. {}\hspace*{1cm}+
\Theta (-x_2)\frac{2s(1-x_1)\bar zx_2+k^2_\perp
(2-x_1-\zeta)}{ x_2\zeta}\right] +
{\cal O}(k^0_\perp) \ . 
\eea
Since the integration is restricted to the region $k_\perp^2 \ll Q^2$ 
the constant terms ${\cal O}(k^0_\perp)$ 
can be omitted. Note the $+i\epsilon$ prescription 
to go around the singularity as a ``memory'' of the direction 
in which the analytical continuation to the complex $x_1$ plane is performed, 
which, in turn, has its origin in the Feynman $+i\epsilon$ prescription 
in the quark propagator $1/[(\zeta p_2 -k-\bar zp_1)^2+i\epsilon]$.  

Calculation of the other diagrams in Fig.~\ref{fig:vector} is similar.
Here we cite only the final result, cf.  \cite{BFGMS94,Rad96a}:
\beq{rho}
{\cal M}=\frac{4\pi \sqrt{4\pi\alpha}\, e_q\alpha_s f_V}{N_cQ}\int\limits^1_0 dz \frac{\phi_V (z)}{
z\bar z}
\int\limits^1_0 dx_1 \frac{\sqrt{1-\zeta} {\cal F}_\zeta^g
(x_1)}{(x_1-i\epsilon)(x_2 +i\epsilon)} \ ,
\eeq
where
\beq{g-skewed}
{\cal F}_\zeta^g (x)=\frac{\al_S}{2\pi}C_F
\left[
\frac{1+(1-x)^2-\zeta}{1-\zeta}-\Theta (\zeta-x)\frac{(\zeta -
x)(2-x-\zeta)}{\zeta (1-\zeta)}
\right]\int \frac{dk^2_\perp}{k^2_\perp} \,
\eeq
can be identified with the (perturbative) generalized gluon distribution of
a quark%
\footnote{Eq.~\re{rho} is written assuming a single light quark 
flavor. For $\rho^0$ meson one has to substitute $e_q f_V \to f_\rho/\sqrt{2}$
with $f_\rho \simeq 200$~MeV.}. 
Owing to the $+i\epsilon$ prescription to go around the singularity
at $x_2=0$ the amplitude acquires an imaginary part 
\beq{rhoim}
{\rm Im}\ {\cal M}=
\frac{-4\pi^2 \sqrt{4\pi\alpha}\, e_q\alpha_s f_V s}{N_c Q^3}
\int\limits^1_0 \!dz\,\frac{\phi_V
(z)}{ z\bar z} {\cal F}_\zeta^g (\zeta)\sqrt{1-\zeta} \, .
\eeq
Note that the $1/k^4_\perp$ terms and the double pole, 
$\sim 1/(x_2+i\epsilon)^2$ present in Eq.~(\ref{a1}) cancel 
in the gauge invariant sum of Feynman diagrams, as expected.

{}For comparison, let us do the same calculation in the light-cone 
gauge, with the propagator 
\beq{propagator}
{D}_{\mu\nu}(k) = \frac{-i}{k^2+i\epsilon}
\left[
g_{\mu\nu}-\frac{k_\mu p_{1\nu}+ k_\nu p_{1\mu}}{(k p_1)}
\right] \ .
\eeq
{}For the moment we do not specify a particular prescription to go around 
the singularity at $k p_1 =0$, this choice will be discussed in detail 
in what follows.

Using this expression we obtain the following result 
for the radiation factor (\ref{Rfac}) in the light-cone gauge:
\beq{Nlc}
{\cal R}^{\rm LC}_{\mu\nu}= \frac{ 2}
{x_1\,x_2}\,
\frac{{\cal N}_{\mu\nu}}
{[k^2+i\epsilon][(k-\zeta
p_2)^2+i\epsilon][(p_2-k)^2+i\epsilon]}
\ ,
\eeq
where the numerator ${\cal N}_{\mu\nu}$ is equal to 
\bea{w}
{\cal N}_{\mu\nu} &=&
(2-\zeta)k_{\mu}^\perp k_{\nu}^\perp + \frac12 \alpha s x_1 x_2
g^\perp_{\mu\nu}
+p_{1\mu}p_{1\nu}[6\alpha k^2_\perp/s +8\alpha^2(1-x_1)]
\nonumber \\
&& {}
+ k^\perp_{\mu}p_{1\nu}[2k^2_\perp/s +\alpha
(4-3x_1)]+k^\perp_{\nu}p_{1\mu}[2k^2_\perp/s +\al (4-3x_1-\zeta)]
\eea
and the prefactor $1/(x_1x_2)$ comes from $1/(kp_1)\cdot 1/[(k-\zeta p_2)p_1]$
in the propagators (\ref{propagator}) of the $t$-channel gluons.

The integration over the Sudakov variable $\alpha$ (\ref{Sudakov}) converges
at values $\alpha \sim k_\perp^2/s$. The numerator in (\ref{Nlc}) is,
therefore, of order ${\cal O}(k^2_\perp)$ and the 
radiation factor (\ref{Rfac}) in the light-cone gauge has at most a 
$1/k_\perp^2$ singularity at small transverse momenta, compared to 
$1/k_\perp^4$ in the Feynman gauge. Because of this, the QCD factorization 
in light-like gauge is valid for each diagram in Fig.~\ref{fig:vector}
separately: In the upper parts of the diagrams one can 
substitute $k = x_1 p_2 + \alpha p_1 +k_\perp \to x_1 p_2$ and neglect 
the gluon virtuality and transverse momentum altogether. This is how 
the parton picture emerges:  The amplitude for hard exclusive meson 
production is given by the convolution of 
the scattering amplitude off the on--shell transverse
gluon $\gamma^* g\to V g$ and the gluon distribution in the target quark.
In perturbation theory the gluon distribution is given by the integral
\beq{fa}
\int d\alpha \int d^2 k_\perp\, {\cal R}^{\rm LC}_{\mu\nu}\sim 
g_{\mu\nu}^\perp\, \frac{{\cal F}_\zeta^g (x_1)}{x_1x_2}\,.
\eeq
In fact, only the two first terms in (\ref{w}) contribute.    
Performing the integral over $\alpha$ and 
averaging over the directions of $k_\perp$ we indeed reproduce
the result in Eq.~(\ref{g-skewed}).

The argument presented above is standard and tacitly assumes that the
singularities at $x_2\to 0$ and $x_1\to 0$ play no r\^ole%
   \footnote{The singularity at $x_1\to 0$ is in fact irrelevant since the 
   gluon distribution ${\cal F}_\zeta^g(x_1)$ 
   (\ref{naiveFg}) vanishes at this point \cite{Rad96a}, cf. 
   Eq.~(\ref{g-skewed}).}. 
In order to 
recover the correct prescription in Eq.~(\ref{rho}) we have to be 
more careful. By the explicit calculation of the six diagrams in  
Fig.~\ref{fig:vector} we obtain the following expression:
\beq{l.c.}
{\cal M}=
\frac{i\alpha_s}
{4\pi^3}
\int 
\frac{
d(\alpha s)\,dx_1\,d^2k_\perp \,[(2-\zeta ){ k}^2_\perp
-(\alpha s)x_1x_2]\sqrt{1-\zeta }
}
{ x_1x_2\,
[(\alpha s)x_1-k^2_\perp +i\epsilon ]
[(\alpha s)x_2-k^2_\perp +i\epsilon ]
[(\alpha s)(x_1-1)-k^2_\perp +i\epsilon ]
} \cdot J
\eeq
where
\bea{hm}
J&=&\frac{2\pi \sqrt{4\pi\alpha}\, e_q\alpha_sf_V}{ N_c Q}\int\limits^1_0
dz\, \phi_V(z)\left[
\frac{x_2s}
{[(\alpha +\bar z)s x_2-{k}^2_\perp +i\epsilon]}-
\frac{x_1s}
{[(\alpha - z)s x_1-{k}^2_\perp
+i\epsilon]} \right.  \nonumber \\
&&\left.
{}-
\frac{{k}^2_\perp (\zeta s)}
{[(\alpha +\bar z)s x_2-{k}^2_\perp +i\epsilon][(\alpha -
z)s x_1-{k}^2_\perp
+i\epsilon]} + (z\leftrightarrow \bar z)
\right] \ . 
\eea
The three terms in the square brackets in (\ref{hm}) correspond to 
the diagrams in Fig.~\ref{fig:vector}a,b,c, respectively,  while 
the symmetric contributions  $(z\leftrightarrow \bar z)$ originate from the 
remaining three diagrams. If $x_1, |x_2| \gg k^2_\perp/s$ then both 
$\alpha$ and $k^2_\perp$ can be neglected in this expression and the factor $J$
is easily recognized as the $\gamma^*g\to V g$ on-shell amplitude
\beq{Jlc}
J\to {\cal M}_{\gamma^* g\to V g} \sim  \int\limits^1_0 dz\, \frac{\phi_V
(z)}{
z\bar z} \,.
\eeq
Note that in this case the diagrams in Fig.~\ref{fig:vector}c and 
Fig.~\ref{fig:vector}f do not contribute.
Performing the remaining integration and collecting all factors we recover
the result in (\ref{rho}) obtained earlier in the Feynman gauge.
On the other hand, the contribution of the singularity at $x_2=0$, or, 
equivalently, of the imaginary part of the amplitude requires some 
attention. First note that the expression for $J$ vanishes 
identically when $x_1\to 0$ or $x_2\to 0$. As a consequence, 
the $1/(x_1 x_2)$ factor appearing in Eq.~(\ref{l.c.}) does not in fact 
produce any singularity 
{\em in the gauge invariant sum of all Feynman diagrams}.
The result of the calculation, therefore, does not depend on a particular 
prescription to go around the auxilary singularity in the gluon 
propagator in the light-cone gauge. Any prescription produces the 
same result. Since the poles at $x_2=0$ and $x_1=0$ are spurious, we are 
left with the imaginary parts corresponding to $+i\epsilon$ prescription
in the Feynman propagator in (\ref{hm}) and invoking the standard argument 
with the contour deformation recover the $x_2+i\epsilon$ in (\ref{rho}).    
The gauge prescription becomes important, however, if one insists 
on QCD factorization in each Feynman diagram separately. 
Consider $x_2\to 0$ and two different prescriptions 
for the factor $1/[(k-\zeta p_2)p_1\pm i\epsilon]$ in the gluon propagator. 
The first term in (\ref{hm}) corresponding to the diagram in 
Fig.~\ref{fig:vector}a contains a factor $x_2$ in the numerator and is not
affected. In the second term, corresponding to Fig.~\ref{fig:vector}b,  
one can still take the light-cone limit and neglect $\alpha$ and $k_\perp$.
The contribution of this diagram then reads 
\beq{bb}
{\cal M}_{(b)}\sim \int dx_1\frac{{\cal F}_\zeta^g(x_1)}{x_1(x_2\pm
i\epsilon)}\frac{1}{z} \,.
\eeq
It has both real and imaginary parts and the sign of the imaginary part 
depends on the gauge prescription. Note that in covariant gauges this 
`crossed box' diagram has no imaginary part in the physical region.
The third contribution in Eq.~(\ref{hm}) originates from 
the diagram in Fig.~\ref{fig:vector}c and is more delicate.
This contribution vanishes in the `naive' light-cone limit. 
In the vicinity of the point $x_2=0$ the small
factor ${k}^2_\perp$ in the numerator is compensated, however,
by the small denominator:
\beq{c}
{\cal M}_{(c)}\sim \int dx_1\frac{{\cal F}_\zeta^g(x_1)}{ x_1(x_2\pm
i\epsilon)}\frac{1}{z}\frac{{k}^2_\perp}{[(\alpha+\bar z)x_2s-{
k}^2_\perp +i\epsilon]} \, .
\eeq
At small ${k}^2_\perp$ and $\alpha \sim {k}^2_\perp/s$ this reduces to 
\beq{cc}
{\cal M}_{(c)}\sim \int dx_1\frac{{\cal F}_\zeta^g(x_1)}{x_1}\frac{1}{z}
\left( -\frac{1}{[x_2\pm i\epsilon]}+\frac{1}{[x_2 + i\epsilon]} \right) \,.
\eeq     
We see that first, the gauge poles $[x_2\pm i\epsilon]$ indeed cancel in the
sum of diagrams in Fig.~\ref{fig:vector}b and Fig.~\ref{fig:vector}c and,
second, the  diagram  in Fig.~\ref{fig:vector}c is equal to zero in the
light--cone limit if and only if one chooses the  $+i\epsilon$ 
prescription for the gauge pole.   The reason
becomes obvious if one tries to deform the integration contour in $x_1$ away
from the singularity at $x_1=\zeta (x_2=0)$, as shown in 
Fig.~\ref{fig:pinch}.
If the  $+i\epsilon$ prescription for the 
gauge pole is used,  both the gauge pole and 
the pole of the quark propagator lie below the real axis.
Hence the contour deformation is not obstructed and  
taking the limit $k_\perp\to 0$ we obtain zero, the parton model 
result for the contribution of this diagram.
On the other hand, if the $-i\epsilon$ prescription
is used, the gauge pole appears to be above the real axis and in the 
$k_\perp\to 0$ limit is pinched with the pole of the quark propagator.
In this case one cannot move the contour away from the singularity.
As the result, the diagram  in  Fig.~\ref{fig:vector}c
acquires a non-zero imaginary part which has no parton model interpretation
and, most importantly, is missed in the `naive' calculation when 
$k_\perp$ and $\alpha$ are put to zero at the beginning.   

The net outcome of our discussion is the verification of the ansatz
in Eq.~(\ref{Rad+}) for the leading-order contribution in the strong 
coupling: One can calculate the hard coefficient function as  
on-shell $\gamma^* g \to V g$ amplitude, and use the prescription 
in (\ref{Rad+}) for the definition of the gluon distribution in 
Eq.~(\ref{naiveFg}). In order to assemble the all-order proof, 
one has to show that to arbitrary order in perturbation theory 
the poles of Feynman propagators corresponding 
to soft gluon exchanges all lie below the $x_2$ real axis.  The corresponding 
discussion can be found in \cite{CF98}. The physical interpretation is that 
all soft gluon exchanges can be thought of as part of the final state 
interaction between the outgoing hadrons; they are reduced to eikonal 
factors that enter the (gauge-invariant) definition of the gluon 
distribution. Pinching of the singularities in this case cannot occur since 
initial state soft interactions are not present.  The interpretation of 
the $\pm i\epsilon$ prescriptions in (\ref{Rad+}) as due to soft final state 
interactions allows for an alternative derivation of this result, 
by noticing that the suitable boundary condition for the gluon field 
in the light-cone gauge is such that \cite{IKR85,Rad96a}    
\beq{time+}
A_\mu (y)=p_1^\nu \int\limits^{\infty}_0 G_{\mu\nu}(y+\sigma
p_1)e^{-\epsilon \sigma }d\sigma \ 
\eeq
with the integration extended to plus infinity in time. We remind that 
$p_1$ is the pion momentum. It is easy to 
check \cite{Rad96a} that the ansatz in (\ref{Rad+}) is an immediate 
consequence of this relation. 
A still another  interpretation of this ansatz is that the amplitude of 
hard exclusive meson production only has a $s$-channel discontinuity in 
the physical region. Its energy dependence  has, therefore, to be a 
function of $s+i\epsilon$ which translates to $\zeta \to \zeta -i\epsilon$   
and $x_2 \to x_1-\zeta +i\epsilon = x_2+i\epsilon$, respectively. 
In any case, we see that the ansatz in (\ref{Rad+}) is specific for the 
considered process and cannot be taken over for the hard dijet production 
without a careful analysis. In fact, the very existence of such an ansatz
is a consequence of QCD factorization%
\footnote{
An instructive  example is provided by  the  process  
$V p\to \gamma^*(Q^2) p$ where the photon in the final state 
has positive virtuality. This reaction is similar to electroproduction
and can be treated in the same way.
In this case, however, the leading-order calculation yields the prescription 
$[x_2-i\epsilon]$, i.e. opposite to the ansatz in (\ref{Rad+}).
The reason is that the soft interaction is in the initial state
in this case, and in order to preserve the parton picture one has
to use a different gauge condition.  
Another difference is that the amplitude $V p\to \gamma^*(Q^2) p$ 
has nonzero dispersive parts in both variables $s$ and $Q^2$. 
One can check that the contributions of $s$--channel singularities all cancel 
in the light--cone limit with part of the singularities in the 
$Q^2$--channel. Hence one is left with  $Q^2$--channel
singularities alone, and the dependence of the amplitude on 
$[x_1-\zeta-i\epsilon]$ 
can be understood as the replacement
$Q^2\to Q^2+i\epsilon$. }.  

\subsection{Dijet production: Dispersion approach}

Since the theoretical status of hard dijet production continues to be 
controversial \cite{NSS99,FMS00,BISS01,Che01}, in this paper 
we will present our calculation using two different 
techniques and show that both lead to the same result. 
In this section we  calculate the imaginary part 
of the $\pi\to 2$~jets scattering amplitude on a quark target in the 
high energy limit.
The imaginary part is interesting in several respects. First, we have 
found in Sec.~2 that the imaginary part of the quark exchange 
is affected by the pinch singularities and 
contains logarithmic end-point divergencies
which destroy collinear factorization. Since Feynman diagrams 
describing quark and gluon contributions have similar topologies, the same 
problem is expected for the gluon contribution as well.
Second, at high energies the scattering amplitude corresponding to Pomeron
exchange is dominated by its imaginary part, so that this contribution 
is numerically the most important one. Last but not least, the cut 
diagrams appearing in the calculation of the imaginary part are     
built of tree-level on-shell scattering amplitudes and their form is 
strongly constrained by gauge invariance, see below. This simplification 
has been widely used in the literature in the calculations of high-energy 
asymptotics of scattering amplitudes starting from \cite{ChWu,LiFr}.  
The idea to use this approach for the dijet production was 
suggested in \cite{FMS00}.   

\begin{figure}[htbp]
\centerline{\epsfxsize16.0cm\epsffile{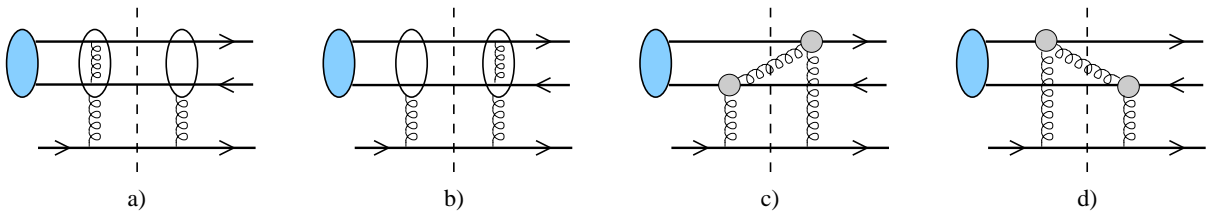}}
\caption[]{\small 
The decomposition of the imaginary part of the amplitude $\pi q\to (\bar q q)q$
into four gauge-invariant contributions.   
 }
\label{fig:3}
\end{figure}

\begin{figure}[htbp]
\centerline{\epsfxsize12.0cm\epsffile{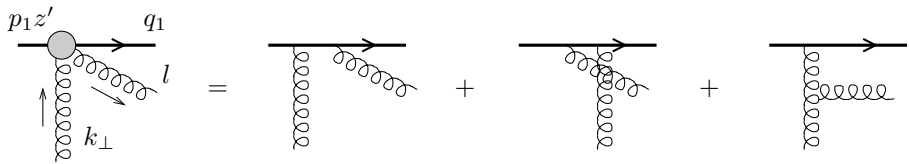}}
\caption[]{\small 
The effective vertex.  
 }
\label{fig:4}
\end{figure}

The $s$-channel discontinuty of the amplitude $\pi q \to 2~\mbox{\rm jets}+q$
is described by the cut diagrams shown in Fig.~\ref{fig:3}. They can be  
grouped into the  four gauge-invariant contributions in Fig.~\ref{fig:3}a--d
%
%
which differ by the position of the hard gluon that provides the large 
momentum transfer to the jets. 
The corresponding contributions to the imaginary part of the amplitude 
will be denoted by 
${\cal C}_{(a)}$, ${\cal C}_{(b)}$, ${\cal C}_{(c)}$ and 
${\cal C}_{(d)}$, respectively.  For example, in  Fig.~\ref{fig:3}a it is 
assumed that the hard gluon exchange appears to the left of the cut.
This contribution is given by the sum of 10 Feynman diagrams, 
for further details see our letter \cite{BISS01}. 
Similarly, the contribution in    
Fig.~\ref{fig:3}b is given by the sum of 10 diagrams with the hard gluon 
exchange appearing to the right of the cut.
The two remaining contributions in Fig.~\ref{fig:3}c 
and Fig.~\ref{fig:3}d take into account the possibility of 
real gluon emission in the intermediate state. The filled circles 
stand for the effective vertices describing gluon 
radiation, see Fig.~\ref{fig:4}. Each of the two contributions in   
Fig.~\ref{fig:3}c,d corresponds to a sum of 9 different Feynman diagrams. 
%
%

\begin{figure}[t]
\centerline{\epsfxsize 7.5cm\epsffile{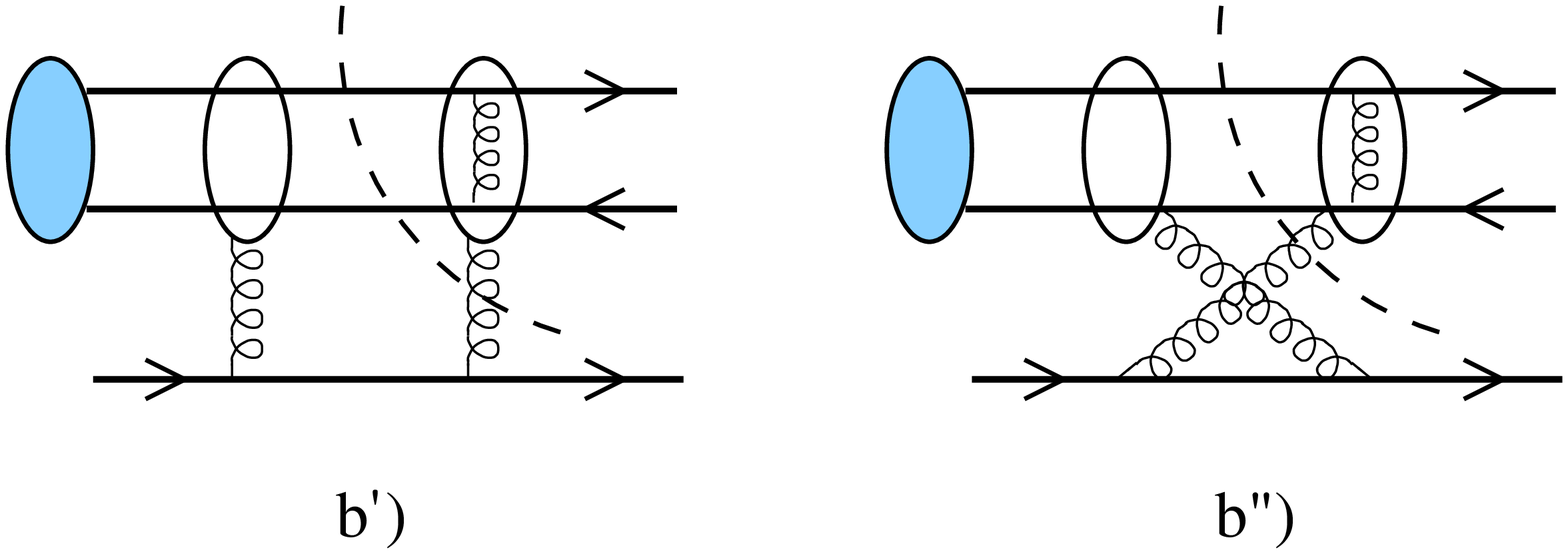}}
\caption[]{\small 
The $M^2$-discontinuity of the amplitude $\pi q\to (\bar q q)q$.
 }
\label{M2cut}
\end{figure}

At first sight, we have to consider the discontinuity in the invariant mass
$M^2$ of the jets as well. It is easy to check that the corresponding cut 
diagrams have to have the hard gluon exchange to the right of the cut in
order that the transition from the intermediate to the final state corresponds
to a physical process. On the other hand, unlike for the $s$-discontinuity, 
the $t$-channel gluon lines can be crossed in this case. 
The two possible cut diagrams are shown in Fig.~\ref{M2cut}.
We denote the corresponding contributions by  
${\cal C}_{(b')}$ and  ${\cal C}_{(b'')}$, respectively. 
They differ by the interchange of the Mandelstam variables $s$ and $u$. 

%
%
It is known for a long time that at high-energies 
any amplitude with vacuum quantum numbers exchange in the $t$-channel 
has to be crossing-symmetric, or in other words has positive signature.
In our case we are dealing with a two-gluon exchange that grows linearly
with energy and consider a one-loop amplitude which adds a logarithm. 
The crossing symmetry then implies that the contribution to the
scattering amplitude of which ${\cal C}_{(b)}$, ${\cal C}_{(b')}$ and  
${\cal C}_{(b'')}$
are the three nonvanishing dispersion parts in the physical region 
of the $s$-channel has the following schematic structure:
\beq{signature}
{\cal M} \sim s\ln \left(\frac{-s}{-M^2}\right)+u\log \left(\frac{-u}{-M^2}\right) +
{\cal O}\left( s^0\right) \,.
\eeq
At high energies $u \simeq -s$ and the above structure implies 
that that the $s$-channel and the $M^2$-channel discontinuities 
are related: ${\cal C}_{(b)}=-{\cal C}_{(b')}$ and 
${\cal C}_{(b')}=-{\cal C}_{(b'')}$.
The first equality  ${\cal C}_{(b)}=-{\cal C}_{(b')}$  which amounts to
the cancellation of the $s$- channel and
$M^2$- channel discontinuities is not specific for high energies, but 
rather a general property of the scaling limit:
In so far as the amplitude is a function of the dimensionless ratio
$(-s)/(-M^2)$, the $s$- and $M^2$- discontinuities cancel each other 
identically%
\footnote{We have already met with an example of such cancellation
for the quark contribution, see the discussion of cut diagrams 
in Fig.~\ref{fig:quark_cuts}(2c-I) and 
Fig.~\ref{fig:quark_cuts}(2c-II).}.
It follows that the total imaginary part of the amplitude 
$\pi q\to (\bar q q)q$ in the scaling limit is given by the sum of the
three $s$-channel cut diagrams in Fig.~\ref{fig:3}a,c,d and the 
$M^2$-cut diagram with crossed $t$-channel gluon lines 
in Fig.~\ref{M2cut}b$^{\prime\prime}$. However, at high energies    
the crossing symmetry implies that ${\cal C}_{(b'')}={\cal C}_{(b)}$  
and the net result is that the $M^2$-discontinuity can be neglected 
altogether. We are left, therefore, with the set of cut diagrams shown in 
Fig.~\ref{fig:3}.

The general strategy of the calculation is the following.
As is easily checked by inspection, in any cut diagram 
two internal lines are on the mass shell.
The corresponding two on-shellness conditions fix $\alpha$ and $x_1$
variables in the Sudakov parametrisation \re{Sudakov} of the gluon
momentum and relate the variables $z^\prime$ and $x_1$ 
(see Fig.~\ref{fig:1} for the notations) to one other. 
Since the $\alpha$ and $x$-variables for both gluons are of the order 
of $1/s$,  
the $1/k_{1,2}^2$ factors in the propagators of the $t-$channel gluons 
can be approximated by using $k^2 \simeq - k^2_{\perp}$, to the 
${\cal O}(1/s)$ accuracy. 
In this calculation we use the Feynman gauge and perform the usual 
substitution 
$
g_{\mu\nu} \rightarrow {p_2^\mu\,p_1^\nu}/{(p_1p_2)}
$
in  the numerators of the $t-$channel gluon propagators, which is valid 
to the same accuracy.
Using  the on-shellness conditions for the contributions
in Fig.~\ref{fig:3}a and Fig.~\ref{fig:3}b one obtains 
$x_1=\zeta$, $x_2=0$, for any $z^\prime $. 
For Fig.~\ref{fig:3}d one finds  
$x_1={\zeta z^\prime \bar z}/{(z^\prime -z)}$,
$x_2={\zeta z \bar z^\prime }/{(z^\prime -z)}$ and $z^\prime > z$, where
the last condition ensures that the energy of the cut gluon
is positive.
Finally, for the set of cut-diagrams corresponding to  
Fig.~\ref{fig:3}c  we obtain 
$x_1={\zeta z \bar z^\prime}/{(z-z^\prime)}$,
$x_2={\zeta z^\prime  \bar z}/{(z-z^\prime)}$ and $z > z^\prime$. 

After the on-shellness conditions are used, 
a single integration over the gluons transverse momentum $k_\perp$ remains:
\beq{impact-rep}
{\rm Im}\;{\cal M} \sim \int\;
\frac{d^2k_\perp}{(k^2_\perp
)^2}\;J_{up}(k_\perp,q_{\perp})\,J_{down}(k_\perp,q_{\perp})\,,
\eeq
where $k^4_\perp$ comes from the product of the two gluon propagators. 
$J_{up}$ and $J_{down}$ are dubbed impact factors and 
stand for the upper and the lower parts of the diagrams in 
Fig.~\ref{fig:3}a--d that are connected by the two-gluon exchange. 
The representation \re{impact-rep} is  well known \cite{ChWu,LiFr} from 
studies of QED scattering at high energies.

Properties of the impact-factors $J_{up}$ and
$J_{down}$ as a functions of $k_{\perp}$ at $k_\perp\to 0$ 
are of crucial importance. 
Since $J_{down}$ is the impact-factor of a point-like target quark, 
$J_{down}(k_\perp,q_\perp)\sim const$. On the other hand,
$J_{up}(k_\perp,q_\perp)$ stands for the scattering of the colorless
$q\bar q$ (Fig.~\ref{fig:3}a--b) or $q\bar q G$ (Fig.~\ref{fig:3}c--d)
state having a transverse size $\sim 1/q_{\perp}$ and 
has to vanish at small $k_\perp \ll q_{\perp}$,
$J_{up}(k_\perp,q_\perp)\sim k^2_{\perp}$, as a consequence
of the color neutrality of the quark-antiquark pair: 
A gluon with a large wave length $\sim 1/k_\perp$ cannot resolve 
a color dipole of the small size $\sim 1/q_{\perp}$.
Since in our case there are two gluons, $J_{up}$ is proportional to 
the product $k_{\perp}\cdot k_{\perp}=k^2_{\perp}$%
\footnote{The ${\cal O}(k_\perp^2)$ behavior can be traced 
to the gauge invariance of the amplitude, see \cite{LiFr} for the details.}. 
In the opposite limit of large transferred momenta, 
$k_\perp\gg q_{\perp}$, the two 
$t-$channel gluons are forced to 
couple to the same parton (quark or gluon) in the 
upper block in Fig.~\ref{fig:3}a--d. It follows that at large $k_\perp$
$J_{up}(k_\perp,q_\perp)\sim const$.  

Taking into account the properties of the impact-factors discussed above, we 
conclude that the transverse momentum integration in \re{impact-rep}
diverges logarithmically at small $k_\perp$ and the integral can be 
estimated by ${\cal M} \sim \int^{q_\perp^2} dk_\perp^2/k_\perp^2 \sim 
\ln q_\perp^2$, as expected.    
The region of $k_\perp^2 > q_{\perp}^2$ 
does not produce the large logarithm and can be neglected.
Note that the correct small $k_\perp$ behavior of the impact factors 
is only recovered in the sum of cut diagrams for the gauge 
invariant amplitudes 
${\cal C}_{(a)}$, ${\cal C}_{(b)}$, ${\cal C}_{(c)}$ and 
${\cal C}_{(d)}$, but not for each diagram separately.   

In addition to the diagrams discussed so far, the amplitude 
$\pi q \to (\bar q q)q$ receives a contribution from the three-gluon 
exchange in the t-channel. Such terms can be viewed as belonging to the cut 
diagrams shown in Fig.~\ref{fig:3}a in which the hard gluon in the blob 
is attached  to the bottom quark line. We have checked that this extra  
contribution does not contain the large
collinear logarithm $\ln q_\perp^2$ and therefore we neglect it.

The calculation of ${\cal C}_{(d)}$ proceeds as follows.
Let $l^\mu=\alpha_lp_1^\mu+x_lp_2^\mu+l^\mu_\perp$ be the momentum of the 
(real) gluon in the intermediate state and let $e^\mu(l)$ be one
of the two physical polarization vectors.  The two conditions 
$(e\cdot p_2)=0$ and $(e\cdot l)=0$ fix the gauge and result in
$e^\mu(l) = e_\perp^\mu +
2p_2^\mu\,\left(e_\perp l_\perp\right)/(\alpha_l\,s)$.

The effective vertex corresponding to the sum of the three diagrams 
in  Fig.~\ref{fig:4} has the form
\beq{e-vertex}
i \frac{g^2\,z\,(z'-z)}
{q_{\perp}^2\, z'}\,
\left[\frac{1}{z'}\,\left(t^l\,t^a \right)_{ i \,j}-
\frac{1}{ z}\,\left(t^a\,t^l \right)_{ i \,j}  \right]
 {\bar u}(q_1)
\left[\!\not\! b\,\!\not\! e_\perp \ -2\,\frac{ z}{ z' -
z}(e_\perp b)\right]\,\frac{\!\not\! p_2}{s}\,u(z'p_1) \,. 
\eeq
Here $t^l$ and $t^a$ are the $SU(3)$ generators. The color indices $l$ and 
$a$ belong to the emitted gluon and the $t-$channel gluon,
respectively. We have also
introduced an auxiliary two-dimensional vector $b^\mu$ defined as:
\beq{bvec}
b^\mu = k_\perp^\mu - 2\,\frac{(k_\perp
q_{\perp})}{q_{\perp}^2}\,q_{\perp}^\mu\, ,\;\;\;\; b^2=k_\perp^2\,.
\eeq
Note that the effective vertex, in the limit of small
$k_\perp$,  is proportional to $b\propto k_\perp$. The constant terms
cancel in the gauge invariant sum of the diagrams in Fig.~\ref{fig:4}.

The second effective vertex in Fig.~\ref{fig:3}d has a similar form.
Combining both of them  and
performing the sum over the polarizations of the emitted gluon we obtain the 
impact-factor $J_{up}^{(d)}$.  
Since each effective vertex is proportional to $k_\perp$, it follows that 
$J_{up}^{(d)}\sim k^2_\perp$, as expected. The result for the amplitude 
${\cal C}_{(d)}$ is obtained using the representation in \re{impact-rep}.
The calculation of ${\cal C}_{(c)}$ is very similar. 
The result for their sum reads: 
\bea{Mcd}
\lefteqn{
{\cal C}_{(c)}+{\cal C}_{(d)} = D\,C_F^2\,\int\,\frac{d k_\perp^2}{k_\perp^2}
\,\int\limits_0^1dz'\,
\phi_\pi(z')\,\left(\frac{z\,\bar
z}{z'\,\bar z'}+1 \right)\times}
\nonumber \\
&\times&\left[\left(\frac{z\,\bar z}{z'\,\bar z'}+1\right)
+\frac{1}{(N_c^2-1)}\left(\frac{z}{z'}+ \frac{\bar z}{\bar z'}  \right)
\right]
 \left[ \frac{\Theta(z'-z)}{(z'-z)} + \frac{\Theta(z-z')}{(z-z')} \right],
\eea
where
\beq{D}
D= -i\,s\,f_\pi\,\alpha_s^3\,\frac{4\,\pi^2}{N_c^2\,q_{\perp}^4}\,{\bar
u}(q_1)\gamma_5 \frac{{\!\not\! p}_2}{s}v(q_2)\,
\delta_{i\,j}\,\delta_{c\,c'}\,.
\eeq
The color indices $(i,j)$ correspond to the 
produced quark-antiquark pair (jets) and $(c,c')$ stand for the color 
indices of the target quark in the initial and the final state.
The contributions $\sim \Theta(z'-z)$ and $\sim\Theta(z-z')$ 
belong to ${\cal M}_{(d)}$ and ${\cal M}_{(c)}$, respectively.

For the  cut diagrams in Fig.~\ref{fig:3}a and Fig.~\ref{fig:3}b 
we obtain:
\bea{Mab}
{\cal C}_{(a)} &=& - D\,C_F^2\,\int\frac{d k_\perp^2}{k_\perp^2}
\,\int\limits_0^1 dz'\, \phi_\pi(z')
\left( \frac{\bar z}{z'}+ \frac{z}{\bar z'} \right),
\nonumber\\
{\cal C}_{(b)} &=& D\,C_F^2 \,\int\frac{d k_\perp^2}{k_\perp^2}
\,\int\limits_0^1 dz'\,
\frac{\phi_\pi(z')}{z'\bar z'}\,
\left[z \bar z\left(\frac{\bar z}{z'}+\frac{z}{\bar z'}
\right) +\frac{1}{(N^2-1)}\left(\frac{z\bar z}{z'\bar z'}+1
\right)   \right].
\eea

The transverse momentum integrals 
in \re{Mcd} and \re{Mab} can be identified
with the small $x$ limit, $x,\zeta \ll 1$, of the (perturbative) 
generalized gluon distributions of a quark defined in 
Eq.~(\ref{g-skewed}): 
${\cal F}_\zeta^g (x)\simeq \frac{\alpha_s}{\pi}C_F 
\int dk_\perp^2/k_\perp^2$. 
We obtain, in this approximation \cite{BISS01}
\beq{ImM}
\frac{1}{\pi} {\rm Im}\,[i{\cal M}_{\rm gluon}] = 
 \,s\,f_\pi\,\alpha_s^2\,\frac{4\,\pi^2}{N_c^2\,q_{\perp}^4}\,{\bar
u}(q_1)\gamma_5 \frac{{\!\not\! p}_2}{s}v(q_2)\,\tilde {\cal I}\, 
\delta_{i\,j}\,
\eeq
with 
\bea{I}
\tilde {\cal I} &=& 
 \int\limits_0^1\!dz'\, \phi_\pi(z', \mu^2)
\left\{\left[C_F\left(\frac{z\bar z}{z'\bar z'}-1\right)
\left(\frac{\bar z}{z'}+\frac{z}{\bar z'}\right)
+\frac{1}{2N_c\,z' \bar z'}\left(\frac{z\bar z}{z'\bar z'}+1\right) 
\right]\,{\cal F}_\zeta^g(\zeta,\mu^2) \right. 
\nonumber \\
&&\left. +\left(\frac{z\bar z}{z'\bar z'}+1  \right)
\left[C_F\left(\frac{z\bar z}{z'\bar z'}+1\right)+\frac{1}{2N_c}\left(\frac{z}{z'}
+\frac{\bar z}{\bar z'} \right)\right] \right. \nonumber \\
&&\left. \times\left[\frac{\Theta(z'-z)}{(z'-z)}\,{\cal F}_\zeta^g
\left(\frac{\zeta\,z'\bar
z}{z'-z},\mu^2
\right) + \frac{\Theta(z-z')}{(z-z')}\,{\cal F}_\zeta^g
\left(\frac{\zeta\,\bar z' z}{z-z'},\mu^2
  \right)    \right] \right\}.
\eea

The expression in Eq.~(\ref{I}) presents the main result of this section. 
Note that the integrand in \re{I} is singular at $z'=z$ so that 
there is a logarithmic enhancement of the  contribution 
of the integration region $\zeta \ll |z'-z| \ll 1$. In addition, 
there is a logarithmic divergence at the end-points   
$z'\to 0$ and $z'\to 1$ which signals that the collinear factorization
is broken, as expected. In what follows we will  discuss the 
contributions from these regions in some detail. Before doing this, however, 
in the next section we derive the complete result for the amplitude
in the scaling limit (both real and imaginary parts, and including 
${\cal O}(1/s)$ corrections) using a different approach. 

\subsection{Dijet production: Factorization and the light-cone limit} 

Our aim in this section is to derive the complete result for the leading-order
contribution to the kernel $T^g_H(z,z^\prime, x_1,x_2)$ such that  
\beq{G-result}
{\cal M}_{\rm gluon} =\frac{ 4\pi^2 \alpha_s^2
s\, i f_\pi}{N_c^2 q^4_\perp}\sqrt{1\!-\!\zeta}\,
\bar u(q_1)\gamma_5 \frac{\!\not\! p_2}{ s}
v(q_2)\, \delta_{i j}\int\limits^1_0 \!dz^\prime
\, \phi_{\pi}(z^\prime)\!\!
\int\limits^1_0 dx_1\, {\cal F}^{g}_\zeta (x_1)\,
T_{\rm gluon}(z,z^\prime, x_1,x_2)\,,
\eeq
cf. Eq.~(\ref{factor}). For high energies, $\zeta\to 0$, we expect to recover
in this way the result of the direct calculation of the 
imaginary part given in (\ref{I}).  

We will mainly be concerned with the singularity structure of 
$T_{\rm gluon}(z,z^\prime, x_1,x_2)$ at $x_2\to0$.
The singularity structure 
at $x_1\to 0$ can, in principle, be found from similar considerations.
This is in fact not necessary since it can
be established using the crossing 
symmetry $T_{\rm gluon}(z,z^\prime, x_1,x_2) = T_{\rm gluon}
(z,z^\prime, -x_2,-x_1)$ that corresponds to the interchange of 
$s$- and $u$-channels for the corresponding pion-gluon amplitude. 
 Apart from 
this issue, the calculation of $T_{\rm gluon}(z,z^\prime, x_1,x_2)$ 
is straightforward
and can most easily be done by considering pion scattering from on-shell 
transversely polarized gluons, cf. \cite{Che01}. For the 
simpler case of hard exclusive production of vector mesons we have argued
in Sect.~3.1 that the light-cone calculation (with on-shell gluons) is 
actually sufficient since the singularity structure in Eq.~(\ref{rho})
can be restored using the prescription (\ref{Rad+}) in the definition 
of the off-forward gluon distribution (\ref{naiveFg}). One way to understand 
this result was that for hard exclusive production of vector mesons 
it is possible to choose an axial (light-cone) gauge in such a way that 
gauge singularities of the t-channel gluons and causal singularities of 
Feynman propagators lie on the same side of the integration contour in $x_1$ 
so that there are no pinches in the Glauber region. In the present case,
a similar simplification is not expected since soft gluon exchanges 
occur both in the initial and in the final state.

\begin{figure}[hbtp]
\centerline{\epsfxsize14.0cm\epsffile{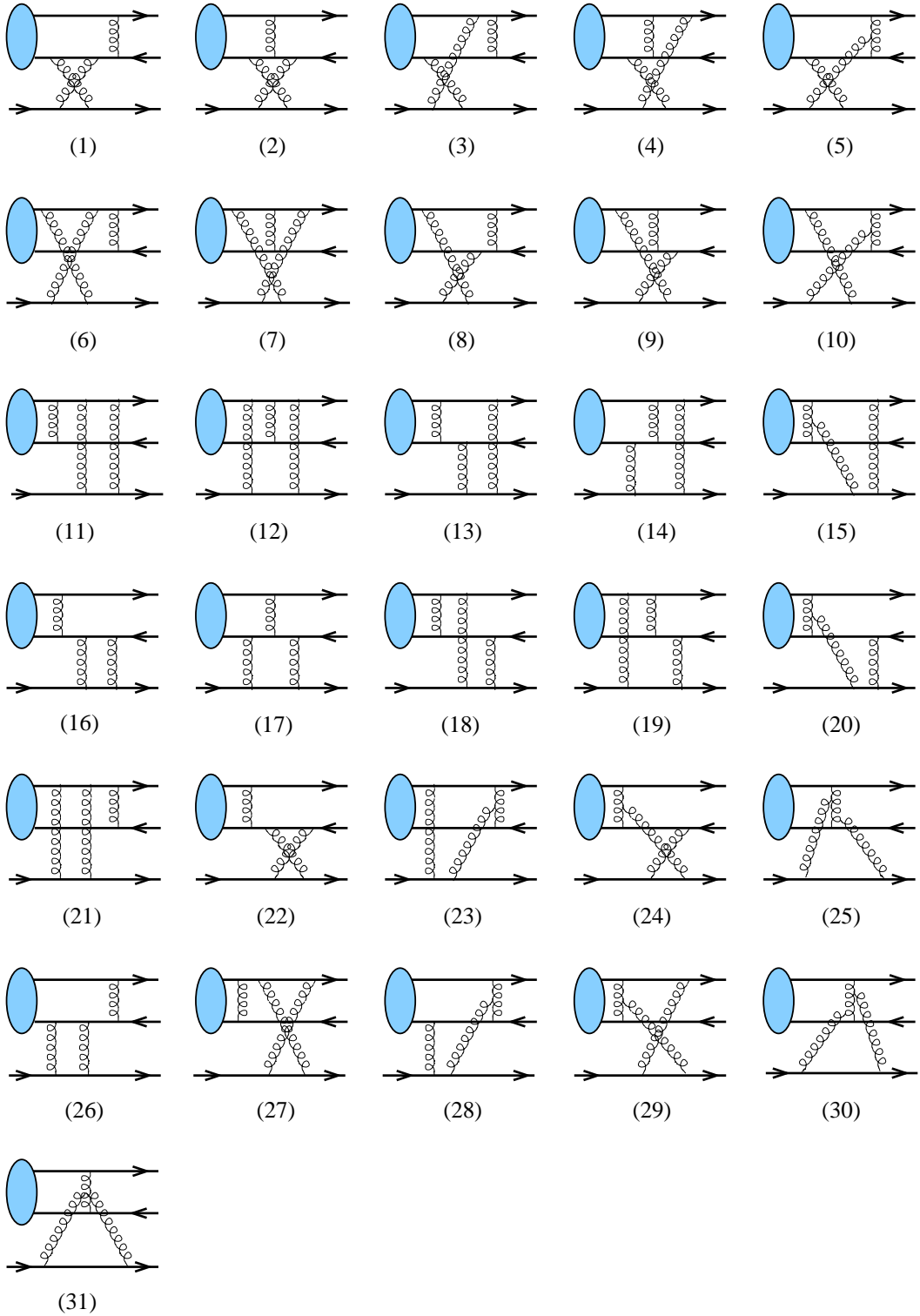}}
\caption[]{\small 
 The complete set of Feynman diagrams for the two-gluon exchange contribution
 to the process $\pi q \to (\bar q q)q$ to 
 the leading order in the strong coupling.     
 }
\label{31}
\end{figure}

The complete set of relevant Feynman diagrams is shown in Fig.~\ref{31}.
%
%
{}For 20 out of in total 31 existing diagrams, Fig.~\ref{31}(1-20), 
 it is kinematically possible that the $t$-channel gluon with 
vanishing momentum fraction $x_2$ couples 
either to the initial or to the final on-shell quark lines. 
The corresponding quark propagators produce a singularity in the 
$x_2\to0$ limit. In particular, in the 5 diagrams in Fig.~\ref{31}(11-15)
where the soft gluon is attached to the outgoing  quark,
the quark propagator is
\beq{prop1}
1/[(q_1+k-\zeta p_2)^2+i\epsilon]=1/[(z+\alpha)(x_1-\zeta
z)s-(q_{\perp}+k_\perp)^2+i\epsilon] \to 1/[zs(x_2+i\epsilon) ]
\eeq
in the limit $\alpha , k_\perp\to 0$. Similarly, in the 
5 diagrams in Fig.~\ref{31}(16-20) where the gluon is attached 
to the outgoing  antiquark, one obtains
\beq{prop2}
1/[(q_2+k-\zeta p_2)^2+i\epsilon]=1/[(\bar z+\alpha)(x_1-\zeta
\bar z)s-(q_{\perp}+k_\perp)^2+i\epsilon] \to 1/[\bar zs(x_2+i\epsilon)]
\eeq                                                        
in the same limit. However, in the 5 diagrams in Fig.~\ref{31}(1-5) 
where the soft gluon is emitted by
the ingoing  antiquark, the denominator of the quark propagator is
\beq{prop3}
1/[(\bar z^\prime p_1-k+\zeta p_2)^2+i\epsilon]=
 1/[(\bar z^\prime -\alpha)(\zeta - x_1)s
-k_\perp^2+i\epsilon] \to  1/[-\bar z^\prime s (x_2-i\epsilon)]
\eeq
and similar for the diagrams in in Fig.~\ref{31}(6-10) 
\beq{prop4}
1/[(z^\prime p_1-k+\zeta p_2)^2+i\epsilon]=1/[(z^\prime -\alpha)(\zeta - x_1)s
-k_\perp^2+i\epsilon] \to  1/[-z^\prime s (x_2-i\epsilon)]\,.
\eeq
We see that both types of singularities are produced, $(x_2\pm i\epsilon)$,
depending on whether the soft interaction occurs in the initial or in 
the final state \cite{CSS83}. As a consequence, it is not possible to 
choose an axial gauge in which the light-cone limit exists for each Feynman 
diagram separately. 

For definiteness, let us define the light-cone gauge in the same way as
this was convenient for hard exclusive production of vector mesons,
$1/x_2 \to 1/(x_2+i\epsilon)$. In this case the diagrams with soft 
interaction in the final state Fig.~\ref{31}(11-20) and the diagrams 
that are non-singular in the $x_2\to0$ limit Fig.~\ref{31}(21-31) can be 
calculated in the light-cone limit, with on-shell $t$-channel gluons,
and the result multiplied by the gluon distribution (\ref{naiveFg})
using the prescription in (\ref{Rad+}). On the other hand, in the   
diagrams with soft interaction in the initial state Fig.~\ref{31}(1-10)
the integration contour over $x_1$ will be pinched. We can
calculate their contribution by deforming the integration contour 
in the same direction as for the other diagrams $x_2=x_1-\zeta+i\epsilon$, 
and adding the contribution of the pole of the quark propagator   
\re{prop3} and \re{prop4}. In what follows we refer to these two 
contributions as `naive' light-cone $T_i^{\rm LC}$ and pole 
$T_i^{\rm pole}$ contributions, respectively. The subscript `$i$' 
numerates the diagrams according to Fig.~\ref{31}. The additional pole 
contributions are present (with our gauge choice) in the diagrams 
in Fig.~\ref{31}(1-10) so that 
\beq{naive+pole}
T_{\rm gluon}(z,z^\prime, x_1,x_2)= T^{\rm LC} + T^{\rm pole} =   
\sum_{i=1}^{31} T_i^{\rm LC} + 
 \sum_{i=1}^{10} T_i^{\rm pole}\,.
\eeq
 $T^{\rm LC}$ can be calculated as the gauge-invariant on-shell pion-gluon
scattering amplitude, times the gauge factor \re{Rad+} 
\beq{pion-gluon}
  T^{\rm LC} = \frac{\zeta}{(x_1-i\epsilon)(x_2+i\epsilon)} \sum_{i=1}^{31} 
 {\cal A}^{\pi g\to (\bar q q)g}_i 
\eeq 
where $A_i$ correspond to the upper parts of the diagrams in Fig.~\ref{31},
not including the target quark and the propagators of the $t$-chanel gluons.
Since ${\cal A}^{\pi g\to (\bar q q)g} = \sum_{i=1}^{31} 
 {\cal A}^{\pi g\to (\bar q q)g}_i $ is the amplitude of a physical process, 
it is gauge invariant and can be calculated in any gauge. 
 In the Feynman gauge we find   
\bea{feyresults}
&&{} {\cal A}_1=C_F\frac{\bar zz^2}{z^\prime\bar z^\prime}\,,  \quad 
{\cal A}_2=-\frac{1}{2N_c}\frac{ z^2(x_1-z\zeta)}{z^\prime\bar
z^\prime[x_1-i\epsilon]}\, , \quad 
{\cal A}_{16}=C_F\frac{z^2(2\zeta\bar z-x_1)}{z^\prime[x_2+i\epsilon]}\, , 
\nonumber \\
&&{} 
{\cal A}_3=-C_F\frac{z^2\bar z^2\zeta}{z^\prime \bar
z^\prime[(z^\prime-z)x_1-z^\prime \bar z \zeta +i\epsilon]}\, , \quad
{\cal A}_4=-\frac{1}{2N_c}\frac{z^2\bar z^2\zeta^2}{ \bar        
z^\prime [x_1-i\epsilon][(z^\prime-z)x_1-z^\prime \bar z \zeta +i\epsilon]}\, ,
\nonumber \\
&&{}
{\cal A}_5=\frac{N_c}{4}\frac{\bar z z^2(2x_1+\zeta(1-z^\prime -2z))}{z^\prime
\bar        
z^\prime[(z^\prime-z)x_1-z^\prime \bar z \zeta +i\epsilon]} \, , \quad
{\cal A}_{20}=-\frac{N_c}{4}\frac{z^2(2x_1^2-3\zeta x_1\bar z+\zeta^2
\bar z(3-4z))}
{z^\prime [x_2+i\epsilon][(z^\prime-z)x_1-z^\prime \bar z \zeta
+i\epsilon]} \, ,
\nonumber \\
&&{}
{\cal A}_{31}=-N_c\frac{z\bar z}{z^\prime \bar
z^\prime}\, , \quad 
{\cal A}_{25}=-\frac{N_c}{2}\frac{z\bar z(\zeta
x_1(4+z^\prime-z)-4x_1^2-\zeta^2(2-5z+zz^\prime+4z^2))}
{z^\prime\bar z^\prime
\zeta[(z^\prime-z)x_1-z^\prime \bar z \zeta
+i\epsilon]} \, , \nonumber \\
&&{}
{\cal A}_{18}=C_F\frac{z\bar z\zeta (\zeta^2 z\bar z-(x_1-\zeta \bar
z)^2)}{[x_1-i\epsilon][x_2+i\epsilon][(z^\prime-z)x_1-z^\prime \bar z \zeta
+i\epsilon]}\, .
\eea 
The remaining contributions are given by the replacements
\bea{repl}
{\cal A}_{1}(z,z^\prime,x_1,x_2)&=&{\cal A}_{6}(\bar z,\bar z^\prime,x_1,x_2)=
{\cal A}_{26}(z,z^\prime,-x_2,-x_1)={\cal A}_{21}(\bar z,\bar
z^\prime,-x_2,-x_1)\, , \nonumber \\
{\cal A}_{2}(z,z^\prime,x_1,x_2)&=&{\cal A}_{7}(\bar z,\bar z^\prime,x_1,x_2)=
{\cal A}_{17}(z,z^\prime,-x_2,-x_1)={\cal A}_{12}(\bar z,\bar
z^\prime,-x_2,-x_1)\, , \nonumber \\
{\cal A}_{4}(z,z^\prime,x_1,x_2)&=&{\cal A}_{9}(\bar z,\bar z^\prime,x_1,x_2)=
{\cal A}_{14}(z,z^\prime,-x_2,-x_1)={\cal A}_{19}(\bar z,\bar
z^\prime,-x_2,-x_1)\, , \nonumber \\
{\cal A}_{5}(z,z^\prime,x_1,x_2)&=&{\cal A}_{10}(\bar z,\bar z^\prime,x_1,x_2)=
{\cal A}_{28}(z,z^\prime,-x_2,-x_1)={\cal A}_{23}(\bar z,\bar
z^\prime,-x_2,-x_1)\, , \nonumber \\
{\cal A}_{16}(z,z^\prime,x_1,x_2)&=&{\cal A}_{11}(\bar z,\bar z^\prime,x_1,x_2)=
{\cal A}_{22}(z,z^\prime,-x_2,-x_1)={\cal A}_{27}(\bar z,\bar
z^\prime,-x_2,-x_1)\, , \nonumber \\
{\cal A}_{20}(z,z^\prime,x_1,x_2)&=&{\cal A}_{15}(\bar z,\bar z^\prime,x_1,x_2)=
{\cal A}_{24}(z,z^\prime,-x_2,-x_1)={\cal A}_{29}(\bar z,\bar
z^\prime,-x_2,-x_1)\, , \nonumber \\
{\cal A}_{25}(z,z^\prime,x_2,x_1)&=&{\cal A}_{30}(z,
z^\prime,-x_2,-x_1)\, , \ \ \
{\cal A}_{3}(z,z^\prime,x_1,x_2)={\cal A}_{8}(z,z^\prime,-x_2,-x_1) \, , \nonumber
\\
{\cal A}_{18}(z,z^\prime,x_1,x_2)&=&{\cal A}_{13}(z,
z^\prime,-x_2,-x_1)\, . 
\eea
For the sum of all diagrams we obtain 
\bea{G-coeff-naive}
T^{\rm LC}(z,z^\prime, x_1,x_2)&=&C_F
\left(
\frac{\bar z}{z^\prime}+\frac{z}{\bar z^\prime}
\right)
\left(
\frac{\zeta}{[x_1-i\epsilon]^2}+
\frac{\zeta}{[x_2+i\epsilon]^2} -
\frac{\zeta}{[x_1-i\epsilon][x_2+i\epsilon ]}
\right) \nonumber \\
&&{}
+
\left(
\frac{z\bar z}{z'\bar z'}+1
\right)
\left[
C_F
\left(
\frac{z\bar z}{z'\bar z'}+1
\right)
+\frac{1}{2N_c}
\left(
\frac{z}{z'}
+\frac{\bar z}{\bar z'}
\right)    
\right]    
\nonumber \\
&&{}
\times
\left(
\frac{1}{[(z-z^\prime)x_1-z\bar z^\prime \zeta+i\epsilon ] }
+
\frac{1}{[(z^\prime -z)x_2-z\bar z^\prime \zeta+i\epsilon ]}
\right)    
  \\
&&{}
-
\left[
C_F
\frac{z\bar z}{z'\bar z'}
\left(\frac{\bar z}{z'}+\frac{z}{\bar z'}\right)
+\frac{1}{2N_c\,z' \bar z'}\left(\frac{z\bar z}{z'\bar z'}+1\right)
\right]    
\frac{\zeta}{[x_1-i\epsilon ][x_2+i\epsilon ]}  \ . \nonumber
\eea
Taking the imaginary part of \re{G-coeff-naive} we reproduce the answer 
for the imaginary part of the coefficient function obtained 
in Ref.~\cite{Che01}%
\footnote{
          In order to make the comparison it is necessary to take into 
          account that the answer given in \cite{Che01} is rewritten 
          in a different form using the symmetry 
          $\phi_\pi(z) = \phi_\pi(1-z)$ of the pion
          distribution amplitude.}.   

As explained above, this is not the whole story, however, and one  
has to add the pole contributions. For example, consider the diagram 
in Fig.~\ref{31}(3). In the notation of Sect.~3.1 we obtain the 
corresponding contribution as   
\bea{M3}
{\cal M}_{3} &=& -g^6 f_\pi \sqrt{1-\zeta}
\frac{C_F^2}{8N_c^2} 
\int \frac{d^4 k}{(2 \pi)^4} \int\limits^1_0 dz^\prime \, \phi_\pi (z^\prime )
{\cal R}_{\mu_1\mu_2}^{\rm LC} \,
N_{\nu_1\nu_2}(k+z^\prime p_1-q_1) \\
&& {}
\times 
\bar u(q_1)\gamma_{\nu_1} \frac{(z^\prime\!\not\! p_1+ \!\not\!
k)}{(z^\prime p_1+k)^2+i\epsilon}
 \gamma_{\mu_1}\gamma_5\!\not\! p_1 \gamma_{\mu_2} 
\frac{(\!\not\! k - \zeta \!\not\! p_2 
-\bar z^\prime  \!\not\! p_1)}{(k - \zeta p_2 -\bar z^\prime  p_1)^2+i\epsilon}
\gamma_{\nu_2} v(q_2)\,,  
\nonumber
\eea
where ${\cal R}_{\mu_1\mu_2}^{\rm LC}$ is defined in Eq.~(\ref{Nlc})
and $N_{\nu_1 \nu_2}(k+z^\prime p_1-q_1) $ is the nominator of the 
hard gluon propagator in axial gauge.
In the present case only the first two terms in Eq.~(\ref{w}) contribute
and according to our procedure we choose the prescription 
$1/x_2\to 1/[x_2+i\epsilon]$ in the corresponding factor in Eq.~(\ref{Nlc}). 
This singularity is due to gauge fixing and is pinched with 
the quark propagator $[(k - \zeta p_2 -\bar z^\prime
p_1)^2+i\epsilon ]$, cf. \re{prop3}. Although at the end we have to isolate 
the quark propagator pole contribution, it is convenient to rewrite
the gauge pole as 
\beq{gaugepole}
\frac{1}{[x_2+i\epsilon]}=\frac{1}{[x_2-i\epsilon]}-(2\pi i)\delta (x_2)
\eeq
and find the corresponding two contributions to \re{M3} separately  
\beq{M3ab}
{\cal M}_{3}={\cal M}_{3}^a+{\cal M}_{3}^b 
\eeq
as an intermediate step.

Calculation of the first term, ${\cal M}_{3}^a$, is straightforward 
since poles of the
quark propagator and $1/[x_2-i\epsilon]$ do not pinch. One can neglect 
$k_\perp $ and $\alpha$ in the integrand of  (\ref{M3}) everywhere except
in ${\cal R}_{\mu_1\mu_2}^{\rm LC}$. The integral of 
${\cal R}_{\mu_1\mu_2}^{\rm LC}$  over $\alpha $ and 
$k_\perp$ reproduces the generalized gluon distribution, cf. (\ref{fa}),
and gets factored out. The remainder gives the contribution to the 
coefficient function (in the light-cone gauge)  
\beq{T3a}
T^a_{3}=C_F\frac{z^2\bar z^2\zeta^2 
[\zeta (2z^\prime -z)-x_1]}{z^\prime \bar z^\prime 
x_1[x_2-i\epsilon][(z^\prime -z)x_1-\zeta z^\prime
\bar z+i\epsilon](x_1-\zeta \bar z)} \, .
\eeq 
The factor $(x_1-\zeta \bar z)$ in the denominator of the above expression
originates from the hard gluon exchange propagator in the light cone gauge,
the factor $N_{\nu_1\nu_2}(k+z^\prime p_1-q_1)$ in Eq.~(\ref{M3}).        

In the second term, ${\cal M}_{3}^b$, we use the $\delta$-function
to put $x_1\to \zeta$ and need to expand the 
quark propagators in Eq.~(\ref{M3}) in the limit $k_\perp\to 0$ and 
$\alpha\sim k^2_\perp/s$. To the required accuracy
\bea{M3b}
{\cal M}_{3}^b&=&\frac{ 4\pi^2 \alpha_s^2
 s\, i f_\pi}{ N_c^2 q^4_\perp}\sqrt{1-\zeta}\,
\bar u(q_1)\gamma_5 \frac{ \!\not\! p_2}{ s}
v(q_2)\, \delta_{i j}\int\limits^1_0 dz^\prime
\, \phi_{\pi}(z^\prime)\! \\
&& {}
\times \frac{C_F}{2\pi }
\int \frac{dx_1 d(\alpha s) dk_\perp^2  \delta(x_2)(2-\zeta )}{[(\alpha
s)\zeta -k^2_\perp +i\epsilon][(\alpha
s)(\zeta -1) -k^2_\perp +i\epsilon]}
\left(
\frac{C_F \bar z^3}{\bar z^{\prime 2}}+{\cal O}(k^2_\perp) \right ) \, .  
\nonumber
\eea 
The integration over $\alpha$ produces the expression 
which can be identified with the perturbative leading-order generalized
gluon distribution of a quark at the point 
${\cal F}^{g}_\zeta (x_1=\zeta)$, 
cf. Eq.~(\ref{g-skewed}). The corresponding coefficient function takes the form
\beq{T3b}
T^b_{3}=(2\pi i)\delta (x_2)C_F\frac{\bar z^3}{\bar z^{\prime 2}} \,.
\eeq
As the final step, we go from $1/[x_2-i\epsilon]$ in eq. (\ref{T3a}) 
to $1/[x_2+i\epsilon]$ using (\ref{gaugepole}). The result equals  
the contribution of this diagram in the `naive' light-cone limit,
$T_3^{\rm LC}$, and the extra terms $\sim \delta(x_2)$ correspond to
the `pole' contribution $T_3^{\rm pole}$ in the sense 
of Eq.~\re{naive+pole}. We obtain   
\beq{T3-LC}
{} T_3^{\rm LC}=C_F\frac{z^2\bar z^2\zeta^2
[\zeta (2z^\prime -z)-x_1]}{z^\prime \bar z^\prime
x_1[x_2+i\epsilon][(z^\prime -z)x_1-\zeta z^\prime
\bar z+i\epsilon](x_1-\zeta \bar z)}\, ,
\eeq 
and
\beq{T3-pole}                                                       
T_{3}^{\rm pole}=(2\pi i)\delta(x_2)        
C_F\frac{\bar z^2(1+z)}{z^\prime \bar z^{\prime }}\,.
\eeq       
Note that in difference to the expressions collected in \re{feyresults} 
the result in \re{T3-LC} is obtained in the light-cone gauge.

Calculation of the other diagrams is similar. We obtain
\bea{Tpole1-5}                        
T_{1}^{\rm pole}&=&(2\pi i)\delta(x_2)                       
C_F\frac{\bar z z^2}{z^\prime \bar z^{\prime }}\,,
\nonumber\\                        
T_{2}^{\rm pole}&=&-(2\pi i)\delta(x_2)                       
\frac{1}{2N_c}\frac{\bar z z^2(2z^\prime-1)}{z^\prime \bar z^{\prime }}
-\delta(x_2)\frac{1}{2N_c}\eta \frac{\bar z z^2}{z^\prime} \,,
\nonumber\\                                              
T_{4}^{\rm pole}&=&-(2\pi i)\delta(x_2)                          
\frac{1}{2N_c}\frac{z \bar z^2 (3-2z^\prime)}{\bar z^{\prime 2}}
+\delta(x_2)\frac{1}{2N_c}\eta \frac{z \bar z^2}{\bar z^\prime} \,,
\nonumber\\                                                            
T_{5}^{\rm pole}&=&(2\pi i)\delta(x_2) \frac{N_c}{2}
\frac{\bar z(z^\prime -z^2-1)}{z^\prime \bar z^{\prime 2}}                 
+\delta(x_2)\frac{N_c}{2}\eta \frac{z \bar z(2z-1)}{2z^\prime 
\bar z^\prime} \,,
\eea   
and
\bea{Tpole6-10}
T_{6}^{\rm pole}(z,z^\prime)&=&T_{1}^{\rm pole}(\bar z,\bar z^\prime)\,,  
\quad T_{7}^{\rm pole}(z,z^\prime)=T_{2}^{\rm pole}(\bar z,\bar z^\prime) \,,
\quad T_{8}^{\rm pole}(z,z^\prime)=T_{3}^{\rm pole}(\bar z,\bar z^\prime) \,,
\nonumber \\
\quad T_{9}^{\rm pole}(z,z^\prime)&=&T_{4}^{\rm pole}
 (\bar z,\bar z^\prime) \,,
\quad T_{10}^{\rm pole}(z,z^\prime)=T_{5}^{\rm pole}
 (\bar z,\bar z^\prime) \,.
\eea
In these expressions
\beq{eta}
\eta= \frac{2\zeta}{(2-\zeta )}
\int \frac{d(\alpha s) (k^2_\perp +2(\alpha s)(1-\zeta))}{[(\alpha
s)\zeta -k^2_\perp +i\epsilon][(\alpha
s)(\zeta -1) -k^2_\perp +i\epsilon]} \, .
\eeq
The corresponding contributions originate from the last term in Eq.~(\ref{w})
and at first sight are dangerous because an 
additional factor $\alpha$ in the numerator of (\ref{eta}) 
makes the integral divergent at $\alpha\to \infty$. This difficulty is,
however, spurious since the terms $\sim \eta$ actually cancel in the sum 
of all ten diagrams. For the sum of pole terms we obtain
\beq{Tpole}
T^{\rm pole }=\sum\limits_{i=1,\dots , 10} T^{\rm pole}_i 
 =-(2\pi i)\delta(x_2)\left[C_F
\frac{z\bar z}{z'\bar z'}
\left(\frac{\bar z}{z'}+\frac{z}{\bar z'}\right)
+\frac{1}{2N_c\,z' \bar z'}\left(\frac{z\bar z}{z'\bar z'}+1\right)
\right] \,, 
\eeq 
and the final answer for the gluon coefficient function \re{G-result}
can be written as 
\bea{G-coefffun}
T_{\rm gluon} (z,z^\prime, x_1,x_2)&=&C_F
\left(
\frac{\bar z}{z^\prime}+\frac{z}{\bar z^\prime}
\right)
\left(
\frac{\zeta}{[x_1-i\epsilon]^2}+
\frac{\zeta}{[x_2+i\epsilon]^2} -
\frac{\zeta}{[x_1-i\epsilon][x_2+i\epsilon ]}
\right) \nonumber \\
&&{}
+
\left(
\frac{z\bar z}{z'\bar z'}+1
\right)
\left[
C_F
\left(
\frac{z\bar z}{z'\bar z'}+1
\right)
+\frac{1}{2N_c}
\left(
\frac{z}{z'}
+\frac{\bar z}{\bar z'}
\right)    
\right]    
\nonumber \\
&&{}
\times
\left(
\frac{1}{[(z-z^\prime)x_1-z\bar z^\prime \zeta+i\epsilon ] }
+
\frac{1}{[(z^\prime -z)x_2-z\bar z^\prime \zeta+i\epsilon ]}
\right)    
  \\
&&{}
-
\left[
C_F
\frac{z\bar z}{z'\bar z'}
\left(\frac{\bar z}{z'}+\frac{z}{\bar z'}\right)
+\frac{1}{2N_c\,z' \bar z'}\left(\frac{z\bar z}{z'\bar z'}+1\right)
\right]    
\frac{\zeta}{[x_1+i\epsilon ][x_2-i\epsilon ]}  \ . \nonumber
\eea
We see that, as we already discussed before, in this expression there
are terms with different prescriptions for the poles at $x_1=0$, $x_2=0$. 
In particular, in comparison to the light-cone result in 
\re{G-coeff-naive}, there are different $\pm i\epsilon$
prescriptions in the last line.
 
The corresponding imaginary part is  
\bea{imagine}
 {\cal I} &\equiv& \int\limits_0^1\!dz'\, \phi_\pi(z', \mu^2)   
   \int\limits^1_0 \!dx_1 \,{\cal F}^{g}_\zeta (x_1)\,
    \Big[-\frac{1}{\pi} {\rm Im}\,T_{\rm gluon}(z,z^\prime, x_1,x_2)\Big]    
\nonumber \\ 
&=&
 \int\limits_0^1\!dz'\, \phi_\pi(z', \mu^2)
\left\{\left[C_F\left(\frac{z\bar z}{z'\bar z'}-1\right)
\left(\frac{\bar z}{z'}+\frac{z}{\bar z'}\right)
+\frac{1}{2N_c\,z' \bar z'}\left(\frac{z\bar z}{z'\bar z'}+1\right) 
\right]\,{\cal F}_\zeta^g(\zeta,\mu^2) \right. 
\nonumber \\
&&\left. +\left(\frac{z\bar z}{z'\bar z'}+1  \right)
\left[C_F\left(\frac{z\bar z}{z'\bar z'}+1\right)+\frac{1}{2N_c}\left(\frac{z}{z'}
+\frac{\bar z}{\bar z'} \right)\right] \right. \nonumber \\
&& \times\left[\frac{\Theta(z'-z)}{(z'-z)}\,{\cal F}_\zeta^g\left(\frac{\zeta\,z'\bar
z}{z'-z},\mu^2
\right) + \frac{\Theta(z-z')}{(z-z')}\,{\cal F}_\zeta^g\left(\frac{\zeta\,\bar z'
z}{z-z'},\mu^2
  \right)    \right]
\nonumber\\
&&\left. 
  + C_F \left(\frac{\bar z}{z'}+\frac{z}{\bar z'}\right)\frac{\zeta}{2}
\left[
{\cal F}^{g\prime}_\zeta(\zeta+0,\mu^2)+{\cal F}^{g\prime}_\zeta(\zeta-0,\mu^2)-
{\cal F}^{g\prime}_\zeta(0,\mu^2) 
\right]
   \right\}.    
\eea 
In the approximation  that only the (large) imaginary part of the amplitude 
is taken into account, the differential cross section summed over the 
polarizations and the color of quark jets is given by 
\beq{cross}
\frac{d\sigma_{\pi\to 2\,{\rm jets}}}{d q_\perp^2 dt dz} 
= \frac{\alpha_s^4 f_\pi^2 \pi^3}{8N_c^3 q_\perp^8} |{\cal I}|^2\,.
\label{crosection}
\end{equation}
The factorization scale $\mu^2$ has to be 
of order of the transverse momentum of the exchanged gluon.

The expressions in (\ref{G-coefffun}) and \re{imagine} 
present the main result of this paper. The expression for the 
imaginary part in \re{imagine} indeed reproduces the result in \re{I}
apart from the extra last term containing a derivative of the gluon 
distribution.
Using the explicit expression for the perturbative gluon distribution 
function of a quark in Eq.~\re{g-skewed} it is easy 
to check that the derivative term is of order ${\cal O}(\zeta)$ 
and this is the reason why it was absent in the first calculation. 

Note also that $T^{\rm pole}$ given in Eq.~\re{Tpole} is nothing 
else but the $M^2$-channel discontinuity ${\cal C}_{(b^{\prime\prime})}$ given
by the crossed diagram in Fig.~\ref{M2cut}: 
\beq{corres}
 D C_F \frac{\pi}{\alpha_s}\int\! dz^\prime\, 
 \phi_\pi(z^\prime) \int dx_1\, {\cal F}_\zeta^g(x_1)
\Big[-\frac{1}{\pi}{\rm Im}\, T^{\rm pole}\Big]\, = 2\, {\cal C}_{(b^{\prime\prime})}
 \simeq 2\, {\cal C}_{(b)}\,,
\eeq
cf. Eq.~\re{Mab}, 
up to ${\cal O}(\zeta)$ corrections. The imaginary part corresponding to
the `naive' light-cone limit is given therefore by the expression
$\tilde{\cal I}^{\rm LC} \sim 
{\cal C}_{(a)}- {\cal C}_{(b)}+{\cal C}_{(c)}+{\cal C}_{(d)}$, i.e. 
reversing the sign of the contribution in Fig.~\re{fig:3}b.    

\subsection{Discussion}

The leading end-point behavior of the coefficient function  
$T_{\rm gluon} (z,z^\prime, x_1,x_2)$ \re{G-coefffun} at $z'\to 0$ and 
$z'\to 1$ is given by the following expressions:
\bea{THend}
T_{\rm gluon}|_{z^\prime\to 0} &\to & -(2\pi i)\frac{z\bar z}{z^{\prime 2}}
\left(
C_F\bar z+\frac{1}{2N_c} 
\right)\left[ \delta (x_1)+\delta(x_2)\right] \, , \nonumber \\
T_{\rm gluon}|_{z^\prime\to 1} &\to & -(2\pi i)\frac{z\bar z}{\bar z^{\prime 2}}
\left(
C_F z+\frac{1}{2N_c}     
\right)\left[ \delta (x_1)+\delta(x_2)\right] \,. 
\eea
Since ${\cal F}_\zeta^g(x)$ vanishes at $x=0$ \cite{Rad96a}, only 
the terms $\sim \delta(x_2)$ contribute.
Note that the leading $1/z^{\prime 2}$, $1/\bar z^{\prime 2}$  
asymptotics is imaginary, it comes entirely from  
the term which we denoted $T^{\rm pole}$, see Eq. (\ref{Tpole}).
The contribution of $T^{\rm LC}$ is at most ${\cal O}(1/z^\prime)$, or
${\cal O}(1/\bar z^\prime)$, see also \cite{Che01}. 
Eqs.~(\ref{THend}) have to be compared with Eqs.~\re{d-limit},\re{au-limit}
for the quark contribution. Similar to the latter case, the collinear 
factorization is violated due to the pinching of $M^2$-channel and 
$s$-channel singularities in the Glauber region. 
Summing the both contributions in Eq.~(\ref{THend})
and using the symmetry of the
pion distribution amplitude we obtain \cite{BISS01}
\beq{end}
 {\cal  I}\Big|_{\rm end-points}
  =  \left(N_c+\frac{1}{N_c}\right)
 z \bar z \,  \int\limits^1_0 dz'\,
\frac{\phi_\pi(z',\mu^2)}{z'^2}{\cal F}_\zeta^g(\zeta,\mu^2) \,.
\eeq
Since $\phi_\pi(z')\sim z'$ at $z'\to 0$, the integral over $z'$
diverges logarithmically. 
Remarkably, the integral containing the pion distribution
amplitude does not involve any $z$-dependence.
Therefore, the longitudinal momentum distribution of the jets in the
nonfactorizable contribution is calculable and, as it turns out,
has the shape of the asymptotic pion distribution amplitude
$\phi_\pi^{\rm as}(z) = 6z\bar z$. The corresponding physical process
is the following. The limit $z'\to 1$
corresponds to a kinematics in which one quark carries the entire
momentum of the pion. The fast quark radiates a hard gluon
which carries the fraction $(1-z)$ of quark momentum.
This radiation is perturbative and
is described by the effective vertex \re{e-vertex} at $z'=1$.
In the final step the hard gluon transfers its entire longitudinal and
transverse momentum to the quark jet, and emits a soft antiquark which
interacts nonperturbatively with the target proton
and the pion remnant. It follows that the
divergent logarithm in $\int \phi_\pi(z')/z'^2$ is of the form
$\ln q_\perp^2/\mu_{\rm IR}^2$ where $\mu_{\rm IR}$ is related to the
average transverse momentum of the quarks inside the pion. It is possible
that
in the case of scattering from a heavy nucleus $\mu_{\rm IR}$ may grow as
$\sim A^{1/3}$ because of color filtering. On the other hand, it is 
likely that the modified factorization approach of \cite{Sudakov}
can be developed and applied to coherent diffraction as well.  
A detailed discussion of these possibilities goes beyond the tasks of 
this paper.
The other important integration region is the one when  
$\zeta \ll |z'- z| \ll 1$, i.e. 
when the longitudinal momentum fraction carried by the quark is close
(for high energies) to that of the quark jet in the final state.
The enhancement of this region comes from the (small) 
denominator $1/(z'-z)$ which  is present in the contributions in 
Fig.~\ref{fig:3}c,d with real gluon emission in the intermediate
state. The logarithmic integral $\int dz'/|z-z'| \sim \ln s$ is nothing but 
the usual energy logarithm that accompanies each extra gluon in the 
gluon ladder. Its appearance is due to the fact the the gluon 
in Fig.~\ref{fig:3}c,d can be emitted in a broad rapidity interval and
is not constrained to the pion fragmentation region.
To logarithmic accuracy we can simplify the integrand in \re{I}
by assuming $z'=z$ everywhere except for the small denominators and the 
argument of the gluon distribution, to get 
\beq{z=z'}
{\cal I}\Big|_{\zeta\ll |z'- z|\ll 1}
= 4N_c \,\phi_\pi(z)\,\int\limits^{1}_{z}
  \frac{dz'}{z'-z} {\cal F}_\zeta^g(\zeta\frac{z\bar z}{z'-z},q^2_{\perp})
\simeq 4N_c \,\phi_\pi(z)\!\int\limits_\zeta^1 
 \!\frac{dy}{y}\, {\cal F}_\zeta^g(y,q^2_{\perp})\,.
\eeq
For a flat gluon distribution ${\cal F}_\zeta^g(y) \sim {\rm const}$ at 
$y\to 0$,
and the integration gives $ {\rm const}\cdot \ln 1/\zeta $ which is the
above mentioned energy logarithm.
Note that the color factors combine to produce $C_A =N_c$ signaling that
the relevant Feynman diagrams in Fig.~\ref{fig:3}c,d are those with a
three-gluon coupling. Moreover, the factor $2N_c/y$ appearing in \re{z=z'}
can be interpreted as the relevant limit of the DGLAP gluon splitting function
\cite{Rad96}
\beq{NNN}
q_\perp^2 \frac{\partial}{\partial q_\perp^2}\,{\cal F}_\zeta^g(x=\zeta,q_\perp^2)
 = \frac{\alpha_s}{2\pi} \int\limits_{\zeta}^1 dy\,P_\zeta^{gg}(\zeta,y)\,
{\cal F}_\zeta^g(y,q_\perp^2)
 \simeq \frac{\alpha_s}{2\pi} \int\limits_{\zeta}^1 dy\,\frac{2N_c}{y}\,
{\cal F}_\zeta^g(y,q_\perp^2)\,.
\eeq
The quantity on the l.h.s. of \re{NNN} defines what can be called the
unintegrated generalized gluon distribution and the physical meaning of
Eqs.~\re{z=z'} and \re{NNN} is that in the region $z'\sim z$ hard
gluon exchange can be viewed as a large transverse momentum part of the
gluon distribution in the proton, cf. \cite{NSS99}.
This contribution is proportional to
the pion distribution amplitude $\phi_\pi (z,q^2_{\perp})$
and contains the enhancement factor $\ln 1/\zeta \sim \ln s/q_\perp^2$.

\section{Numerical Analysis} 
\setcounter{equation}{0}

%
\begin{figure}[t]
\centerline{\epsfxsize8.0cm\epsffile{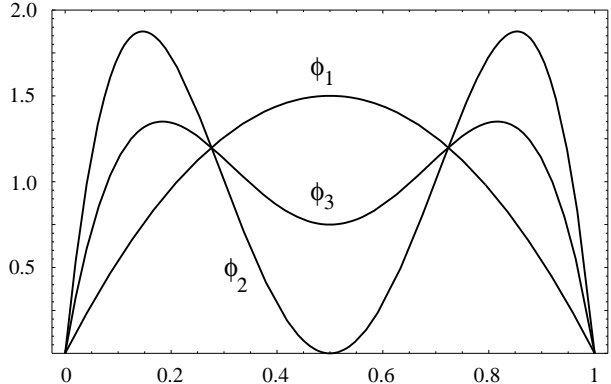}}
\caption[]{\small 
 Different models for the pion distribution amplitude, see Eq.~\re{pionmodel}.
 }
\label{wfmodels}
\end{figure}
%

Since collinear factorization does not hold, the numerical analysis
presented below has to be considered as semiquantitative. We will 
use the simplest prescription  to regularize the infrared divergence
of the end-point contribution by imposing an explicit cutoff 
on the quark momentum fraction in the pion
\beq{ircut}
\int\limits_0^1 dz^\prime \to 
 \int\limits_{\mu^2_{\rm IR}/q_\perp^2}^{1-\mu^2_{\rm IR}/q_\perp^2}
  \!\!dz^\prime 
\eeq 
with $\mu_{\rm IR} =500$~MeV. We use parametrizations of the generalized
quark and gluon parton distributions by Freund and McDermott \cite{GPDs}
that are based on the new MRST2001 leading-order forward distributions
\cite{MRST2001}, see Appendix A,  and the following trial 
pion distribution amplitudes:
\bea{pionmodel}
 \phi_1(z) &=& \phi_\pi^{\rm as}(z)  = 6 z(1-z)\,,\nonumber\\
 \phi_2(z) &=& \phi_\pi^{\rm CZ}(z,\mu = 0.5~\mbox{\rm GeV})  = 
   30 z(1-z)(1-2z)^2\,,\nonumber\\
\phi_3(z) &=& \phi_\pi^{\rm CZ}(z,\mu = 2~\mbox{\rm GeV})  = 
   15 z(1-z)[0.20+(1-2z)^2]\,,
\eea 
see Fig.~\ref{wfmodels}.
The first expression in \re{pionmodel} is the asymptotic distribution 
amplitude at large scales \cite{earlyBL,earlyER}, the second is the 
Chernyak-Zhitnitsky model \cite{CZ} and the third is the Chernyak-Zhitnitsky
model evolved to the typical scales $\mu \sim q_\perp = 2$~GeV
probed by  the E791 experiment \cite{E791a,E791b}. 
$\phi_1(z)$ and $\phi_2(z)$ have very different shape and can be considered 
as two extreme cases. Our aim is 
to find out whether this difference can give rise to sizeable effects
in the momentum fraction distribution of the jets.

Different contributions to imaginary and real parts of 
the amplitude of the hard dijet production at $s = 1000$~GeV$^2$ and 
$q_\perp=2$~GeV from the nucleon target for the two extreme choices 
of the pion distribution  amplitudes are plotted in 
Fig.~\ref{ASamplitude} and Fig.~\ref{CZamplitude}.
The normalization corresponds to Eqs.~\re{crossf}, \re{ampli}, see Appendix B
where we collect all the necessary expressions.
%
\begin{figure}[hbtp]
\epsfxsize8.0cm\epsffile{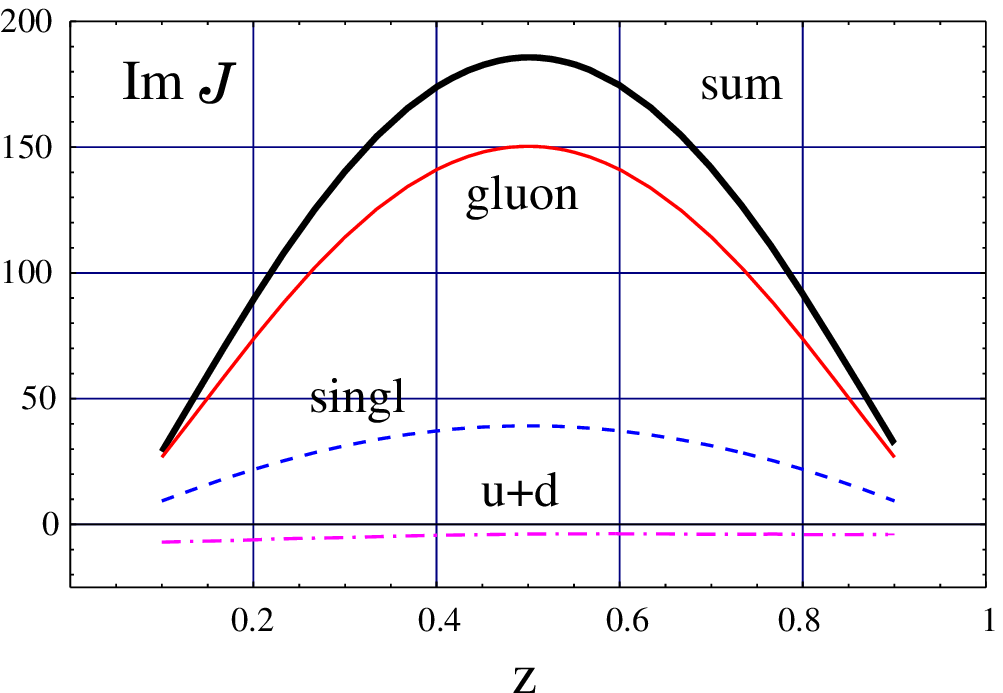}
\epsfxsize8.0cm\epsffile{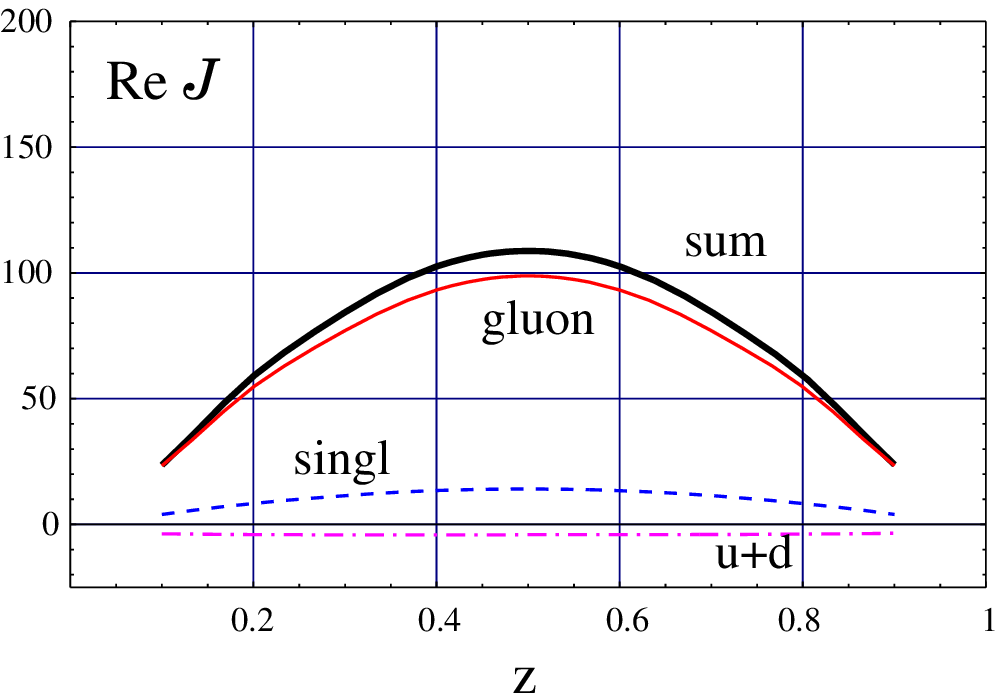}
\caption[]{\small 
Different contributions to imaginary (left) and real (right) parts of 
the amplitude of the hard dijet production at $s = 1000$~GeV$^2$ and 
$q_\perp=2$~GeV from the proton target, assuming the asymptotic 
pion distribution amplitude $\phi_\pi(z)=\phi_1(z)$, Eq.~\re{pionmodel}.
 The normalization corresponds to Eqs.~\re{crossf}, \re{ampli}.
Shown are: the gluon contribution (thin solid curve),
 the quark-singlet contribution (dashed curve) and 
the  $u$ and $d$-quark 
contributions (dash-dotted curve).  In addition,
the sum of all contributions is shown by the thick solid curve. 
 }
\label{ASamplitude}
\end{figure}
%
%
\begin{figure}[hbtp]
\epsfxsize8.0cm\epsffile{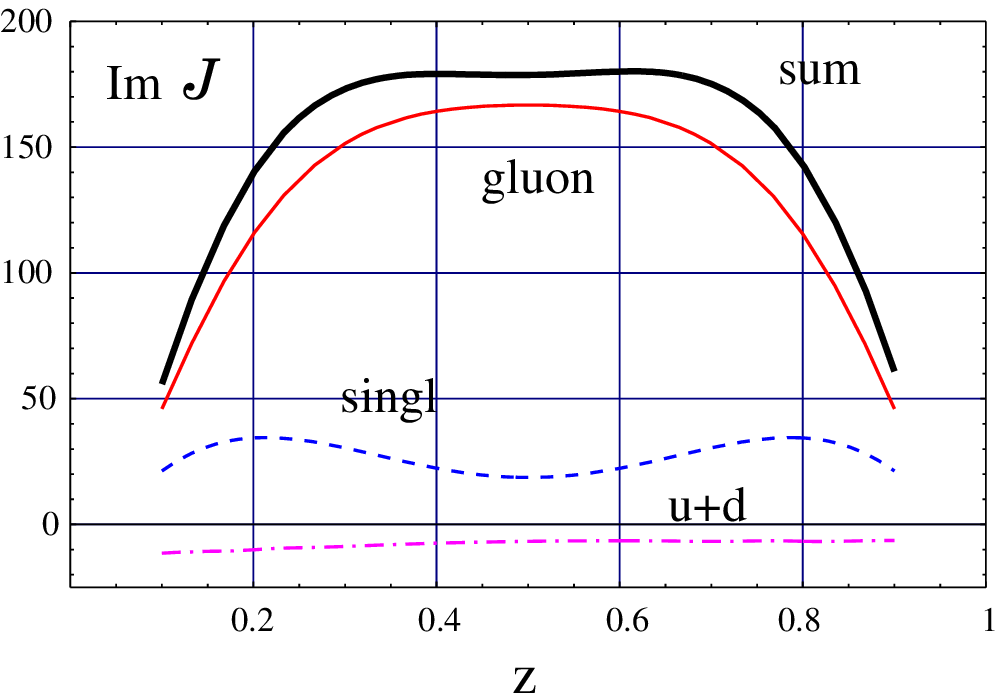}
\epsfxsize8.0cm\epsffile{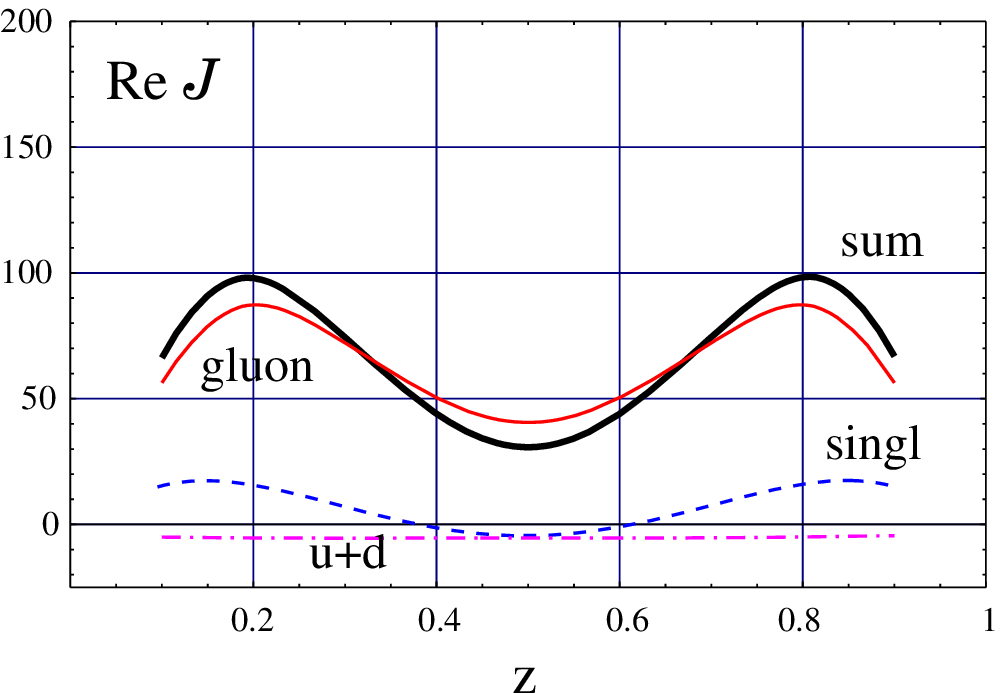}
\caption[]{\small 
The same as in Fig~\ref{ASamplitude}, but with the 
pion distribution amplitude $\phi_\pi(z)=\phi_2(z)$ \re{pionmodel} 
corresponding to the Chernyak-Zhitnitsky model at a low normalization point. 
 }
\label{CZamplitude}
\end{figure}
%
It is seen that both the imaginary and the real part of the amplitude
are dominated by the contribution of the generalized 
gluon distribution. The singlet quark contribution is less than 25\%, 
and the contributions of $u$-quark
annihilation and the $d$-quark exchange are negligible.
In \cite{CG01} the quark
contribution was found to be as important as the gluon one. 
One reason for the difference is that we use a more realistic parametrization 
for the quark distributions in our analysis. 
The second, more important reason, is
that our result for the imaginary part of the 
gluon contribution contains an 
additional non-factorizable term, the last line in Eq.~(\ref{imjgluon}), 
that is related with the discontinuity of the amplitude in the 
$M^2$-channel.
This additional contribution is large and constitutes
approximately one half of the total result for Im$J_{\rm gluon}$.    
Note also that the momentum fraction dependence 
of Im$J_{\rm gluon}$ deviates considerably from the shape of the 
pion distribution amplitude. This means that the large-energy 
approximation in Eq.~(\ref{z=z'}), where $\phi_\pi (z)$ enters as an
overall factor, is not adequate in the energy range of the E791 experiment. 

The real part of the amplitude is significant. 
The ratio Re/Im is not much smaller than unity  and 
depends both on the assumed shape of the pion distribution amplitude 
and the jet momentum fraction.    
Qualitatively, the Re/Im ratio can be estimated using the Gribov-Migdal
formula \cite{GribovMigdal}
\beq{GrMi}
\frac{{\rm Re}\, {\cal M}(s)}{ {\rm Im}\, {\cal M}(s)} =
\frac{\pi}{2}\frac{d}{d\ln s}\ln\frac{{\rm Im} {\cal M}(s)}{s}\,.
\eeq 
In our case, the amplitude is roughly proportional to the value of the 
generalized gluon distribution at the point $x_1=\zeta$, 
${\cal M} \sim \zeta^{-1}{\cal F}^g_\zeta(\zeta)$, and in the 
relevant kinematic range
\beq{porno}
   {\cal F}^g_\zeta(\zeta) \simeq 0.52\, \zeta^{-0.43}\,,
\eeq
cf. Appendix~A. This gives ${\rm Re} {\cal M} /{\rm Im} {\cal M} \sim
 0.43\cdot \pi/2 \simeq 2/3$, in rough agreement with
Figs.~\ref{ASamplitude},\ref{CZamplitude}. 
The large real part is a yet another
indication that the energy of E791 experiment is not
sufficiently high to consider the process as a diffractive one,
mediated by the pomeron exchange.

The discussion in this paper has so far tacitly assumed a nucleon target. 
For nuclei,
one extra effect to be taken into account is the longitudinal momentum 
transfer that is cut off by the nuclear form factor:  
\beq{nucleus}
\frac{d\sigma_{\pi A\to q \bar q A}}{d q_\perp^2 dz} \sim
   \frac{d\sigma_{\pi N \to q \bar q N}}{d q_\perp^2 dz}
 \cdot F^2_A[m^2 \zeta^2]\,, 
\eeq
where $\zeta = q_\perp^2/z\bar z s$, cf. \re{zeta},  
$F_A[t] = \exp[-1/6 R_A^2 t]$ and $R_A$ is the mean square radius of the 
nucleus. We use $R_A=5.27$~fm for platinum \cite{E791a}.
In principle, one also has to take into account that the calculated initial 
momentum fraction distributions 
at the quark level are modified by nonperturbative hadronization corrections
and effects due to the experimental acceptance. These effects lead to a 
certain broadening of the jet distributions compared the quark ones, 
as illustrated in Fig.~2 in Ref.~\cite{E791a}.
For simplicity, we ignore them in what follows. We also ignore a tiny 
QED contribution of Coulomb scattering, calculated in \cite{SI01}.

%
\begin{figure}[hbtp]
\epsfxsize14.0cm\epsffile{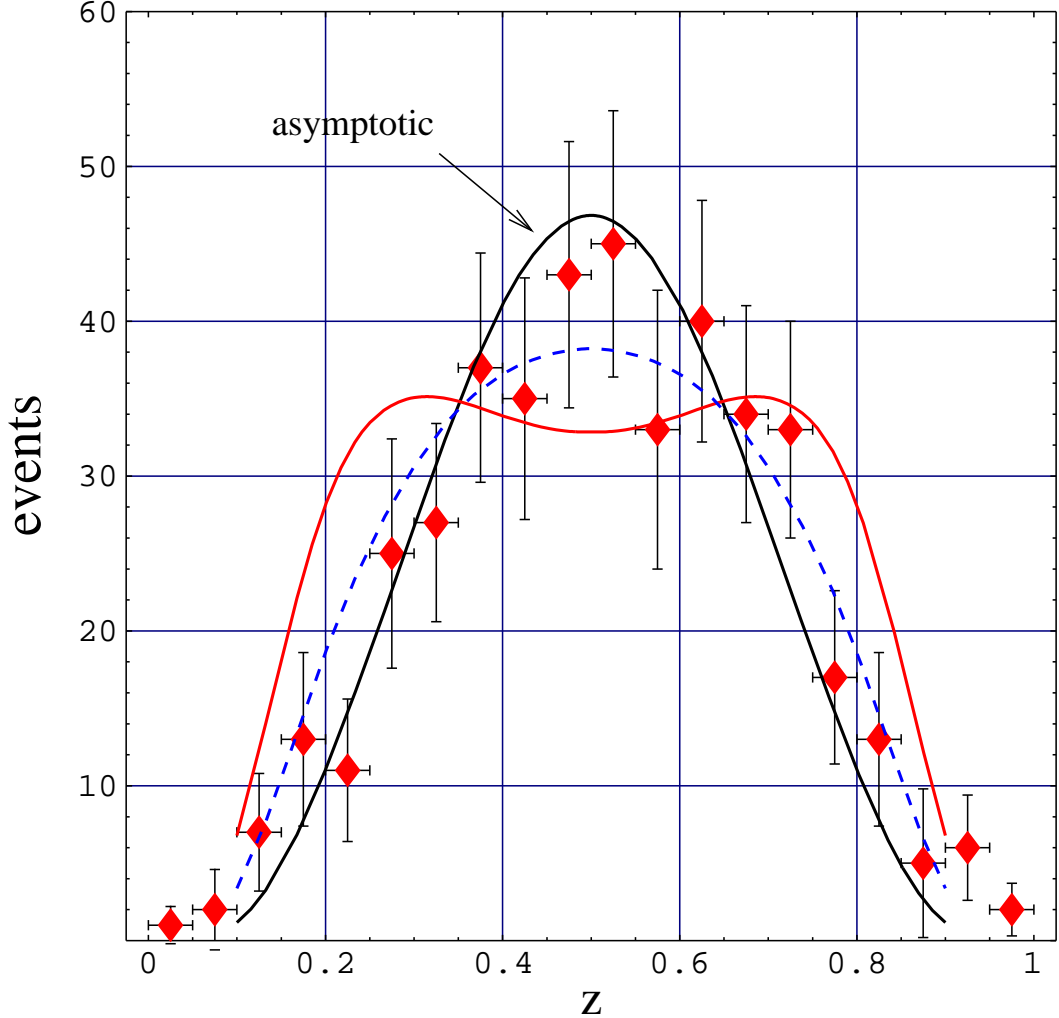}
\caption[]{\small 
The longitudinal momentum fraction distribution of the dijets 
with $1.5 \le q_\perp \le 2.5$~GeV from the platinum target \cite{E791a}.
The two solid curves show the calculations with the two extreme 
pion distribution
amplitudes in Eq.~\re{pionmodel} --- asymptotic  and ``two-humped'', 
respectively. The dashed curve corresponds to the Chernyak-Zhitnitsky 
model evolved to the scale $\mu = 2$~GeV. 
The overall normalization (the same for all curves) is arbitrary.  
 }
\label{E791}
\end{figure}
%

The comparison of the calculated dijet momentum fraction distribution 
with the data \cite{E791a} in the transverse momentum range 
$1.5 \le q_\perp \le 2.5$~GeV is presented in Fig.~\ref{E791}.
The two solid curves correspond to the first two 
choices of the pion distribution
amplitudes in Eq.~\re{pionmodel} --- asymptotic  and ``two-humped'', 
respectively. The dashed curve corresponds to the Chernyak-Zhitnitsky 
model evolved to the scale $\mu = 2$~GeV. 
The overall normalization is arbitrary, but is the same for all three 
choices of the distribution amplitude.  It is seen that experimental
uncertainties do not allow for the separation between the 
distribution amplitudes $\phi_1(z)$ and $\phi_3(z)$ while the extreme
choice $\phi_2(z)$ (see \re{pionmodel} and Fig.~\ref{wfmodels}) is 
not favored. This general conclusion is in agreement with the analysis 
in \cite{CG01}.

%
\begin{figure}[hbtp]
\epsfxsize5.1cm\epsffile{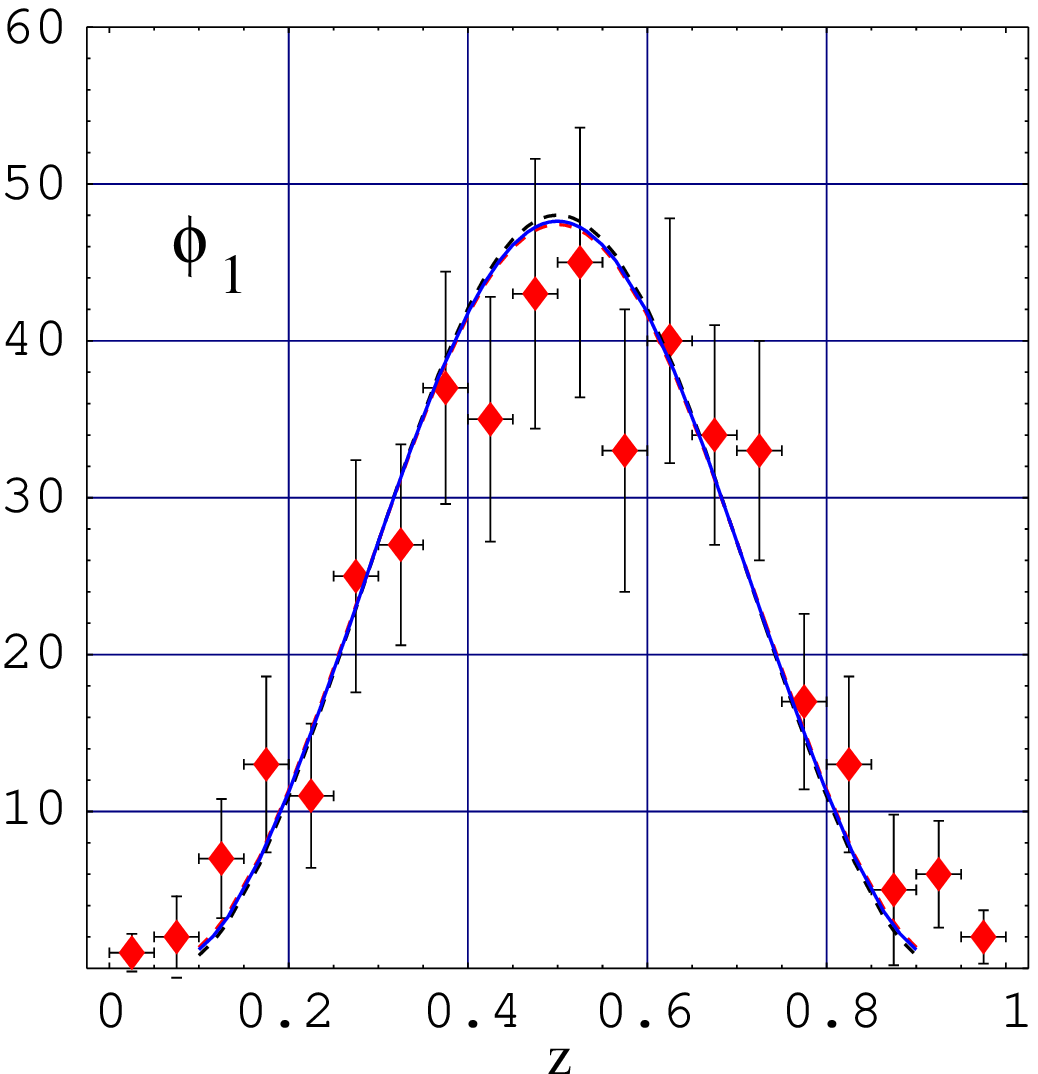}
\epsfxsize5.1cm\epsffile{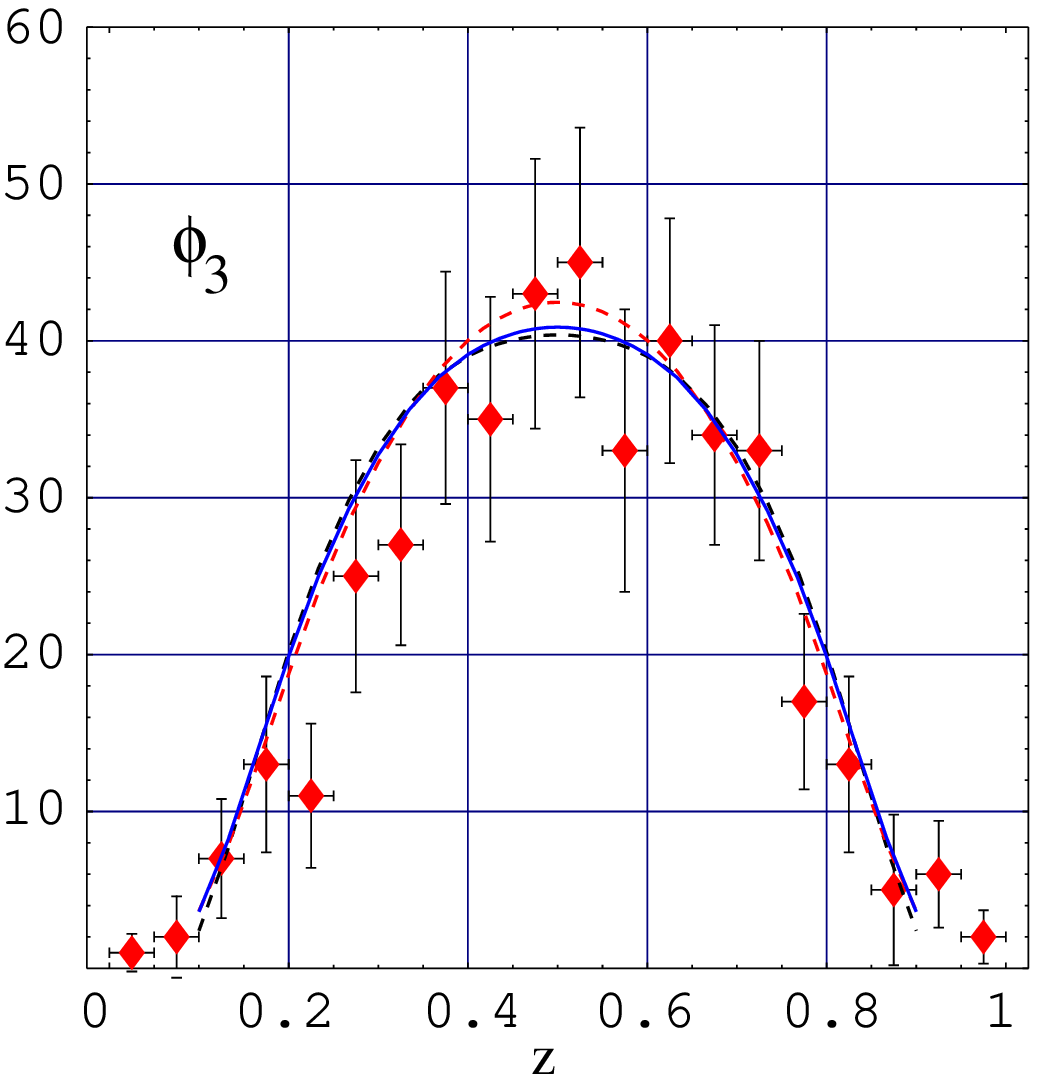}
\epsfxsize5.1cm\epsffile{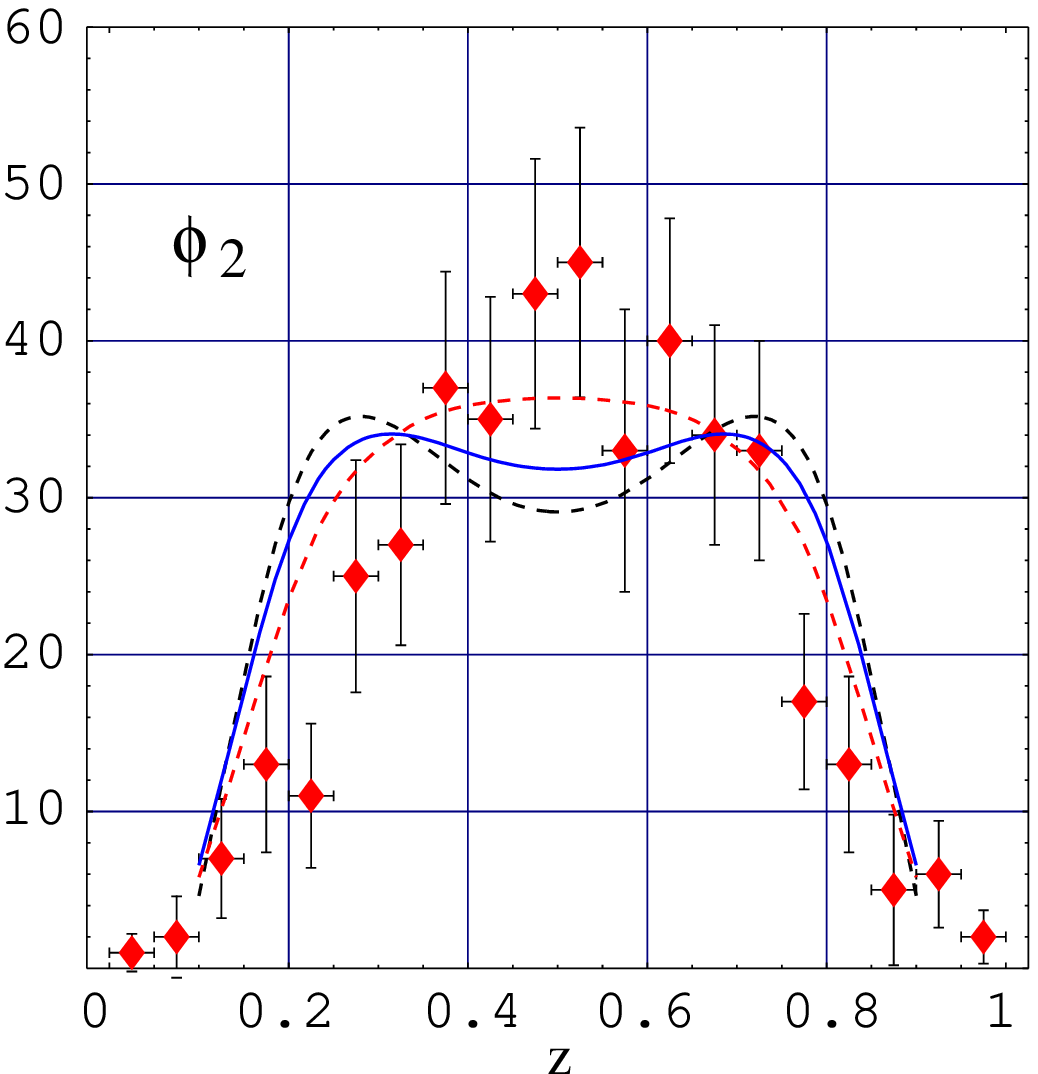}
\caption[]{\small 
Sensitivity of QCD predictions for the shape of the dijet momentum 
fraction distribution to the choice of  the infrared cutoff
for the three different pion distribution amplitudes in Eq.~\re{pionmodel}. 
The solid curves are calculated with 
$z^\prime_{\rm min}=0.0625$, the two dashed curves with 
$z^\prime_{\rm min}=0.0306$ and $z^\prime_{\rm min}=0.09$, 
respectively. The normalization is adjusted to give the same
integrated cross section in all cases. The data points are from 
\cite{E791a}.
 }
\label{obrez}
\end{figure}
%

Finally, one should not forget that the calculation involves a major 
uncertainty in the choice of the infrared cutoff parameter 
in the integration over the quark momentum fraction in the pion, 
Eq.~\re{ircut}. Our choice $\mu_{\rm IR}=500$~MeV corresponds to 
$z^\prime_{\rm min}=1-z^\prime_{\rm max}=0.0625$. 
The sensitivity to the IR cutoff is illustrated in Fig.~\ref{obrez}
where in addition to our ``standard'' choice
(solid curves) we present calculations using $\mu_{\rm IR}=350$~MeV
($z^\prime_{\rm min}=0.0306$) and $\mu_{\rm IR}=600$~MeV
($z^\prime_{\rm min}=0.09$), shown by dashed curves, 
for the three trial pion distribution amplitudes. The normalization
of curves in Fig.~\ref{obrez} is adjusted to give the same integrated cross 
section in all cases.
We see that for the asymptotic pion distribution function $\phi_1(z)$
there is no effect at all (within the line thickness), the uncertainty
is small for $\phi_3(z)$, but more significant for $\phi_2(z)$.
It is worth while to mention that the well-known strong scale dependence of 
the gluon distribution at small $x$ (large energies) by itself leads to  
an effective suppression of the end-point
contribution as compared to the factorizable part of the cross section.
Indeed, the end-point contributions
effectively correspond to large transverse distances and hence involve
a (smaller) gluon distribution at low scales. This effect can be taken
into account using the momentum fraction dependent factorization scale 
$\mu_F^2 \sim q_\perp^2 (z^\prime\bar z^\prime)/(z\bar z)$, which  
may present an attractive alternative to the explicit cutoff. 
It is discussed in our letter \cite{BISS01} and yields results
that are qualitatively similar to the calculation presented above.

\section{Conclusions}
\setcounter{equation}{0}

We have presented a calculation  of the leading contribution to the 
cross section of pion diffraction dissociation into two jets with large 
transverse momenta, originating from a hard gluon exchange. Our main
result
is that collinear factorization is violated. In technical terms, the 
problem is due to pinching of singularities between soft gluon (and quark)
interactions in the initial and in the final state. We have given a
detailed analysis of this phenomenon using different techniques, and also 
explained how the structure of these singularities interferes with gauge 
prescriptions of the $t$-channel gluon propagators. Final expressions 
for the numerous contributions to the amplitude are collected in 
Appendix B, where in difference to the main text we use Ji's conventions 
\cite{Ji97a} for the generalized parton distributions. 
Using a realistic set of the generalized parton distributions \cite{GPDs}
we find that the coherent dijet production is dominated by the gluon
contribution. We also find that the real part of the amplitude is 
quite sizeable and cannot be neglected.     

It happens that the non-factorizable soft contribution to the dijet 
cross section imitates the shape of the asymptotic pion distribution 
amplitude. This implies that the dijet production is unlikely to 
yield fully quantitative constraints on the distribution amplitude. 
On the other hand, if one accepts that the pion distribution is 
close to its asymptotic form, as strongly suggested by the CLEO
measurement \cite{CLEO}
of the $\gamma\gamma^* \pi$ transition form factor, then it appears that 
the QCD calculation explains the E791 data on the momentum fraction
distributions of dijets very well.      
We find that the nonfactorizable contribution is suppressed 
compared to the leading contribution by a logarithm of energy so that
for very large energies, in  the 
double logarithmic approximation  $\ln q_\perp^2 \ln \zeta$,
collinear factorization is restored. 
This limit is achieved  for energies about
two  orders of magnitude larger than the energy in the E791 experiment. 
   
The calculation presented in this work does not address specific
nuclear effects, apart from the minor correction of the dijet 
distribution due to the nuclear form factor. For factorizable contributions
of small transverse distances the dijet longitudinal momentum distribution 
is not expected to exhibit any significant A-dependence. 
The place where nuclear effects can 
play a r\^ole is by making the infrared cutoff introduced in   
\re{ircut} A-dependent. Indeed, it is plausible to assume that 
$\mu^2_{\rm IR}$ rises with the atomic number, e.g.  $\sim A^{1/3}$,  and 
hence nonfactorizable contributions become numerically suppressed
due to color filtering of configurations with a large transverse size.
The corresponding calculation goes beyond the scope of this paper.  

A short comparison with other approaches is in order.
 Our calculation is 
close in spirit to \cite{Che01,CG01} although the conclusions are 
different. A detailed explanation of the differences is given in the text.
In short, the point is that we do not assume the light-cone dominance 
of the cross section from the beginning, but examine the light-cone limit
carefully and argue that the approximation used in \cite{Che01,CG01}
breaks down for Glauber gluons (and quarks). In the double 
logarithmic approximation our result in \re{z=z'} is similar to 
\cite{NSS99} obtained using different methods. We, therefore, 
agree with the interpretation suggested in \cite{NSS99} that in the 
true diffraction limit, for very large energies, the dijet production
can be considered as a probe of the hard component of the pomeron. 
We note, however, that this interpretation breaks down beyond the 
double logarithmic approximation and is not sufficient
for the energy region of the E791 experiment. Finally, we have to 
mention an approach to coherent diffraction suggested 
in \cite{FMS00} that attributes hard dijets to a hard component 
of the pion wave function as in  the original Brodsky-Lepage
approach \cite{earlyBL}. This technique is interesting since it is 
most directly related to the classical view on 
diffraction \cite{dif_classic}, but apparently complicated 
for the discussion of factorization. The general argumentation in
\cite{FMS00}
appears to be in contradiction with our explicit calculations.
To our point of view, the coherent states
formalism of Refs.~\cite{CCM86} can be useful in this context.

\section*{Acknowledgments}     

We are grateful to V.~ Chernyak for the correspondence related to 
the papers \cite{Che01,CG01} and to 
G.~Korchemsky, B.~Pire, A.~Radyushkin and O.~Teryaev for discussions. 
Our special thanks
are due to A.~Freund for providing us with the code for the 
generalized parton distributions. The work of D.I. was supported 
by the Alexander von Humboldt Foundation.

\appendix  
\renewcommand{\theequation}{\Alph{section}.\arabic{equation}}  

\section*{Appendices}  

\section{Generalized parton distributions in symmetric notation}  
\label{app:a}  
\setcounter{equation}{0}  
\setcounter{table}{0}  

In the current literature two different sets of the generalized 
(off-forward, skewed etc.) parton distributions are used, as
introduced in Refs.~\cite{Rad96a, Ji97a}, respectively. 
Our discussion in the text assumed the definitions 
Eqs.~\re{quark-distr} and \re{naiveFg} that correspond to the 
conventions of \cite{Rad96a}. The transition to the notations 
of \cite{Ji97a} is straightforward and corresponds to the redefinition 
\bea{R-Ji}
(1+\xi ){\cal H}_q(y,\xi)&=&{\cal F}_{\zeta}^q(x)-{\cal F}_{\zeta}^{\bar q}
(\zeta-x)\Theta
(-1+\zeta\leq x\leq \zeta) \, ,
\nonumber \\
(1+\xi ){\cal H}_g(y,\xi)&=&{\cal F}_{\zeta}^g(x)+{\cal F}_{\zeta}^g(\zeta-x)\Theta
(-1+\zeta\leq x\leq \zeta) \,.
\eea
The two sets of variables $(x,\zeta)$ and $(y,\xi)$ are related according to 
\bea{R-Ji-var}
&&{} y=\frac{x-\zeta/2}{1-\zeta/2}\, , 
\quad \xi=\frac{\zeta/2}{1-\zeta/2} \, ;
\nonumber \\
&&{}
x=\frac{y+\xi}{1+\xi}\, , \ \ \,\,  \quad \zeta=\frac{2\xi}{1+\xi} \, .
\eea
$y$ varies in the interval $-1\leq y\leq 1$.

%
\begin{figure}[t]
\epsfxsize7.6cm\epsffile{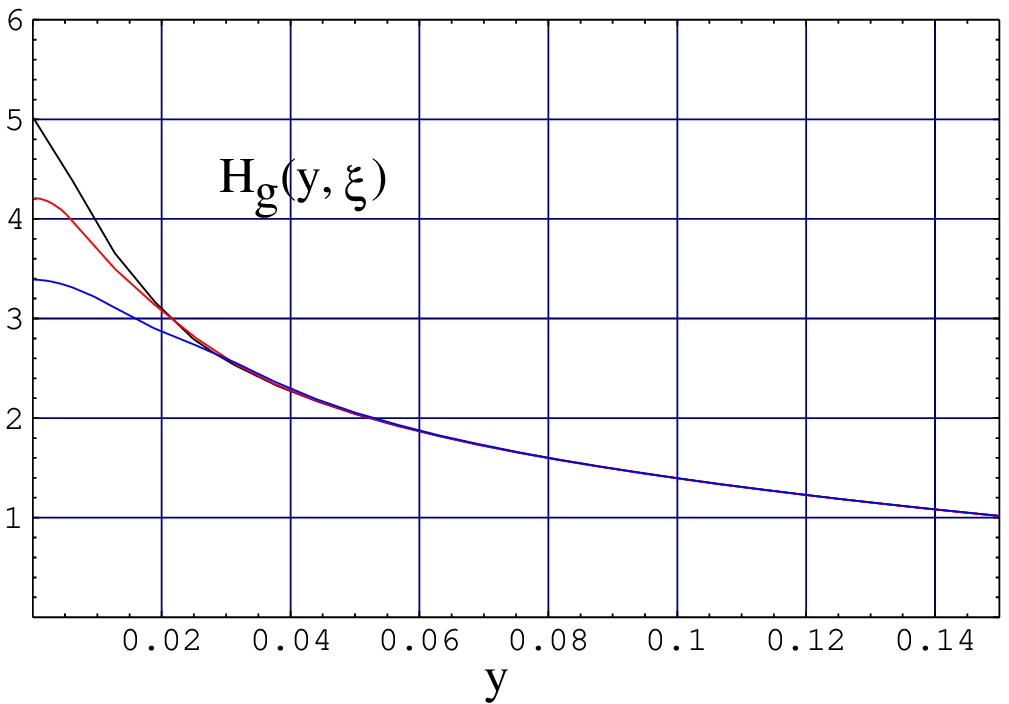}\quad
\epsfxsize8.0cm\epsffile{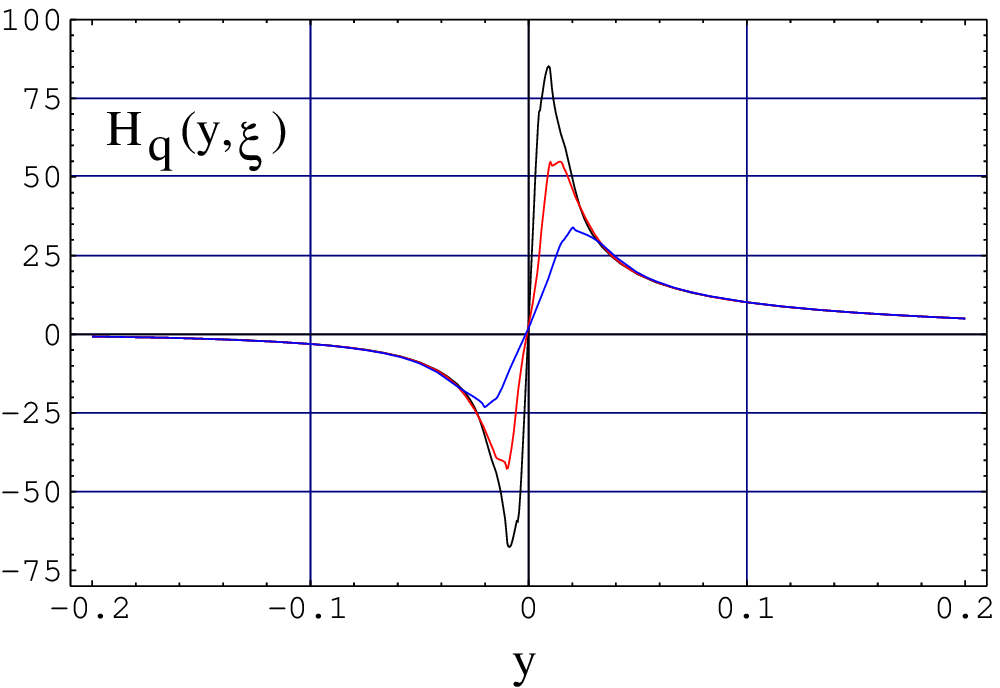}
\caption[]{\small
Gluon (left) and flavor-singlet quark (right)    
generalized parton distributions, for three different values of the
asymmetry parameter $\xi=0.008$,
$\xi= 0.012$ and $\xi= 0.02$ (from top to bottom) at the 
scale $\mu_F=2$~GeV, \cite{GPDs}, based on the LO 
MRST2001 parametrization \cite{MRST2001}.
 }
\label{PDFs}
\end{figure}
%

Rewriting of our results to the notations of \cite{Ji97a}
is straightforward.    
Using the crossing symmetry of the gluon coefficient function 
$T_{\rm gluon}(x)=T_{\rm gluon}(\zeta -x)$, one can easily verify that 
\beq{gl-Ji-R}
{\cal M}_{\rm gluon}\sim \int\limits^1_0 \!dx \,{\cal F}^g_\zeta (x)\,T_{\rm
gluon}(x)=\frac{1}{2}\int\limits^1_{-1}\! dy \,{\cal H}_g(y,\xi)\, T_{\rm
gluon}\left( \frac{y+\xi}{1+\xi} 
\right) \, .
\eeq
Note that the gluon distribution ${\cal H}_g(y,\xi)$ is a symmetric function,
${\cal H}_g(y,\xi)={\cal H}_g(-y,\xi)$. 

In the quark case we have 
\beq{q-Ji-R}
{\cal M}_{\rm quark}\sim \int\limits^1_0 dx\, \left[
{\cal F}^q_\zeta (x)T_{q}(x)
+{\cal F}^{\bar q}_\zeta (x)T_{\bar q}(x)
\right]
=\int\limits^1_{-1} dy \,{\cal H}_q(y,\xi) \,T_{q}\left( \frac{y+\xi}{1+\xi}
\right) \, ,
\eeq
since quark and antiquark coefficient functions are related,
$T_{q}(x)=-T_{\bar q}(\zeta -x)$, cf. \re{rules}.

In the numerical calculations in this paper we have used 
the set of leading-order parton distributions 
by Freund and McDermott based on the LO 
MRST2001 parametrization \cite{MRST2001}, see Fig.~\ref{PDFs}.

\section{Summary of Results}

Here we collect our final results. The hard coherent dijet production 
differential cross section is equal to 
\beq{crossf}
\frac{d\sigma_{\pi\to q\bar q}}{d q_\perp^2 dt dz}{\Bigg|}_{t=0} 
= \frac{\alpha_s^4 f_\pi^2 \pi^3}{8N_c^3 q_\perp^8}\, (1+\xi)^2\, 
|{\cal J}|^2\, ,
\end{equation}
where we separate the quark and gluon contributions
\beq{ampli}
{\cal J}=-\frac{1}{\pi}\left(
J_{\rm gluon}+ J_{\rm quark}
\right)\,.
\eeq
The gluon contribution reads 
\bea{jgluon}
J_{\rm gluon}&=&\frac{1}{2}\int\limits^1_{-1}\! dy \,{\cal H}_g(y,\xi)
\int\limits^1_0\! dz^\prime\,\phi_\pi (z^\prime)
\Bigg\{
C_F\left( \frac{\bar z}{z^\prime}+\frac{z}{\bar z^\prime}\right)
\left( \frac{2\xi}{(y+\xi-i\epsilon)^2}+\frac{2\xi}{(y-\xi +i\epsilon )^2}
\right) \nonumber \\
&&{}
+\left(\frac{z\bar z}{z^\prime\bar z^\prime}+1 \right)\left[
\left(C_F\left(\frac{z\bar z}{z^\prime\bar z^\prime}+1 \right)+
\frac{1}{2N_c}\left(\frac{z}{z^\prime}+\frac{\bar z}{\bar z^\prime} 
\right)\right)
\right. \nonumber \\
&&{}\times
\left(
\frac{1}{y(z-z^\prime)-\xi (z\bar z^\prime+z^\prime \bar z)+i\epsilon}+
\frac{1}{y(z^\prime-z)-\xi (z\bar z^\prime+z^\prime \bar z)+i\epsilon}
\right) \nonumber \\
&&{}
\left.
-\left(C_F\left( \frac{\bar z}{z^\prime}+\frac{z}{\bar z^\prime}\right)
+\frac{1}{2N_cz^\prime\bar z^\prime}\right)
\frac{2\xi}{(y+\xi-i\epsilon)(y-\xi+i\epsilon)}
\right] \nonumber \\
&&{}
-(2\pi i)\!\left( 
C_F\frac{z\bar z}{z^\prime\bar z^\prime }
\left( \frac{\bar z}{z^\prime}+\frac{z}{\bar z^\prime}\right)\!
+\frac{1}{2N_cz^\prime\bar z^\prime}\left(\frac{z\bar z}{z^\prime\bar
z^\prime}+1 \right)\!
\right)
\left(\delta (y\!+\!\xi)+\delta(y\!-\!\xi)\right)
\Bigg\}.
\eea
The contribution in the last line of \re{jgluon} corresponds to the 
pole contribution discussed in the text; the rest of the terms correspond
to the naive light-cone limit $\xi\to \xi-i\epsilon$. The corresponding
imaginary and real parts are  
\bea{imjgluon}
-\frac{1}{\pi}{\rm Im} \left(J_{\rm gluon}\right)&=&
\int\limits^1_0dz^\prime\, \phi_\pi(z^\prime)\Bigg[
C_F\left( \frac{\bar z}{z^\prime}+\frac{z}{\bar z^\prime}\right)
\,2\,\xi\, {\cal H}^\prime_g(\xi,\xi) \nonumber \\
&&{}
+\left(\frac{z\bar z}{z^\prime\bar z^\prime}+1 \right)
\left\{
\left(C_F\left(\frac{z\bar z}{z^\prime\bar z^\prime}+1 \right)+
\frac{1}{2N_c}\left(\frac{z}{z^\prime}+\frac{\bar z}{\bar z^\prime}
\right)\right)\frac{{\cal H}_g(\frac{\xi(z\bar z^\prime+z^\prime \bar
z)}{|z^\prime-z|},\xi)}{|z^\prime-z|}
\right. \nonumber \\
&&{}
\left.
-\left(C_F\left( \frac{\bar z}{z^\prime}+\frac{z}{\bar z^\prime}\right)
+\frac{1}{2N_cz^\prime\bar z^\prime}\right){\cal H}_g(\xi,\xi)
\right\} \nonumber \\
&&{}
+2\left(
C_F\frac{z\bar z}{z^\prime\bar z^\prime }         
\left( \frac{\bar z}{z^\prime}+\frac{z}{\bar z^\prime}\right)
+\frac{1}{2N_cz^\prime\bar z^\prime}\left(\frac{z\bar z}{z^\prime\bar
z^\prime}+1 \right)
\right){\cal H}_g(\xi,\xi)        
\Bigg].
\eea
and
\bea{rejgluon}
{\rm Re} \left(J_{\rm gluon}\right)&=&\int\limits^1_0 dz^\prime 
\int\limits^1_{0} dy\, \phi_\pi (z^\prime)\left\{
C_F\left( \frac{\bar z}{z^\prime}+\frac{z}{\bar z^\prime}\right)
\left( \frac{2\xi {\cal H}_g(y,\xi)}{(y+\xi)^2} \right.\right.
\nonumber \\
&&{}
\left.\hspace*{1cm}
+
\frac{2\xi({\cal H}_g(y,\xi)-{\cal H}_g(\xi,\xi)-(y-\xi)H^\prime(\xi,\xi))}{(y-\xi)^2}
\right)
\nonumber\\
&&{}
+\left(\frac{z\bar z}{z^\prime\bar z^\prime}+1 \right)\left[
\left(C_F\left(\frac{z\bar z}{z^\prime\bar z^\prime}+1 \right)+
\frac{1}{2N_c}\left(\frac{z}{z^\prime}+\frac{\bar z}{\bar z^\prime} 
\right)\right)
\right.
\nonumber \\
&&{}\hspace*{1cm}
\times 2\xi(z\bar z^\prime+z^\prime\bar z)
\frac{
{\cal H}_g(y,\xi)-{\cal H}_g(\frac{\xi(z\bar z^\prime+z^\prime\bar z)}
{|z^\prime-z|},\xi)}
{y^2(z^\prime-z)^2-\xi^2(z\bar z^\prime+z^\prime\bar z)^2}
\nonumber \\
&&{}\hspace*{1cm}
\left.\left.
-\left(C_F\left( \frac{\bar z}{z^\prime}+\frac{z}{\bar z^\prime}\right)
+\frac{1}{2N_cz^\prime\bar z^\prime}\right)
\frac{2\xi\left({\cal H}_g(y,\xi)-{\cal H}_g(\xi,\xi) \right)}
{y^2-\xi^2}
\right]\right\} \nonumber \\
&+&{}
\int\limits^1_0 dz^\prime\, \phi_\pi (z^\prime)
\left\{
C_F\left( \frac{\bar z}{z^\prime}+\frac{z}{\bar z^\prime}\right)
\left( 2\xi {\cal H}^\prime_g(\xi,\xi)\ln\left(\frac{1-\xi}{\xi}\right)
-\frac{2{\cal H}_g(\xi,\xi)}{1-\xi} 
\right)
\right.
\nonumber \\
&&{}
+\left(\frac{z\bar z}{z^\prime\bar z^\prime}+1 \right)\left[
\left(
C_F\left(    
\frac{z\bar z}{z^\prime\bar z^\prime}+1\right)+
\frac{1}{2N_c}
\left(
\frac{z}{z^\prime}+\frac{\bar z}{\bar z^\prime}
\right)
\right)
\right.
\nonumber \\
&&{}\hspace*{1cm}\times
\frac{
{\cal H}_g
\left(
\frac{\xi(z\bar z^\prime+z^\prime\bar z)}
{|z^\prime-z|},\xi
\right)
}
{|z^\prime-z|}
\ln\left(
\left|
\frac{|z^\prime-z|-\xi(z\bar z^\prime+z^\prime\bar z)}
{|z^\prime-z|+\xi(z\bar z^\prime+z^\prime\bar z)}
\right|
\right)
\nonumber \\
&&{}\hspace*{1cm}
\left.\left.
-\left(C_F\left( \frac{\bar z}{z^\prime}+\frac{z}{\bar z^\prime}\right)
+\frac{1}{2N_cz^\prime\bar z^\prime}\right)
{\cal H}_g(\xi,\xi)\ln\left(
\frac{1-\xi}{1+\xi}\right)
\right]\right\},
\eea
where ${\cal H}_g^\prime(\xi,\xi) = (d/dy){\cal H}_g(y,\xi)|_{y=\xi}$.
Although the gluon distribution is generally not an analytic function
at $y=\xi$, we have checked that the first derivative exists for the 
leading order perturbative distribution \re{g-skewed} and one can prove 
that it also exists in the formal $\mu_F\to \infty$ limit.  

For the quark contribution we distinguish between the flavor-singlet 
contribution mediated by the hard two-gluon exchange in the $t$-channel, 
and the separate contributions of $u$ and $d$ quarks: 
\beq{jquark}
J_{\rm quark}=\frac{2\xi z\bar z}{1+\xi}\left(
J_{\rm singl}+J_u+J_d\right) \,.
\eeq
We also introduce the flavor-singlet quark distribution
\beq{hqsingl}
{\cal H}_{q}(y,\xi)=\sum\limits_{p=u,d,s}{\cal H}_p(y,\xi)\,.
\eeq
For the flavor-singlet contribution we obtain 
\beq{jsingl}
J_{\rm singl}=\int\limits^1_{-1}dy\, {\cal H}_{q}(y,\xi) 
\int\limits^1_0 dz^\prime\,\phi_\pi (z^\prime)\, C_F\left(
\frac{z\bar z}{z^\prime \bar z^\prime}+1
\right)\frac{2y}{y^2(z-z^\prime )^2-\xi^2(z\bar z^\prime+z^\prime \bar
z)^2+i\epsilon} \ ,
\eeq
so that 
\bea{imjsingl}
-\frac{1}{\pi}{\rm Im}\left(J_{\rm singl}\right)
&=&\int\limits^1_0
dz^\prime \phi_\pi (z^\prime)\, C_F\left(
\frac{z\bar z}{z^\prime \bar z^\prime}+1
\right)\frac{1}{(z-z^\prime)^2}
\nonumber \\
&&{}\times
\left[
{\cal H}_q\left(\frac{\xi(z\bar z^\prime+z^\prime \bar z)}{|z^\prime -z|},\xi\right)
-{\cal H}_q\left(\frac{-\xi(z\bar z^\prime+z^\prime \bar z)}{|z^\prime
-z|},\xi\right)
\right]
\eea
and
\bea{rejsingl}
{\rm Re} \left(J_{\rm singl}\right)
&=&\int\limits^1_{-1}dy \int\limits^1_0
dz^\prime \phi_\pi (z^\prime) C_F\left(
\frac{z\bar z}{z^\prime \bar z^\prime}+1
\right)\frac{2y \left({\cal H}_q(y,\xi)
-{\cal H}_q\left(\frac{sign(y)\xi(z\bar z^\prime +z^\prime\bar z)}
{|z-z^\prime |},\xi\right)\right)}{y^2(z-z^\prime )^2-\xi^2(z\bar
z^\prime+z^\prime \bar
z)^2}
\nonumber \\
&&{}
+\int\limits^1_0 dz^\prime \phi_\pi(z^\prime)
C_F\left(
\frac{z\bar z}{z^\prime \bar z^\prime}+1
\right)
\ln\left( \left|
\frac{(z-z^\prime )^2 -\xi^2(z\bar z^\prime +z^\prime \bar z)^2}
{\xi^2(z\bar z^\prime +z^\prime \bar z)^2}
\right|\right)
\nonumber \\
&&{}\hspace*{1cm}\times
\left[\frac{
{\cal H}_q\left(\frac{\xi(z\bar z^\prime +z^\prime\bar z)}{|z-z^\prime |},
\xi\right)
-{\cal H}_q\left(-\frac{\xi(z\bar z^\prime +z^\prime\bar z)}{|z^\prime-z|},
\xi\right)
}{(z-z^\prime)^2}\right].
\eea
For the flavor-nonsinglet contributions we obtain
\bea{jquarku}
J_u&=&2C_F\int\limits^1_{-1}dy {\cal H}_u(y,\xi) \int\limits^1_0 dz^\prime 
\phi_\pi(z^\prime)\left[
C_F\left(\frac{z(2-z)}{\bar z\bar z^\prime(y+\xi-i\epsilon)}+
\frac{z}{z^\prime(y-\xi+i\epsilon)}\right)
\right. \nonumber \\
&&{}
+\frac{1}{2N_c}\left(
\frac{z}{\bar z\bar z^{\prime 2}(y+\xi-i\epsilon)}+
\frac{(z-\bar z^\prime)^3}{z^\prime \bar z^{\prime 2}\bar z
(y(\bar z^\prime -z)-\xi(z z^\prime +\bar z\bar z^\prime)+i\epsilon)}
\right. \nonumber \\
&&{}
\left.
+\frac{1}{\bar z(y(z^\prime -z)-\xi(z\bar z^\prime +z^\prime \bar z)+
i\epsilon)}+
\frac{z\bar z}{z^\prime \bar z^{\prime 2}
(y(z-z^\prime )-\xi(z\bar z^\prime +z^\prime \bar z)+
i\epsilon)} \right) \nonumber \\
&&{}
\left.
-(2\pi i)\left(
C_F\frac{z(2-z)}{\bar z\bar z^\prime}+\frac{1}{2N_c}
\frac{z}{\bar z\bar z^{\prime 2}}\right)\delta(y+\xi)
\right],
\eea
\bea{imjquarku}
-\frac{1}{\pi}{\rm Im} \left(J_u\right)
&=&2C_F\int\limits^1_0 dz^\prime 
\phi_\pi(z^\prime)\left[
C_F\left( \frac{z}{z^\prime}{\cal H}_u(\xi,\xi)-
\frac{z(2-z)}{\bar z\bar z^\prime}{\cal H}_u(-\xi,\xi)\right)
\right. \nonumber \\
&&{}
+\frac{1}{2N_c}\left(
\frac{(z-\bar z^\prime)^3}{z^\prime \bar z^{\prime 2}\bar z |\bar z^\prime -z|}
{\cal H}_u\left(\frac{\xi(zz^\prime+\bar z\bar z^\prime)}{1-z-z^\prime},\xi\right)
-
\frac{z}{\bar z\bar z^{\prime 2}}{\cal H}_u(-\xi,\xi)
\right. \nonumber \\
&&{}
\left.
+\frac{1}{\bar z|z^\prime -z|}{\cal H}_u\left(\frac{\xi(z\bar z^\prime+
z^\prime\bar
z)}{z^\prime-z},\xi\right)+
\frac{z\bar z}{z^\prime \bar z^{\prime 2}
|z^\prime -z|}{\cal H}_u\left(\frac{\xi(z\bar z^\prime+
z^\prime\bar
z)}{z-z^\prime},\xi\right) \right) \nonumber \\
&&{}
\left.
+2\left(
C_F\frac{z(2-z)}{\bar z\bar z^\prime}+\frac{1}{2N_c}
\frac{z}{\bar z\bar z^{\prime 2}}\right){\cal H}_u(-\xi,\xi)
\right]
\eea
and
\bea{jquarkd}
J_d&=&2C_F\int\limits^1_{-1}dy \int\limits^1_0 dz^\prime 
{\cal H}_d(y,\xi)\phi_\pi(z^\prime)\left[
C_F\left(\frac{\bar z(1+z)}{z z^\prime(y-\xi+i\epsilon)}+
\frac{\bar z}{\bar z^\prime(y+\xi-i\epsilon)}\right)
\right. \nonumber \\
&&{}
+\frac{1}{2N_c}\left(
\frac{\bar z}{z z^{\prime 2}(y-\xi+i\epsilon)}+
\frac{(z^\prime -\bar z)^3}{\bar z^\prime z^{\prime 2} z
(y(\bar z^\prime -z)-\xi(z z^\prime +\bar z\bar z^\prime)+i\epsilon)}
\right. \nonumber \\
&&{}
\left.
-\frac{1}{z(y(z^\prime -z)-\xi(z\bar z^\prime +z^\prime \bar z)+
i\epsilon)}-
\frac{z\bar z}{\bar z^\prime z^{\prime 2}
(y(z-z^\prime )-\xi(z\bar z^\prime +z^\prime \bar z)+
i\epsilon)} \right) \nonumber \\
&&{}
\left.
+(2\pi i)\left(
C_F\frac{\bar z(1+z)}{z z^\prime}+\frac{1}{2N_c}
\frac{\bar z}{z z^{\prime 2}}\right)\delta(y-\xi)
\right],
\eea
\bea{imjquarkd}
-\frac{1}{\pi}{\rm Im}(J_d)
&=&2C_F\int\limits^1_0 dz^\prime\, 
\phi_\pi(z^\prime)\left[
C_F\left( 
\frac{\bar z(1+z)}{z z^\prime}{\cal H}_d(\xi,\xi)-
\frac{\bar z}{\bar z^\prime}{\cal H}_d(-\xi,\xi)\right)
\right. \nonumber \\
&&{}
+\frac{1}{2N_c}\left(
\frac{(z^\prime -\bar z)^3}{\bar z^\prime z^{\prime 2} z |\bar z -z^\prime|}
{\cal H}_d\left(\frac{\xi(zz^\prime+\bar z\bar z^\prime)}{1-z-z^\prime},\xi\right)
+
\frac{\bar z}{z z^{\prime 2}}{\cal H}_d(\xi,\xi)
\right. \nonumber \\
&&{}
\left.
-\frac{1}{z|z^\prime -z|}{\cal H}_d\left(\frac{\xi(z\bar z^\prime+
z^\prime\bar
z)}{z^\prime-z},\xi\right)-
\frac{z\bar z}{\bar z^\prime z^{\prime 2}
|z^\prime -z|}{\cal H}_d\left(\frac{\xi(z\bar z^\prime+
z^\prime\bar
z)}{z-z^\prime},\xi\right) \right) \nonumber \\
&&{}
\left.
-2\left(
C_F\frac{\bar z(1+z)}{z z^\prime}+\frac{1}{2N_c}
\frac{\bar z}{z z^{\prime 2}}\right){\cal H}_d(\xi,\xi)
\right].
\eea
The expressions in the last lines of Eqs.~\re{jquarku}--\re{imjquarkd}
correspond to the pinch contributions, see text.
We do not present explicit expressions for the real parts of the 
$u$- and $d$-quark contributions because the corresponding contributions 
to the dijet cross section are very small.

\end{document}